\documentstyle{amlts}
\begin{document}
\annalsline{158}{2003}
\received{February 20, 2001}
\startingpage{509}
\def\bye{\end{document}}
 \font\tenrm=cmr10
\def\ritem#1{\item[{\rm #1}]}
\def\nhs{\hskip-8pt}
\catcode`\@=11
\font\twelvemsb=msbm10 scaled 1100
\font\tenmsb=msbm10
\font\ninemsb=msbm10 scaled 800
\newfam\msbfam
\textfont\msbfam=\twelvemsb  \scriptfont\msbfam=\ninemsb
  \scriptscriptfont\msbfam=\ninemsb
\def\msb@{\hexnumber@\msbfam}
\def\Bbb{\relax\ifmmode\let\next\Bbb@\else
 \def\next{\errmessage{Use \string\Bbb\space only in math
mode}}\fi\next}
\def\Bbb@#1{{\Bbb@@{#1}}}
\def\Bbb@@#1{\fam\msbfam#1}
\catcode`\@=12

 \catcode`\@=11
\font\twelveeuf=eufm10 scaled 1100
\font\teneuf=eufm10
\font\nineeuf=eufm7 scaled 1100
\newfam\euffam
\textfont\euffam=\twelveeuf  \scriptfont\euffam=\teneuf
  \scriptscriptfont\euffam=\nineeuf
\def\euf@{\hexnumber@\euffam}
\def\frak{\relax\ifmmode\let\next\frak@\else
 \def\next{\errmessage{Use \string\frak\space only in math
mode}}\fi\next}
\def\frak@#1{{\frak@@{#1}}}
\def\frak@@#1{\fam\euffam#1}
\catcode`\@=12

\renewcommand{\phi}{\varphi}
\newcommand{\const}{\mbox{\rm \small (Const.)}}
\newcommand{\iint}{\mathop{\displaystyle\int\!\!\int}}
\newcommand{\set}[2]{{\left\{#1 : #2\right\}}}
\newcommand{\downskip}[1]{\rule{0pt}{#1pt}}
\newcommand{\upskip}[1]{\rule[-#1pt]{0pt}{#1pt}}
\newcommand{\upchi}[1]{\raise1pt\hbox{$\chi$}_{#1}}
\newcommand{\thetar}{\theta_r}
\newcommand{\thetam}{\theta_{\scriptscriptstyle(0)}}
\newcommand{\thetaplus}{\theta_{\downskip{5}_+}}
\newcommand{\chiplus}[1]{\upchi{#1}^{\upskip{3}+}}
\newcommand{\chiminus}[1]{\upchi{#1}}
\newcommand{\uprho}{\raise1pt\hbox{$\rho$}}
\newcommand{\upgamma}{\raise1pt\hbox{$\gamma$}}
\newcommand{\I}{{\mathord{I}}}
\newcommand{\ea}{{{\cal E}^{\upskip{3}\hbox{\tiny A}}}}
\newcommand{\acn}{A_{\rm cn}}
\newcommand{\etf}[1]{{{\cal E}^{\upskip{3}\TF}_{#1}}}
\newcommand{\ehf}{{{\cal E}^{\upskip{3}\HF}}}
\newcommand{\HF}{{\hbox{\rm \tiny HF}}}
\newcommand{\TF}{{\hbox{\rm \tiny TF}}}
\newcommand{\OTF}{{\hbox{\rm \tiny OTF}}}
\newcommand{\Ex}{{\mathord{{\cal E}\!{{\scriptscriptstyle\cal X}}}}}
\newcommand{\D}{{\mathord{\cal D}}}
\newcommand{\B}[1]{{B(#1)}}
\newcommand{\A}[2]{{A(#1,#2)}}
\newcommand{\CK}{\mathord{C}_K}
\newcommand{\CM}{\mathord{C}_M}
\newcommand{\CPhi}{{\mathord{C_{\Phi}}}}
\newcommand{\Om}[1]{{\Omega(#1)}}
\newcommand{\Tr}{{\mathop{\rm Tr}}}

\newcommand{\R}{{{\Bbb R}}}
\newcommand{\C}{{{\Bbb C}}}
\newcommand{\rC}{{\rm C}}
\newcommand{\supp}{{\mathop{\rm supp\ }}}
\renewcommand{\epsilon}{{\varepsilon}}
\newcommand{\phihf}{\phi^{\upskip{3}\HF}}
\newcommand{\Phihf}[1]{\Phi^{\upskip2 \HF}_{\downskip{6}{#1}}}
\newcommand{\Phitf}[1]{\Phi^{\upskip2 \TF}_{\downskip{6}{#1}}}
\newcommand{\rhohf}{\uprho^{\upskip{3}\HF}}
\newcommand{\phitf}{\phi^{\upskip{3}\TF}}
\newcommand{\mutf}{\mu^{\upskip{3}\TF}}
\newcommand{\mutfr}{\mu^{\OTF} _{\downskip{6}{r}}}
\newcommand{\rhotf}{\uprho^{\upskip{3}\TF}}
\newcommand{\rhotfv}{{\displaystyle\uprho^{\TF}_{\downskip{8}V}}}
\newcommand{\mutfv}{{\displaystyle\mu^{\TF}_{\downskip{8}V}}}
\newcommand{\phitfv}{{\displaystyle\phi^{\TF}_{\downskip{8}V}}}
\newcommand{\rhotfr}{\uprho^{\OTF}_{\!\downskip{6}r}}
\newcommand{\etfr}{{{\cal E}^{\upskip{3}\OTF}_{r}}}
\newcommand{\phitfr}{\phi^{\OTF}_{\!\downskip{6}r}}
\newcommand{\rhohfr}{\uprho^{\HF}_{\!\downskip{6}r}}
\newcommand{\phihfr}{\phi^{\HF}_{\!\downskip{6}r}}
\newcommand{\gammar}{\upgamma^{\HF}_{\!\downskip{6}r}}
\newcommand{\gammahf}{\upgamma^{\upskip{3}\HF}}
\newcommand{\gammam}{\upgamma^{\HF}_{\!\downskip{6}\scriptscriptstyle(0)}}
\newcommand{\gammai}{\upgamma^{\HF}_{\!\downskip{6}-}}
\newcommand{\omegap}[1]{\omega^+_{#1}}
\newcommand{\omegam}[1]{\omega^-_{#1}}
\newcommand{\V}[1]{V_{#1}}
\newcommand{\W}[1]{W_{#1}}
\newcommand{\spec}{{\mathop{\rm spec}}}
\newcommand{\HN}{{H_{\rho}^{(N)}}}
\newcommand{\HMF}{H_{\gammahf}}
\newcommand{\KMF}{{\cal K}_{\gamma^\HF}}
\newcommand{\mfr}[2]{{\textstyle\frac{#1}{#2}}}
\renewcommand{\Re}{{\mathop{\rm Re}}}

\title{The ionization conjecture\\ in Hartree-Fock theory} 
\shorttitle{The ionization conjecture in Hartree-Fock theory} 

  \acknowledgements{Work partially supported by an EU-TMR grant, by a grant from the
Danish Research Council, and  by MaPhySto-Centre for Mathematical
Physics and Stochastics, funded by a grant from the Danish National
Research Foundation.}
 \author{Jan Philip Solovej}
 \institutions{University of Copenhagen, Copenhagen, Denmark\\
{\eightpoint {\it E-mail address\/}: solovej@math.ku.dk
}}

\vglue12pt
\centerline{\bf Abstract}
\vglue12pt
We prove the ionization conjecture within the Hartree-Fock theory of
atoms.  More precisely, we prove that, if the nuclear charge is
allowed to tend to infinity, the maximal negative ionization charge
and the ionization energy of atoms nevertheless remain bounded.
Moreover, we show that in Hartree-Fock theory the radius of an atom
(properly defined) is bounded independently of its nuclear charge.
 
\def\sni#1{\smallbreak\noindent{\phantom{1}#1}. }
\def\dsni#1{\smallbreak\noindent{#1}. }
\vfil 

 \centerline{\bf Contents}
\vfil
\sni{1} Introduction and main results
\sni{2} Notational conventions and basic prerequisites
\sni{3} Hartree-Fock theory
\sni{4} Thomas-Fermi theory
\sni{5} Estimates on the standard atomic TF theory
\sni{6} Separating the outside from the inside
\sni{7} Exterior $L^1$-estimate
\sni{8} The semiclassical estimates
\sni{9} The Coulomb norm estimates
\dsni{10} Main estimate
\dsni{11} Control of the region close to the nucleus: proof of Lemma 10.2
\dsni{12} Proof of the iterative step Lemma 10.3 and of Lemma 10.4
\dsni{13} Proving the main results Theorems 1.4, 1.5, 3.6, and 3.8
\pagebreak

\section{Introduction and main results} 
\vfil
One of the great triumphs of quantum mechanics is that it explains the
order in the periodic table qualitatively as well as quantitatively.
In elementary chemistry it is discussed how quantum mechanics implies
the shell structure of atoms which gives a qualitative understanding
of the periodic table.  In computational quantum chemistry it is found
that quantum mechanics gives excellent agreement with the quantitative
aspects of the periodic table. It is a very striking fact, however,
that the periodic table is much more ``periodic'' than can be
explained by the simple shell structure picture.  As an example it
can be mentioned that e.g., the radii of different atoms belonging to
the {\it same} group in the periodic table do not vary very much,
although the number of electrons in the atoms can vary by a factor of
10. Another related example is the fact that the maximal negative
ionization (the number of extra electrons that a neutral atom can
bind) remains small (possibly no bigger than 2) even for atoms with
large atomic number (nuclear charge).  These experimental facts can to
some extent be understood numerically, but there is no good qualitative
explanation for them.
 
In the mathematical physics literature the problem has been formulated as
follows (see e.g., Problems 10C and 10D in \cite{Simon15} 
or Problems 9 and 10 in \cite{Simon21}). 
Imagine that we consider `the infinitely large periodic
table', i.e., atoms with arbitrarily 
large nuclear charge $Z$; is it then still true that 
the radius
and maximal negative ionization remain bounded?  
This question often referred to as the {\it ionization} conjecture
is the subject of this paper. 

To be completely honest neither the qualitative nor the quantitative
explanations of the periodic table use the full quantum mechanical 
description.
On one hand the simple qualitative shell structure picture ignores the interactions
between the electrons in the atoms. 
On the other hand even in computational quantum chemistry one most often uses
approximations to the full many body quantum mechanical description.
There are in fact a hierarchy of models for the structure of atoms.
The one which is usually considered most complete is the Schr\"odinger
many-particle model. There are, however, even more complicated models, which 
take relativistic and/or quantum field theoretic corrections into account. 

A description which is somewhat simpler than the Schr\"odinger model 
is the Hartree-Fock (HF) model. Because of its greater simplicity it has 
been more widely used  in computational quantum chemistry than the full
Schr\"odinger model. Although, chemists over the years have developed 
numerous generalizations of the Hartree-Fock model,
it is still remarkable how tremendously successful
the original (HF) model has been in describing
the structure of atoms and molecules. 

A model which is again much simpler than the Hartree-Fock model 
is the Thomas-Fermi (TF) model. In this model the problem of finding
the structure of an atom is essentially reduced to solving an ODE. 
The TF model has some features, which are qualitatively wrong. 
Most notably it predicts that atoms do not bind to form 
molecules (Teller's no binding theorem; see \cite{LiebSimon:TF}). 

In this work we shall show that the TF model is, indeed,
a much better approximation to the more complicated HF model than 
generally believed. In fact, 
we shall show that it is only the outermost region of the atom
which is not well described by the TF model. 

As a simple corollary of this improved TF approximation we shall prove
the {\it ionization conjecture} within HF theory.  The corresponding
results for the full Schr\"odinger theory are still open and only much
simpler results are known (see e.g.,
\cite{Fef-Sec}, \cite{LSST}, \cite{Ruskai}, \cite{Sigal},  
\cite{SSS}).  In \cite{BenguriaLieb} the
ionization conjecture was solved in the Thomas-Fermi-von
Weizs{\"a}cker generalization of the Thomas-Fermi model. In
\cite{Solovej:RHF} the ionization conjecture was solved in a
simplified Hartree-Fock mean
field model by a method very similar to the one presented here.
In the simplified model the atoms are entirely spherically symmetric.
In the full HF model, however, the atoms need not be spherically
symmetric. This
lack of spherical symmetry in the HF model is one of the main
reasons for many of the difficulties that have to be overcome in the
present paper, although this may not always be apparent from the presentation.

We shall now describe more precisely the results of this paper. 
In common for all the 
atomic models is that, given the number of electrons
$N$ and the nuclear charge $Z$, they  describe how to find
the electronic ground state 
density $\rho\in L^1(\R^3)$, with $\int\rho=N$. Or more precisely
how to find {\it one} ground state density, since it may not be unique. 
In the TF model the ground state is described 
only by the density, whereas in the Schr\"odinger and HF models the 
density is derived from more detailed descriptions of the ground state.
For all models we shall use the following definitions. 
We distinguish quantities in the different models by adding
superscripts TF, HF. (In this work we shall not be concerned with the 
Schr\"odinger model at all.) Throughout the paper we use units in which
$\hbar=m=e=1$, i.e., {\it atomic units}. 

We shall discuss Hartree-Fock theory in greater detail
in Section \ref{sec:hf} and Thomas-Fermi theory in greater detail in
Section \ref{sec:tf}. For a complete discussion of TF theory we 
refer the reader to the original paper by Lieb and Simon~\cite{LiebSimon:TF}
or the review by Lieb~\cite{Lieb:tf}. In this introduction we
shall only make the most basic definitions and enough remarks in order to state
some of the main results of the paper.

\demo{Definition {\rm 1.1. (Mean field potentials)}}
Let $\rhohf$ and $\rhotf$ be the 
densities of atomic ground states in the HF and TF models respectively. 
We define the corresponding {\it mean field potentials}

\begin{eqnarray}
\noalign{\vskip-16pt}
        \phihf(x)&:=&Z|x|^{-1}-\rhohf*|x|^{-1}=
        Z|x|^{-1}-\int \rhohf(y)|x-y|^{-1}dy\label{eq:phihf}\\
        \phitf(x)&:=&Z|x|^{-1}-\rhotf*|x|^{-1}=
        Z|x|^{-1}-\int \rhotf(y)|x-y|^{-1}dy\label{eq:phitf}
\end{eqnarray}
and for all $R\geq0$ the {\it screened nuclear potentials at radius} $R$
\begin{eqnarray}
        \Phihf{R}(x)&:=&Z|x|^{-1}-\int_{|y|<R} \rhohf(y)|x-y|^{-1}dy 
        \label{eq:Phihf}\\
        \Phitf{R}(x)&:=&Z|x|^{-1}-\int_{|y|<R} \rhotf(y)|x-y|^{-1}dy.
        \label{eq:Phitf}
\end{eqnarray}
This is the  potential from the nuclear charge 
$Z$ screened by the electrons in the region 
$\set{x}{|x|<R}$. The screened nuclear potential 
will be very important in the technical proofs in 
Sections10--13.
\enddemo

\demo{Definition {\rm 1.2.  (Radius)}}
Let again $\rhohf$ and $\rhotf$ be the 
densities of atomic ground states in the HF and TF models respectively. 
We define the  radius $R_{Z,N}(\nu)$ to the $\nu$ last  electrons
by
$$
  \int_{|x|\geq R^{\TF}_{Z,N}(\nu)}\rhotf(x)\,dx =\nu,\quad
  \int_{|x|\geq R^{\HF}_{Z,N}(\nu)}\rhohf(x)\,dx =\nu.
$$
\enddemo

The 
functions $\phitf$ and $\rhotf$ are
the unique solutions to the set of equations
\begin{eqnarray}
        \Delta \phitf(x)&=&4\pi \rhotf(x)-4\pi Z\delta(x)\label{eq:tfsystemphi} \\
        \rhotf(x)&=&2^{3/2}(3\pi^2)^{-1}\left[\phitf(x)-\mutf\right]_+^{3/2}
        \label{eq:tfsystemrho} 
        \\
        \int \rhotf&=&N\label{eq:tfsystemN}.
\end{eqnarray}
Here $\mutf$ is a nonnegative parameter called the {\it chemical potential},
which is also uniquely determined from the equations. 
We have used the notation $[t]_+=\max\{t,0\}$ for all $t\in\R$.
The equations (\ref{eq:tfsystemphi}--\ref{eq:tfsystemN}) only have 
solutions when\break $N\leq Z$. For $N>Z$ we shall let $\phitf$ and $\rhotf$
refer to the solutions for $N=Z$, the neutral case.
Instead of fixing $N$ 
and determining $\mutf$ 
(the `canonical' picture) one could fix $\mutf$ and determine $N$ 
(the `grand canonical' picture).
The equation (\ref{eq:tfsystemphi}) is essentially equivalent to (\ref{eq:phitf})
and expresses the fact that $\phitf$ is the mean field potential generated by
the positive charge $Z$ and the negative charge distribution $-\rhotf$. 
The equations (\ref{eq:tfsystemrho}--\ref{eq:tfsystemN}) state that $\rhotf$ is given by the 
semiclassical expression for the density of an electron gas of $N$ electrons
in the exterior potential $\phitf$. 
For a discussion of semiclassics we refer the reader to Section \ref{sec:sc}.
\advance\theoremcount by 2

\numbereddemo{{R}emark} The total energy of the atom in Thomas-Fermi theory is 
\begin{eqnarray}
  \mfr{3}{10}(3\pi^2)^{2/3}
        \int\rhotf(x)^{5/3}\,dx-Z\int\rhotf(x)|x|^{-1}dx&&\\
        +\mfr{1}{2}\iint\rhotf(x)|x-y|^{-1}\rhotf(y)dx\,dy&\geq&-e_0Z^{7/3}\nonumber
\end{eqnarray}
where $e_0$ is the total binding energy of a neutral TF atom
of unit nuclear charge. 
Numerically \cite{Lieb:tf}, 
\begin{equation}\label{eq:ine0}
 e_0=2(3\pi^2)^{-2/3}\cdot 3.67874=0.7687.
\end{equation}
For a neutral atom, where $N=Z$,
the above inequality is an equality. The inequality states that 
in Thomas-Fermi theory
the energy is smallest for a neutral atom.
\enddemo

We can now state two of the main results in this paper. 
\proclaimtitle{Potential estimate}
\proclaim{Theorem}\label{thm:potentialestimate}
For all $Z\geq1$ and all integers 
$N$ with $N\geq Z$ for which 
there exist Hartree\/{\rm -}\/Fock ground states with $\int\rhohf=N$ 
we have 
\begin{equation}\label{eq:mainresult}
  \left|\phihf(x)-\phitf(x)\right|\leq 
  A_\phi|x|^{-4+\epsilon_0}+A_1,
\end{equation}
where $A_\phi,A_1,\epsilon_0>0$ are universal constants.
\endproclaim
 
This theorem is proved in Section~\ref{sec:provingmain}
on page 535.  
The significance of the power $|x|^{-4}$ is that for $N\geq Z$ we have
$\lim_{Z\to\infty}\phitf(x)=3^42^{-3}\pi^2|x|^{-4}$.
The existence of this limit known as the Sommerfeld asymptotic law 
\cite{Sommerfeld}
follows from Theorem~2.10 in \cite{Lieb:tf}, but we shall also prove it 
in Theorems~\ref{thm:sommerfeldup} and \ref{thm:sommerfeldlow} below.

Note that the bound in 
Theorem~\ref{thm:potentialestimate} is uniform in $N$ and
$Z$. 

The second main theorem is the 
universal bound on the atomic radius mentioned in 
the beginning of the introduction. In fact, not only do we prove 
uniform bounds but we also establish a certain exact 
asymptotic formula for the radius of an ``infinite atom''.
 
\proclaim{Theorem} \proclaimtitle{The radius of an infinite neutral HF atom}
\label{thm:radius}
Both $\liminf\limits_{Z\to\infty}R_{Z,Z}^{\HF}(\nu)$ 
and $\limsup\limits_{Z\to\infty}R_{Z,Z}^{\HF}(\nu)$ 
are bounded and have the asymptotic behavior
$$
  2^{-1/3}3^{4/3}\pi^{2/3}\nu^{-1/3}+ o(\nu^{-1/3})
$$
as $\nu\to\infty$.
\endproclaim

The proof of this theorem can be found in Section \ref{sec:provingmain} 
on page~535.
The universal bound 
on the maximal ionization is given in Theorem~\ref{thm:maxN}.
The proof is given in Section \ref{sec:provingmain} 
on page~534.
A universal bound on the ionization energy (the energy it takes to
remove one electron) is formulated in Theorem~\ref{thm:ie}.
The proof is given in Section ~\ref{sec:provingmain} 
on page~537. 
Theorems~\ref{thm:maxN} and \ref{thm:ie} are as important as 
Theorems~\ref{thm:potentialestimate} and \ref{thm:radius}. We have
deferred the statements of Theorems~\ref{thm:maxN} and \ref{thm:ie}
in order not to have to make too many definitions here in the introduction.

One of the main ideas in the paper is to use the strong universal
behavior of the TF theory reflected in the Sommerfeld asymptotics.  If
we combine (\ref{eq:tfsystemphi}) and (\ref{eq:tfsystemrho}) we see
that for $\mutf=0$ the potential satisfies the equation
$\Delta\phitf(x)= 2^{7/2}(3\pi)^{-1}[\phitf]_+^{3/2}(x)$ for
$x\ne0$. It turns out that the singularity at $x=0$ of any solution to
this equation is either of weak type $\sim Z|x|^{-1}$ for some
constant $Z$ or of strong type $\sim 3^42^{-3}\pi^2|x|^{-4}$ (see
\cite{Veron} for a discussion of singularities for differential
equations of similar type). 
The surprising fact, contained in 
Theorem~\ref{thm:potentialestimate}, is that the same type
of universal behavior holds also for the much more complicated HF
potential. We prove this by comparing with appropriately modified TF
systems on different scales, using the fact that the modifications
do not affect the universal behavior. 
A direct comparison works only in a short
range of scales. This is however enough to use an iterative 
renormalization argument to bootstrap 
the comparison to essentially all scales. 

The paper is organized as follows. 
In Section \ref{sec:notations} we fix our notational 
conventions and give some basic prerequisites. 
In Section \ref{sec:hf} we discuss Hartree-Fock theory. 
In Sections \ref{sec:tf} and \ref{sec:atomictf}
we discuss Thomas-Fermi theory. In particular we show that 
the TF model, indeed, has the universal behavior for large $Z$ 
that we want to establish for the HF model. In the TF model 
the universality can be expressed very precisely 
through the Sommerfeld asymptotics.
 
In Section \ref{sec:io} we begin the more technical work. 
We show in this section that the HF atom in the region $\set{x}{|x|>R}$
is determined to a good approximation, in terms of energy, from knowledge of the 
screened nuclear potential $\Phihf{R}$. It is this crucial step in the 
whole argument that I do not know how to generalize to the
Schr\"odinger model or even to the case of molecules in HF theory. 

For the outermost region of the atom one cannot use the energy 
to control the density. In fact, changing the density of the atom 
far from the nucleus will not affect the energy very much. 
Far away from the nucleus one must use the exact energy
minimizing property of the ground state, i.e., that it satisfies a
variational equation. This is done in Section \ref{sec:l1} to estimate 
the $L^1$-norm of the density in a region of the form $\set{x}{|x|>R}$. 

In Section \ref{sec:sc} we establish the semiclassical estimates that 
allow one to compare the HF model with the TF model. 
To be more precise, there is no semiclassical parameter in our setup, 
but we derive bounds that in a semiclassical limit would be
asymptotically exact. 

It turns out to be useful to use the electrostatic energy 
(or rather its square root) as a norm in which  to control the
difference between the densities in TF and HF theory. 
The properties of this norm, which we call the Coulomb norm,
are discussed in Section \ref{sec:cn}.
Sections \ref{sec:tf}--\ref{sec:cn} can be read almost
independently. 

In Section \ref{sec:me} we state and prove the main technical 
tool in the work. It is a comparison of the screened
nuclear potentials in HF and TF theory. Using a comparison
between the screened nuclear potentials at radius $R$
one may use the result of the separation of the outside from the
inside given in Section \ref{sec:io} to get good control on the outside
region $\set{x}{|x|>R}$. Using an iterative scheme one establishes 
the main estimate for all $R$. 
The two main technical lemmas are proved in Section \ref{sec:smallx}
and Section \ref{sec:iteration} respectively.

Finally the main theorems are proved in Section \ref{sec:provingmain}.

The main results of this paper were announced in
\cite{Solovej:announcement} and a sketch of the proof was given
there. The reader may find it useful to read this sketch as a summary
of the proof. 
 
\section{Notational conventions and basic prerequisites} 
\label{sec:notations}
\advance\eqcount by 10

We shall throughout the paper use the definitions 
\begin{eqnarray}
  \B{r}&:=&\set{y\in\R^3}{|y|\leq r},\label{def:b}\\
  \B{x,r}&:=&\set{y\in\R^3}{|y-x|\leq r},\label{def:bx}\\
  \A{r_1}{r_2}&:=&\set{x\in\R^3}{r_1\leq |x|\leq r_2}.\label{def:a}
\end{eqnarray}
For any $r>0$ we shall denote by $\chiminus{r}$ the characteristic
function of the ball $\B{r}$ and by $\chiplus{r}=1-\chiminus{r}$.
We shall as in the introduction use the notation 
$[t]_\pm=(t)_\pm:=\max\{\pm t,0\}$.

Our convention for the Fourier transform is 
\begin{equation}
        \hat{f}(p):=(2\pi)^{-3/2}\int e^{ipx} f(x)\, dx.
\end{equation}
Then 
\begin{equation}
        \widehat{f*g}=(2\pi)^{3/2}\hat{f}\hat{g},\quad
        \|f\|_2=\|\hat f\|_2,\quad |\hat f(p)|\leq (2\pi)^{-3/2}\|f\|_1
\end{equation}
and
\begin{equation}
        \iint f(x)|x-y|^{-1}\overline{g(y)}dx\,dy
        =2(2\pi)\int \hat{f}(p)\overline{\hat{g}(p)}|p|^{-2}\, dp.
\end{equation}

\demo{Definition {\rm 2.1. (Density matrix)}} Here we shall use the definition that a {\it density matrix}, 
on a Hilbert space $\cal H$, is a positive trace class 
operator satisfying the 
operator inequality $0\leq\gamma\leq I$. 
When $\cal H$ is either $L^2(\R^3)$ or $L^2(\R^3;\C^2)$ we 
write $\gamma(x,y)$ for the integral kernel for $\gamma$. 
It is $2\times 2$ matrix valued
in the case $L^2(\R^3;\C^2)$. We 
define
the {\it density} $0\leq\uprho_\gamma\in L^1(\R^3)$
corresponding to $\gamma$ by
\begin{equation}\label{eq:den}
  \uprho_\gamma:=\sum_j\nu_j|u_j(x)|^2,
\end{equation}
where $\nu_j$ and $u_j$ are the eigenvalues and corresponding
eigenfunctions of $\gamma$. Then $\int\uprho_\gamma=\Tr[\gamma]$.
\enddemo

\advance\theoremcount by 1
\numbereddemo{{R}emark}\label{remark:deltagamma}
Whenever $\gamma$ is a density matrix with eigenfunctions $u_j$ 
and corresponding eigenvalues $\nu_j$ 
on either $L^2(\R^3)$ or $L^2(\R^3;\C^2)$ we shall write
\begin{equation}\label{eq:trdeltagamma}
        \Tr\left[-\Delta\gamma\right]
        :=\sum_j\nu_j\int |\nabla u_j(x)|^2\,dx.
\end{equation}
If we allow the value $+\infty$ then the right side is defined for all 
density matrices. The expression $-\Delta\gamma$ may of course 
define a trace class operator for some $\gamma$, 
i.e., if the eigenfunctions $u_j$ are in the 
Sobolev space $H^2$ and the right side above is finite. 
In this case the left side is well defined and is equal to the 
right side. On the other hand, the right side may be finite 
even though $-\Delta\gamma$ does not even define 
a bounded operator, i.e., if an eigenfunction is in $H^1$, but not in $H^2$.
Then the sum on the right is really 
$$
        \Tr\left[(-\Delta)^{1/2}\gamma(-\Delta)^{1/2}\right]
        =\Tr\left[\nabla\cdot\gamma\nabla\right].
$$
It is therefore easy to see that (\ref{eq:trdeltagamma}) holds
not only for the spectral decomposition, but more generally, whenever
$\gamma$ can be written as 
$\gamma f=\sum_j\nu_j (u_j,f)u_j$, with $0\leq\nu_j$ 
(the $u_j$ need not be orthonormal). The same is also true for the 
expression (\ref{eq:den}) for the density. 
\enddemo

\proclaim{Proposition} The map 
$\gamma\mapsto\Tr[-\Delta\gamma]$ as defined above on all 
density matrices is affine and weakly lower semicontinuous.
\endproclaim

{\it Proof}. Choose a basis $f_1,f_2,\ldots$ for 
$L^2$ consisting of functions from $H^1$. 
Then 
$$\Tr[-\Delta\gamma]=\sum_m (\nabla f_m,\gamma \nabla f_m).$$
The affinity is trivial and the lower semicontinuity 
follows from Fatou's lemma. \phantom{moresnow}
\hfill\qed
\vglue8pt

We are of course abusing notation when we define $\Tr[-\Delta\gamma]$ 
for all density matrices. This is, however, very convenient and
should hopefully not cause any confusion.

If $V$ is a positive measurable function, we always
identify $V$ with a multiplication operator on $L^2$. If 
$V\uprho_\gamma\in L^1(\R^3)$ we abuse notation and write
$$
        \Tr\left[V\gamma \right]:=\int V\uprho_\gamma.
$$
As before if $V\gamma $ happens to be trace class then the left side 
is well defined and finite and is equal to the right side. Otherwise,
we really have
$\int V\uprho_\gamma=\Tr\left[[V]_+^{1/2}\gamma[V]_+^{1/2}\right]
-\Tr\left[[V]_-^{1/2}\gamma[V]_-^{1/2}\right]$.
\enddemo

\proclaimtitle{The IMS formulas}
\proclaim{Lemma}\label{lm:IMS}
If $u$ is in the Sobolev space $H^1(\R^3;\C^2)$ or 
$H^1(\R^3)$ and if $\Xi\in C^1(\R^3)$ is real{\rm ,} bounded{\rm ,} and has 
bounded derivative then\footnote{We denote by $u^*$ 
the complex conjugate of $u$. In the case when $u$ takes values in ${\scriptstyle\C}^2$ 
this refers to the complex conjugate matrix.} 
\begin{equation}\label{eq:IMSformula}
        \Re\int\nabla\left(\Xi^2{u}^*\right)\cdot
        \nabla u
        =\int \left|\nabla(\Xi u)\right|^2-\int\left|\nabla \Xi\right|^2
        |u|^2.
\end{equation}
If $\gamma$ is a density matrix on $L^2(\R^3;\C^2)$ or $L^2(\R^3)$
and if $\Xi_1,\ldots,\Xi_m\in C^1(\R^3)$ are real{\rm ,} bounded{\rm ,} have   
bounded derivatives{\rm ,} and satisfy $\Xi_1^2+\ldots+\Xi_m^2=1$
then
\begin{eqnarray}\label{eq:IMS}
        \Tr\left[-\Delta\gamma\right]&=&
        \Tr\left[-\Delta(\Xi_1\gamma\Xi_1)\right]
        -\Tr\left[\left(\nabla\Xi_1\right)^2\gamma\right]
        +\ldots\\
        &&{}+\Tr\left[-\Delta(\Xi_m\gamma\Xi_m)\right]
        -\Tr\left[\left(\nabla\Xi_m\right)^2\gamma\right].\nonumber
\end{eqnarray}
Note that $\Xi_j\gamma\Xi_j$ again defines a density matrix 
{\rm (}\/where we identified $\Xi_j$ with a multiplication operator\/{\rm )}.
\endproclaim

\demo{Proof}
The identity (\ref{eq:IMSformula}) follows from a simple 
computation. If we sum this identity and use
$\Xi_1^2+\ldots+\Xi_m^2=1$ we obtain
$$
        \int|\nabla u|^2=
        \int \left|\nabla(\Xi_1 u)\right|^2-\int\left|\nabla \Xi_1\right|^2
        |u|^2+\ldots+
        \int \left|\nabla(\Xi_m u)\right|^2-\int\left|\nabla \Xi_m\right|^2
        |u|^2.
$$
If we allow the value $+\infty$ this identity holds for all functions 
$u$ in $L^2$. Thus (\ref{eq:IMS}) is a simple consequence of the definition
(\ref{eq:trdeltagamma}).
\enddemo

\proclaimtitle{Lieb-Thirring inequality} 
\proclaim{Theorem}
We have the Lieb\/{\rm -}\/Thirring inequality 
\begin{equation}\label{eq:LTdensity}
        \Tr\left[-\mfr{1}{2}\Delta \gamma\right]
        \geq K_1\int\uprho_\gamma^{5/3},
\end{equation}
where $K_1:=20.49$.  Equivalently{\rm ,} 
If $V\in L^{5/2}(\R^3)$ 
and if $\gamma$ is any density matrix such that $\Tr[-\Delta\gamma]<\infty$
we have 
\begin{equation}\label{eq:LT}
        \Tr\left[-\mfr{1}{2}\Delta\gamma\right] -\Tr\left[V\gamma\right]
        \geq -L_1\int [V]_+^{5/2},
\end{equation}
where $L_1:=\mfr{2}{5}\left(\mfr{3}{5K_1}\right)^{2/3}
=0.038$. 
\endproclaim

The original proofs of these inequalities can be found in 
\cite{LiebThirring}. The constants here are taken from 
\cite{HundertmarkLaptevWeidl}.
{F}rom the min-max principle it is clear that the right side of (\ref{eq:LT})
is in fact a lower bound on the sum of the negative eigenvalues of 
the operator $-\mfr{1}{2}\Delta -V$.

\proclaimtitle{Cwikel-Lieb-Rozenblum inequality}
\proclaim{Theorem}
If $V\in L^{3/2}(\R^3)$ then the number of nonpositive eigenvalues 
of $-\mfr{1}{2}\Delta -V${\rm ,} i.e.{\rm ,} 
$$\Tr\left[\chi_{(-\infty,0]}\left( -\mfr{1}{2}\Delta
    -V\right)\right],$$
where $\chi_{(-\infty,0]}$ is the characteristic function of the
interval $(-\infty,0]${\rm ,} satisfies the bound
\begin{equation}\label{eq:CLR}
        \Tr\left[\chi_{(-\infty,0]}\left( -\mfr{1}{2}\Delta -V\right)\right]
        \leq L_0\int[V]_+^{3/2},
\end{equation}
where $L_0:=2^{3/2}0.1156=0.3270$.
\endproclaim

The original (independent) proofs can be found in Cwikel~\cite{Cwikel},
Rozenblum~\cite{Rozenblum}, and Lieb~\cite{Lieb:CLR}.
The constant is from Lieb~\cite{Lieb:CLR}.

\section{Hartree-Fock theory}
\label{sec:hf}
\advance\eqcount by 23

In Hartree-Fock theory, as opposed to Schr\"odinger theory,
one does not consider the 
full $N$-body Hilbert space $\bigwedge^NL^2(\R^3;\C^2)$. 
One rather restricts attention to the pure wedge products
(Slater determinants)
\begin{equation}\label{eq:Slater}
        \Psi=(N!)^{-1/2}u_1\wedge\ldots\wedge u_N,
\end{equation}
where $u_1,\ldots,u_N\in L^2(\R^3;\C^2)$. 
Then one minimizes the energy expectation
$$
        \frac{\left(\Psi, H_{N,Z}\Psi\right)}{\left(\Psi, \Psi\right)}
$$ 
of the Hamiltonian
\begin{equation}\label{eq:NHamiltonian}
        H_{N,Z}:=\sum_{i=1}^N \left(-\mfr{1}{2}\Delta
        -\frac{Z}{|x|}\right)+\sum_{1\leq i<j\leq N}
        \frac{1}{|x_i-x_j|}
\end{equation}
over wave functions $\Psi$ of the form (\ref{eq:Slater}) only.

If $\gamma$ is the projection onto the $N$-dimensional space 
spanned by the 
functions $u_1,\ldots,u_N$, the energy depends only on $\gamma$. 
In fact, 
$$
        \frac{\left(\Psi, H_{N,Z}\Psi\right)}{\left(\Psi, \Psi\right)}
        =\ehf(\gamma).
$$
Here we have defined the {\it Hartree-Fock} energy functional
\begin{eqnarray}\label{eq:ehf}
  \ehf(\gamma):&=&
  \Tr\left[\left(-\mfr{1}{2}\Delta-Z|x|^{-1}\right)\gamma\right]
  +\D(\gamma)-\Ex(\gamma)\\
  &=&\Tr\left[-\mfr{1}{2}\Delta\gamma\right]
  -\int Z|x|^{-1}\uprho_\gamma(x)\,dx
  +\D(\gamma)-\Ex(\gamma),\nonumber
\end{eqnarray}
where we have introduced the {\it direct Coulomb energy}, 
defined in terms of the Coulomb inner product $D$ (see also
(\ref{eq:Coulombinner}) below),
by 
\begin{equation}\label{eq:direct}
        \D(\gamma):=D(\uprho_\gamma,\uprho_\gamma)=
        \mfr{1}{2}\iint \uprho_\gamma(x)|x-y|^{-1}\uprho_\gamma(y)
        dx\,dy
\end{equation}
and the {\it exchange Coulomb energy}
\begin{equation}\label{eq:exchange}
        \Ex(\gamma):=\mfr{1}{2}
        \iint\Tr_{\C^2}\left[|\gamma(x,y)|^2\right]|x-y|^{-1}
        dx\,dy.
\end{equation}

\demo{Definition {\rm 3.1. (The Hartree-Fock ground state)}}
Let $Z>0$ be a real number and $N\geq 0$ be an integer. 
The Hartree-Fock ground state energy is 
$$
        E^\HF(N,Z):=\inf\set{\ehf(\gamma)}{\gamma^*=\gamma,\ 
        \gamma=\gamma^2,\ \Tr[\gamma]=N}.
$$
If a minimizer $\gammahf$
exists we say that the atom has an HF ground state
described by $\gammahf$. In particular, its density is
$\rhohf(x)=\uprho_{\gammahf}(x)$.
\enddemo
\advance\theoremcount by 1

\proclaimtitle{Bound on the Hartree-Fock energy}
\proclaim{Theorem}
\label{thm:HFenergy}
For $Z>0$ and any integer $N>0$ we have
$$
E^\HF(N,Z)\geq -3(4\pi L_1)^{2/3}Z^{2}N^{1/3},
$$ 
where $L_1$ is the constant in the
Lieb\/{\rm -}\/Thirring inequality {\rm (\ref{eq:LT}).}
\endproclaim

\demo{Proof}
Let $\gamma$ be an $N$ dimensional projection. 
Since the last term in $H_{N,Z}$  is positive we see that 
$\ehf(\gamma)\geq
\Tr\left[\left(-\mfr{1}{2}\Delta-Z|x|^{-1}\right)\gamma\right]$. 
It the follows from the Lieb-Thirring inequality (\ref{eq:LT}) that
for all $R>0$ we have 
$$
   \ehf(\gamma)\geq -L_1\int_{|x|<R}Z^{5/2}|x|^{-5/2}\,dx
   -ZNR^{-1}.
$$
The estimate in the theorem follows by evaluating the integral
and choosing the optimal value for $R$. 
\enddemo

\numbereddemo{{R}emark} The function $N\mapsto E^\HF(N,Z)$ is nonincreasing.
This can be seen fairly easily by constructing a trial
$N+1$-dimensional projection from any $N$-dimensional projection by 
adding an extra dimension corresponding to a function $u$ concentrated far
from the origin and with very small   kinetic energy $\int|\nabla
u|^2$. This trial projection can be constructed such that it has an
energy arbitrarily close to the original $N$-dimensional projection. 
Therefore we also have that 
\begin{equation}\label{eq:rel}
E^\HF(N,Z)=\inf\set{\ehf(\gamma)}{\gamma^*=\gamma,\ 
  \gamma^2=\gamma,\ \Tr\gamma\leq N}.
\end{equation}
\enddemo
\vglue-12pt
 
This Hartree-Fock minimization problem was studied by Lieb and Simon in 
\cite{LiebSimon:HF}. They proved the following about the existence of minimizers.

\proclaimtitle{Existence of HF minimizers}
\proclaim{Theorem} 
If $N$ is a positive integer such that $N<Z+1$ then there exists
an $N$\/{\rm -}\/dimensional projection $\gammahf$ minimizing the functional
$\ehf$ in {\rm (\ref{eq:ehf}),} i.e.{\rm ,} $E^\HF(N,Z)=\ehf(\gammahf)$ is a minimum. 
\endproclaim

In the opposite direction the following result was proved by Lieb
\cite{Lieb:2z+1}.

\proclaimtitle{Lieb's bound on the maximal ionization}
\proclaim{Theorem} 
\label{thm:2z+1} 
If $N$ is a positive integer such that $N>2Z+1$ there are 
no minimizers for the Hartree\/{\rm -}\/Fock 
functional among $N$\/{\rm -}\/dimensional projections{\rm ,} i.e.{\rm ,} 
there does not exist an $N$\/{\rm -}\/dimensional projection $\gamma$ such that 
$\ehf(\gamma)=E^\HF(N,Z)$. 
\endproclaim

This theorem will, in fact, follow from the proof of
Lemma~\ref{lm:extl1} below (see
page~503).
Although this result is very good for $Z=1$ it is far from optimal
for large $Z$. In particular the factor $2$ should rather be $1$. 
This fact known as the ionization conjecture
is one of the of the main results of the present work.

\proclaimtitle{Universal bound on the maximal ionization charge}
\proclaim{Theorem} 
\label{thm:maxN} There exists 
a  universal constant $Q>0$ such that for all positive integers
satisfying $N\geq Z+Q$ there are no minimizers for the Hartree\/{\rm -}\/Fock 
functional among $N$\/{\rm -}\/dimensional projections.
\endproclaim

\numbereddemo{{R}emark} Although, it is possible to calculate an exact value for the 
constant $Q$ above it is quite tedious to do so. Moreover, 
the present work does not attempt to optimize this constant.  
The result of this work is mainly to establish that such a 
{\it finite} constant exists. This of course raises the very
interesting question of finding a {\it good} estimate 
on the constant, but we shall not address this here. 
\enddemo

The proof of Theorem~\ref{thm:maxN} is given in Section \ref{sec:provingmain} 
on page~534.

\proclaimtitle{Bound on the ionization energy}
\proclaim{Theorem} 
\label{thm:ie} 
The ionization energy of a neutral atom $E^\HF(Z-1,Z)-E^{\HF}(Z,Z)$
is bounded by a universal constant {\rm (}\/in particular{\rm ,} independent of
$Z${\rm ).}
\endproclaim

This theorem is proved in Section \ref{sec:provingmain} on page
\pageref{proof:ie}.

The variational equations (Euler-Lagrange equations) for the minimizer
was also given in  \cite{LiebSimon:HF}.
Since the Hartree-Fock variational equations shall be used later in this 
work, we shall derive them in Theorem~\ref{thm:HFequations} below.

We first note that the Hartree-Fock functional $\ehf$ may be extended
from projections (i.e., density matrices with $\gamma^2=\gamma$)
to all density matrices. 
If $\Tr\left[-\Delta\gamma\right]<\infty$
all the terms of $\ehf$ are finite. In fact, 
$\Tr\left[Z|x|^{-1}\gamma\right]$ is finite by the Lieb-Thirring 
inequality (\ref{eq:LTdensity}) since $Z|x|^{-1}\in
L^\infty(\R^3)+ L^{5/2}(\R^3)$. The term 
$\D(\gamma)$ is finite by the Hardy-Littlewood-Sobolev
inequality since $\uprho_\gamma\in 
L^1(\R^3)\cap L^{5/3}(\R^3)\subset L^{6/5}(\R^3)$. 
Finally, $\Ex(\gamma)\leq D(\gamma)$ since
$$
        D(\gamma)-\Ex(\gamma)
        =\mfr{1}{4}\sum_{i,j}\nu_i\nu_j\iint 
        \frac{\left\|u_i(x)\otimes u_j(y)
        -u_j(x)\otimes u_i(y)\right\|_{\C^2\otimes\C^2}^2}
        {|x-y|}dx\,dy\geq 0,
$$
when $\nu_i$ are the eigenvalues of $\gamma$
with $u_i$ being the corresponding eigenfunctions. 
If $\Tr\left[-\Delta\gamma\right]=\infty$ we 
set $\ehf(\gamma):=\infty$. 
It is clear that $\lim_n\ehf(\gamma_n)=\infty$ 
if $\lim_n\Tr\left[-\Delta\gamma_n\right]\to\infty$.

\numbereddemo{{R}emark} It is important to realize that
although $D(\gamma)-\Ex(\gamma)$ is positive it is {\it not} 
a convex functional on the set of density matrices. 
In particular, the Hartree-Fock minimizer need not be unique.
(A simple example of nonuniqueness occurs for the case $N=1$. 
For a one-dimensional projection $\gamma$, it is 
clear that  $D(\gamma)-\Ex(\gamma)=0$, hence the minimizer in this case
is simply the projection onto a ground state of
the operator $-\mfr{1}{2}\Delta-Z|x|^{-1}$ on the space $L^2(\R^3;\C^2)$. 
There are many ground states since the spin can point in any direction.)
\enddemo

Another fact related to the nonconvexity of the Hartree-Fock functional
is the important observation first made by Lieb in \cite{Lieb:Var}
that the infimum of the Hartree-Fock functional is 
not lowered by extending the functional 
to all density matrices. For a simple proof of this see
\cite{Bach}.

\proclaimtitle{Lieb's variational principle}
\proclaim{Theorem} \label{thm:liebvar}
For all nonnegative integers $N$ we have
\begin{eqnarray*}
&&
    \hskip-36pt    \inf\set{\ehf(\gamma)}{\gamma^*=\gamma,\ 
        \gamma=\gamma^2,\ \Tr[\gamma]=N}\\
&&\qquad =
        \inf\set{\ehf(\gamma)}{0\leq\gamma\leq \I,\ \Tr[\gamma]=N} 
\end{eqnarray*}
and if the infimum over all density matrices {\rm (}the {\rm inf} on the right\/{\rm )}
is attained then so is the 
infimum over projections {\rm (}the {\rm inf} on the left\/{\rm ).}\/
\endproclaim

We now come to the properties of the Hartree-Fock
minimizers, especially that they satisfy the Hartree-Fock equations.
These equations state that a minimizing $N$-dimensional projection
$\gammahf$ 
is the projection onto the 
$N$-dimensional space spanned by eigenfunctions with lowest possible
eigenvalues for the {\it HF mean field operator} 
\begin{equation}\label{eq:HMF}
        \HMF:=-\mfr{1}{2}\Delta-Z|x|^{-1}+\rhohf*|x|^{-1}
        -\KMF.
\end{equation} Here $\KMF$ is the {\it exchange
operator} defined by having the $2\times2$-matrix valued integral
kernel
$$
        \KMF(x,y):=|x-y|^{-1}\gammahf(x,y).
$$
Thus $\gammahf(x,y)=\sum_{i=1}^Nu_i(x)u_i(y)^*$, where
$\HMF u_i=\epsilon_i u_i$, 
and $\epsilon_1,\epsilon_2,\ldots,\epsilon_N\break\leq 0$ are 
the $N$ lowest eigenvalues of 
$\HMF$ counted with multiplicities.

This self-consistent property of 
a minimizer $\gammahf$ may equivalently be stated as 
in the theorem below.

\proclaimtitle{Properties of HF minimizers}
\proclaim{Theorem} \label{thm:HFequations} \hskip-8pt
If $\gammahf$ with density $\rhohf$
is a projection  minimizing the {\rm HF} functional $\ehf$ under the
constraint $\Tr\left[\gammahf\right]=N$ then 
$\rhohf\in L^{5/3}(\R^3)\cap L^1(\R^3)$ and 
$\HMF$ defines a semibounded self\/{\rm -}\/adjoint operator with form domain 
$H^1(\R^3;\C^2)$ having at least $N$ nonpositive eigenvalues. 
Moreover{\rm ,} $\gammahf$ 
is the $N$\/{\rm -}\/dimensional projection 
minimizing the map $\gamma\mapsto\Tr\left[\HMF\gamma\right]$.
\endproclaim

\numbereddemo{{R}emark}The reader may worry that, because of
degenerate eigenvalues of $\HMF$, the $N$-dimensional projection
$\gamma$ minimizing $\Tr\left[\HMF\gamma\right]$
may not be unique. 
That it is, indeed, unique was proved in \cite{BLLS}.
\enddemo

\demo{Proof  of Theorem~{\rm \ref{thm:HFequations}}}
We note that  $\Tr\left[\gammahf\right]=N$, 
$\Tr\left[-\Delta\gammahf\right]<\infty$, and the Lieb-Thirring
inequality (\ref{eq:LTdensity}) implies that  
$\rhohf\in L^{5/3}(\R^3)\cap L^1(\R^3)$. 
{F}rom this it is easy to see that 
$\rhohf*|x|^{-1}$ is a bounded function (in fact, it is continuous and tends
to $0$ as $|x|\to\infty$). Moreover, in the operator 
sense $\KMF\leq \rhohf*|x|^{-1}$. This follows, since 
for $f\in L^2(\R^3;\C^2)$ we have
\begin{eqnarray*}
    &&\hskip-36pt    \int \rhohf*|x|^{-1}|f(x)|^2dx- \iint f(x)^*\KMF(x,y)
        f(y) dx\,dy\\[4pt]
    &=&    \sum_{i=1}^N\mfr{1}{2}\iint
        \frac{\left\|u_i(x)\otimes f(y)
        -f(x)\otimes u_i(y)\right\|_{\C^2\otimes\C^2}^2}{|x-y|}dx\,dy, 
\end{eqnarray*}
where $u_1,\ldots,u_N$ is a complete set of 
eigenfunctions of $\gammahf$.
It is therefore clear that $\HMF$ defines a semibounded operator
with form domain $H^1(\R^3;\C^2)$. 
Thus it makes sense to compute $\Tr\left[\HMF\gamma\right]$ 
if and only if $\Tr\left[-\Delta\gamma\right]<\infty$.

Let $\gamma'$ be an $N$-dimensional projection with 
$\Tr\left[-\Delta\gamma'\right]<\infty$.
We shall prove that 
$$
        \Tr\left[\HMF\gamma'\right]\geq 
        \Tr\left[ \HMF\gammahf\right].
$$

For $0\leq t\leq1$, consider the density matrix $\gamma_t=
(1-t)\gammahf+t\gamma'$. It satisfies $\Tr[\gamma_t]=N$.
By the Lieb variational principle, Theorem~\ref{thm:liebvar},
we have that $\ehf(\gammahf)=\ehf(\gamma_0)\leq\ehf(\gamma_t)$,
for all $0\leq t\leq 1$. Hence
$$
0\leq\left.\frac{d\ehf(\gamma_t)}{dt}\right|_{t=0}=
\Tr\left[\HMF\gamma'\right]-\Tr\left[\HMF\gammahf\right].
$$

The fact that $\Tr\left[\HMF\gamma\right]$ is minimized among $N$-dimensional 
projections implies in particular that $\HMF$ has at least $N$ nonpositive 
eigenvalues.
\enddemo

\section{Thomas-Fermi theory} 
\label{sec:tf}

\advance\eqcount by 30

In this section we discuss the facts needed from Thomas-Fermi 
theory. 
We focus only on the results that we shall use in our study 
of Hartree-Fock theory. 

\demo{Definition {\rm 4.1. (Thomas-Fermi functional)}} 
Let $V\in L^{5/2}(\R^3)+L^\infty(\R^3)$ with 
$$
        \inf\set{\|W\|_{L^\infty(\R^3)}}{V-W\in L^{5/2}(\R^3)}=0.
$$
Corresponding to $V$ we define the {\it Thomas\/{\rm -}\/Fermi {\rm (TF)} 
energy functional}
$$
        \etf{V}(\rho)
        =\mfr{3}{10}(3\pi^2)^{2/3}
        \int\rho^{5/3}-\int V\rho
        +\mfr{1}{2}\iint\rho(x)|x-y|^{-1}\rho(y)dx\,dy,
$$
on functions $\rho$
with $0\leq\rho\in L^{5/3}(\R^3)\cap L^1(\R^3)$.
\enddemo
\advance\theoremcount by 1

Note that the Hardy-Littlewood-Sobolev inequality
implies that 
$D(\rho,\rho)=
\mfr{1}{2}\iint\rho(x)|x-y|^{-1}\rho(y)dx\,dy$ 
is finite for functions $\rho\in 
L^{5/3}\cap L^1\subset L^{6/5}$.
Hence $\etf{V}$ is finite on these functions. 
\pagebreak

The proof of existence and uniqueness of minimizers to the TF functional 
and the characterization of their properties can be found in 
the work of Lieb and Simon
\cite{LiebSimon:TF} (see also \cite{Lieb:tf}). 
We state the properties that we need in the following 
theorem.  

\proclaimtitle{The TF minimizer}
\proclaim{Theorem} \label{thm:tf}
Let $V$ be as in Definition~{\rm 4.1.}
For all $N'\geq0$ there exists a unique nonnegative
$\rhotfv\in L^{5/3}(\R^3)$ such that 
$\int\rhotfv\leq N'$ and 
\begin{equation}\label{eq:tfvariation}
        \etf{V}(\rhotfv)
        =\inf\set{\etf{V}(\uprho)}{\uprho\in L^{5/3}(\R^3),\
        \int\rho\leq N'}.
\end{equation}

On the other hand there exists a {\rm (}\/unique\/{\rm )} 
{\rm chemical potential} {\rm (}\/Lagrange multiplier\/{\rm )} $\mutfv(N')${\rm ,}
with $0\leq\mutfv(N')\leq\sup V${\rm ,} such that 
$\rhotfv$ is uniquely characterized by 
\begin{eqnarray}
&&\hskip-36pt \etf{V}(\rhotfv)+\mutfv(N')\int\rhotfv \label{eq:tfmuvariation}\\
        &=& \inf\set{\etf{V}(\uprho)+\mutfv(N')\int\uprho}
        {0\leq \uprho\in L^{5/3}(\R^3)\cap L^1(\R^3)} .\nonumber
\end{eqnarray}

Moreover{\rm ,} 
$\rhotfv$ is the unique solution in 
$L^{5/3}\cap L^1$ to the {\rm Thomas-Fermi equation}
{\rm (}\/the Euler\/{\rm -}\/Lagrange equation for the variational problem 
{\rm (\ref{eq:tfmuvariation}))}
\begin{equation}\label{eq:tfeqgeneral} 
        \mfr{1}{2}(3\pi^2)^{2/3}\left(\rhotfv(x)\right)^{2/3}
        =\left[V(x)-\rhotfv*|x|^{-1}-\mutfv(N')\right]_+.
\end{equation}
If $\mutfv(N')>0$ then $\int \rhotfv=N'$. Therefore
$
        \mutfv(N')\int\rhotfv = \mutfv(N') N'.
$

For all $0<\mu$ there is a unique minimizer $0\leq \rho\in L^{5/3}\cap L^1$
to $\etf{V}(\rho)+\mu\int\rho$. {\rm (}\/If $\mu\geq \sup V$ then $\rho$ is
simply zero\/{\rm .)}
\endproclaim

We shall be interested in properties of the {\it Thomas-Fermi potential}
\begin{equation}\label{eq:tfpotgeneral}
        \phitfv:=V(x)-\rhotfv*|x|^{-1}.
\end{equation}
The Thomas-Fermi equation (\ref{eq:tfeqgeneral}) can 
be turned into the {\it Thomas-Fermi differential equation}
\begin{equation}\label{eq:tfdiffeqgeneral}
        \Delta \phitfv={2}^{7/2}(3\pi)^{-1}\left[\phitfv
        -\mutfv(N')\right]_+^{3/2}
        +\Delta V,
\end{equation}
which holds in distribution sense. 

\proclaimtitle{Maximal ionization}
\proclaim{Theorem} \label{thm:maxion} 
There exists a nonnegative real number $N_c$, possibly equal to $+\infty${\rm ,} 
such that $\mutfv(N')>0$ if and only if $N'<N_c$.
Moreover{\rm ,}
\begin{equation}\label{eq:ncupper}
        N_c\geq \liminf_{r\to\infty}(4\pi)^{-1}\int_{{\cal S}^2}rV(r\omega)
        d\omega,
\end{equation}
where $d\omega$ is the surface measure on the unit $2$\/{\rm -}\/sphere ${\cal S}^2$.
\endproclaim

\demo{Proof}
Since $\etf{V}$ is a convex functional of $\uprho$ it is clear that 
$\etf{V}\left(\rhotfv\right)$ is a convex and decreasing
function of $N'$. Hence there is a value $N_c$ such that 
$\etf{V}\left(\rhotfv\right)$ is strictly decreasing for 
$N'<N_c$ and constant for $N'\geq N_c$.
Thus if $N'\geq N_c$ then $\int\rhotfv=N_c$. 
Since $\int\rhotfv=N'$ if $\mutfv>0$ we must have 
$\mutfv(N')=\mutfv(N_c)=0$ for $N'\geq N_c$. 
On the other hand since $\int\rhotfv=N'$ if $N'<N_c$ 
we cannot have $\mutfv(N')=0$ in this case. 
This proves the first assertion. 

In order to prove the second assertion we may of course assume
that $N_c<\infty$. Since $\mutfv(N_c)=0$ we have for the
corresponding Thomas-Fermi minimizer that
\begin{eqnarray*}
        \int_{{\cal S}^2}\rhotfv(r\omega)d\omega
        &=&\const\int_{{\cal S}^2}\left[
        V(r\omega)-\rhotfv*|r\omega|^{-1}\right]_+^{3/2}d\omega\\[5pt]
        &\geq&\const
        \left[(4\pi)^{-1}\int_{{\cal S}^2}V(r\omega)d\omega
        -r^{-1}\int_{\R^3}\rhotfv\right]_+^{3/2},
\end{eqnarray*}
where the last estimate follows from Jensen's inequality
and Newton's theorem. 
Since we are considering a TF minimizer $\rhotfv$ such that 
$\int\rhotfv=N_c$ it is clear that if
(\ref{eq:ncupper}) is violated then 
$\int_{{\cal S}^2}\rhotfv(r\omega)d\omega>c r^{-3/2}$ 
for some positive constant~$c$ and all large enough $r$.
Hence $N_c =\int\rhotfv=\infty$ in contradiction with 
our assumption.
\enddemo

Proving a bound on $N_c$ in the opposite direction is 
in general more difficult. We shall return to a partial 
converse to (\ref{eq:ncupper})
in Corollary~\ref{cl:uppermaxion} below.

Usually the Thomas-Fermi model is studied for the potential 
$V$ being the 
Coulomb potential, i.e., $Z|x|^{-1}$. In this case we denote  
$\rhotfv$, $\phitfv$, and $\mutfv$ simply by 
$\rhotf$, $\phitf$, and $\mutf$. 
These are the functions discussed in the introduction. In fact, 
the equations (\ref{eq:tfsystemphi}) and (\ref{eq:tfsystemrho}) 
correspond to (\ref{eq:tfeqgeneral}) and (\ref{eq:tfdiffeqgeneral}).

{F}rom Theorem~\ref{thm:maxion} we see that in this case $N_c\geq Z$. 
We shall see below after Corollary~\ref{cl:uppermaxion} that indeed $N_c=Z$. 

The first 
mathematical study of the atomic TF equation was done 
by Hille~\cite{Hille}; a much more complete analysis  can be found in 
\cite{LiebSimon:TF}. 
\vglue2pt
The function $\phitf$ satisfies the asymptotics 
$\phitf(x)\approx 3^42^{-3}\pi^2|x|^{-4}$ for large~$x$.
The important thing to note about this
asymptotics, first discovered by Sommerfeld~\cite{Sommerfeld},
is that it is independent of $Z$. 
The Sommerfeld asymptotics is central to the present work 
and we shall prove a strong version of  it
in Theorems~\ref{thm:sommerfeldup} and
\ref{thm:sommerfeldlow}
below.  
Similar asymptotic
estimates may be derived for the density using the TF equation 
(\ref{eq:tfeqgeneral}).
We shall more generally prove asymptotic bounds for $\phitfv$, 
in the case when the potential $V$ is harmonic 
in certain regions of space.

We now come to the main technical lemma in this section, 
which is a version of the Sommerfeld estimate.\footnote{A 
version of 
this Sommerfeld estimate was stated 
in the announcement~\cite{Solovej:announcement}. The result stated 
was weaker than here in the sense that the exponents in the error terms
were different for the upper and lower bounds. The result in the 
announcement also contained a minor error because the lower bound had been 
stated incorrectly. The better and correct version is the one stated 
and proved here.} 

\proclaimtitle{Sommerfeld estimate}
\proclaim{Lemma} 
\label{lm:sommerfeld}  
Assume that $\phi\geq0$ is a smooth function on $|x|>R$ and
satisfies the differential equation 
$$
         \Delta\phi(x)=2^{7/2}(3\pi)^{-1}\phi(x)^{3/2},\quad\hbox{for
        $|x|>R$},
$$
for some $R\geq0$.
Let $\zeta:=(-7+\sqrt{73})/2\approx0.77$. Define 
\begin{eqnarray*}
        a(R)&:=&\liminf_{r\searrow R}\sup_{|x|=r}
        \left[\left(\frac{\phi(x)}{3^42^{-3}\pi^2r^{-4}}\right)^{-1/2}-1\right]
        r^\zeta \\
\noalign{\hbox{and}}
        A(R)&:=&
        \liminf_{r\searrow R}\sup_{|x|=r}
        \left[\frac{\phi(x)}{3^42^{-3}\pi^2r^{-4}}-1\right]
        r^\zeta.
\end{eqnarray*} 
Then for $|x|>R$ we have 
\begin{equation}\label{eq:tfestimate}
        \left(1+a(R)|x|^{-\zeta}\right)^{-2}\leq
        \frac{\phi(x)}{3^42^{-3}\pi^2|x|^{-4}}\leq 
        \left(1+A(R)|x|^{-\zeta}\right).
\end{equation}
\endproclaim

\numbereddemo{{R}emark} It is important to realize that 
we are not assuming that $\phi$ is spherically symmetric. 
The lemma above can therefore not be proved by ODE techniques. 
By elliptic
regularity the smoothness of $\phi$ would of course be a consequence of
a much weaker assumption. 
\enddemo

\demo{Proof  of Lemma~{\rm \ref{lm:sommerfeld}}}
We first prove that $\phi(x)\to0$ as $|x|\to\infty$.
For this purpose consider $L>4R$ and for $L/4<|x|<L$ the function 
$f(x)=C[(|x|-L/4)^{-4}+(L-|x|)^{-4}]$. 
We compute 
\begin{eqnarray*}
  \Delta f&=& C\Bigl[20\left((|x|-L/4)^{-6}+(L-|x|)^{-6}\right)\\&&{}
    +8|x|^{-1}(L-|x|)^{-5}-8|x|^{-1}(|x|-L/4)^{-5}\Bigr]\\
  &\leq& 44C(L-|x|)^{-6}+20C(|x|-L/4)^{-6}.
\end{eqnarray*}

On the other hand, $f(x)^{3/2}\geq C^{3/2}\left((|x|-L/4)^{-6}
+(L-|x|)^{-6}\right)$. 
It is therefore clear that we can choose $C$ independently of $L$ 
such that 
$\Delta f\leq 2^{7/2}(3\pi)^{-1}f^{3/2}$.
We claim that $\phi(x)\leq f(x)$ for $L/4<|x|<L$. 
This is trivial for $|x|$ close to $L/4$ or close to $L$ 
since here $f(x)$ diverges whereas $\phi(x)$ remains bounded. 
Consider the set $\set{L/4<|x|<L}{\phi(x)>f(x)}$. 
This is an open set on which $\Delta(\phi-f)\geq 
2^{7/2}(3\pi)^{-1}(\phi^{3/2}-f^{3/2})>0$; i.e., 
$\phi-f$ is subharmonic on the set and is zero on its boundary. 
Hence $\phi(x)\leq f(x)$ on the set which is a contradiction unless 
the set is empty. 
Thus for all $L>4R$ we have 
$
        \sup_{|x|=L/2}\phi(x)\leq C\left((1/4)^{-4}+(1/2)^{-4}\right) L^{-4}.
$       
Hence,
$\phi(x)|x|^4$ is bounded.      
\vglue2pt
Next we turn to the proof of the main estimate. 
Let $R'>R$ and set $A'=A(R')$ and $a'=a(R')$. Then 
$a'$ and $A'$ are finite. 
We consider the two functions 
\begin{eqnarray*}
\noalign{\vskip-4pt}
        \omegap{A'}(x)&:=&3^42^{-3}\pi^2|x|^{-4}(1+A'|x|^{-\zeta})\\
\noalign{\noindent and} 
        \omegam{a'}(x)&:=&3^42^{-3}\pi^2|x|^{-4}(1+a'|x|^{-\zeta})^{-2}.
\end{eqnarray*}
Note that by the definition of $a'$ and $A'$ both functions 
are well-defined and positive for $|x|>R'$. 
We claim that 
\begin{equation}\label{eq:omegainequality}
        \Delta \omegap{A'}(x)\leq 2^{7/2}(3\pi)^{-1}\omegap{A'}(x)^{3/2}
\enspace\hbox{and}\enspace
        \Delta \omegam{a'}(x)\geq 2^{7/2}(3\pi)^{-1}\omegam{a'}(x)^{3/2}.\hskip6pt
\end{equation}

As we shall first show the lemma is a simple consequence of the estimates in 
(\ref{eq:omegainequality}).
We give the proof for the upper bound. The lower bound is similar.
Let
$$
        \Omega_+:=\set{|x|>R'}{\phi(x)>\omegap{A'}(x)}.
$$
On $\Omega_+$, $\phi-\omegap{A'}$ is subharmonic. On 
the boundary of $\Omega_+$, $\phi-\omegap{A'}$ vanishes.
For the subset $\partial\Omega_+\cap\set{x}{|x|=R'}$ this follows 
from the choice of $A'$. Since $\phi(x)$ and $\omegap{A'}(x)$ 
both tend to zero as $|x|$ tends to infinity we
conclude that $\Omega_+=\emptyset$.

Therefore $\phi(x)\leq \omegap{A(R')}(x)$ for $|x|>R'$. 
For $|x|>R$ we get
$\phi(x)\leq\liminf_{R'\searrow R}\omegap{A(R')}(x)=\omegap{A(R)}(x)$.

It remains to check (\ref{eq:omegainequality}).
For $\omegam{a'}$ we get
\begin{eqnarray*}
        \Delta \omegam{a'}(x)
        =2^{7/2}(3\pi)^{-1}\omegam{a'}(x)^{3/2}
        &\biggl(&1+\left(1-\mfr{1}{6}\zeta(\zeta+7)\right)a'|x|^{-\zeta}\\
        &&+\mfr{1}{2}(1+a'|x|^{-\zeta})^{-1}(\zeta a' |x|^{-\zeta})^2\biggr).
\end{eqnarray*}
Since $\zeta(\zeta+7)=6$ and $1+a'|x|^{-\zeta}>0$ we see that 
$\Delta \omegam{a'}(x)\geq 2^{7/2}(3\pi)^{-1}\omegam{a'}(x)^{3/2}$.

For $\omegap{A'}$ we have
\begin{eqnarray*}
        &&\hskip-30pt\Delta \omegap{A'}(x)\\
        &&=2^{7/2}(3\pi)^{-1}\omegap{A'}(x)^{3/2}
        (1+A'|x|^{-\zeta})^{-3/2}\!\!\left(1+\left(1+\frac{\zeta(\zeta+7)}{12}
        \right)A'|x|^{-\zeta}\right)\\
        &&\leq2^{7/2}(3\pi)^{-1}\omegap{A'}(x)^{3/2},
\end{eqnarray*}
where we have used that 
\vglue12pt
\hfill ${\displaystyle (1+A'|x|^{-\zeta})^{3/2}\geq 1+\mfr{3}{2}
A'|x|^{-\zeta}=1+(1+\mfr{1}{12}\zeta(\zeta+7))A'|x|^{-\zeta}.
}$
\enddemo
 
We can immediately use this lemma to get estimates on $\phitfv$ when
$\mutfv=0$. For general $\mutfv$ the result can be generalized 
as follows.
 
\proclaimtitle{Sommerfeld estimate for general $\mutfv$}
\proclaim{Theorem} \label{thm:sommerfeldmu}
Assume that $V$ is continuous and harmonic for $|x|>R$ and satisfies
$\lim_{|x|\to\infty}V(x)=0$. Consider the corresponding Thomas\/{\rm -}\/Fermi
potential 
$\phitfv${\rm ,} which satisfies the\break\vglue-11pt\noindent {\rm TF}
differential equation {\rm (\ref{eq:tfdiffeqgeneral}).} 
Assume that $\mutfv<
\liminf\limits_{r\searrow R}\inf\limits_{|x|=r}\phitfv(x)
$.
Define 
\begin{eqnarray}
        a(R)&:=&\liminf_{r\searrow R}\sup_{|x|=r}
        \left[\left(\frac{\phitfv(x)}{3^42^{-3}\pi^2r^{-4}}\right)^{-1/2}-1
        \right]
        r^\zeta \label{eq:armu}\\
\noalign{\hbox{and}}
        A(R,\mutfv)&:=&
        \liminf_{r\searrow R}\sup_{|x|=r}
        \left[\frac{\phitfv(x)-\mutfv}{3^42^{-3}\pi^2r^{-4}}-1\right]
        r^\zeta.\label{eq:Armu}
\end{eqnarray} 
Then again{\rm ,} with $\zeta=(-7+\sqrt{73})/2\approx0.77${\rm ,}
we find for all $|x|>R$ 
\begin{equation}
        \phitfv(x) \leq  3^42^{-3}\pi^2|x|^{-4}
        \left(1+A(R,\mutfv)|x|^{-\zeta}\right)+\mutfv\label{eq:tfupestimate}\\
\end{equation}
\vglue-3pt\noindent 
and
\begin{equation}
        \phitfv(x) \geq \max\left\{ 
        3^42^{-3}\pi^2|x|^{-4}\left(1+a(R)|x|^{-\zeta}\right)^{-2},\,
        \nu(\mutfv)|x|^{-1}\right\},\qquad
        \label{eq:tflowestimate}
\end{equation}
\vglue-4pt\noindent 
where
\begin{equation}
        \nu(\mutfv):=\inf_{|x|\geq R}\max\Bigl\{
        3^42^{-3}\pi^2|x|^{-3}\left(1+a(R)|x|^{-\zeta}\right)^{-2},
        \mutfv|x|\Bigr\}.\qquad  \label{eq:nudef}
\end{equation}
\endproclaim

{\it Proof}. Since $\rhotfv\in L^{5/3}\cap L^1$ it is easy to see that 
$\rhotfv*|x|^{-1}$ is continuous and tends to zero as
$x$  tends to infinity. Thus from the assumption on $V$
it follows that $\phitfv$ is continuous on $|x|>R$ 
and satisfies $\phitfv(x)\to0$ 
as $|x|\to\infty$. 

Let $R'>R$ and set $A'=A(R',\mutfv)$ and $a'=a(R')$. Then 
$a'$ is well-defined if $R'$ is close enough to $R$ 
since then we may assume that $\phitfv(x)>\mutfv\geq 0$ for all $|x|=R'$. Both 
$a'$  and $A'$ are finite. 
Using the notation from the proof
of Lemma~\ref{lm:sommerfeld} we define 
$$
        \omegap{\mutfv,A'}(x):=\omegap{A'}(x)+\mutfv
        \quad\hbox{and}\quad
        \omegam{\mutfv,a'}(x):=
        \max\left\{\omegam{a'}(x),\ \nu'|x|^{-1}\right\},
$$
where 
$$
        \nu':=
        \min_{|x|\geq R'}\max\left\{
        |x|\omegam{a'}(x),\, |x|\mutfv\right\}.
$$
Note that, since we assume that $\phitfv(x)>\mutfv$ for $|x|=R'$, 
we have that both $\omegap{A'}(x)$ and $\omegam{a'}(x)$ are positive
for all $|x|>R'$.
We also have that $\omegam{a'}(x)>\mutfv$ for $|x|=R'$
and hence that $\omegam{a'}(x_0)=\mutfv $ at points $x_0$ where the 
minimum, defining $\nu'$, is attained.
(Note that $|x|\omegam{a'}(x)$ is a radially decreasing function for 
$|x|>R'$.)

The proof of the present lemma 
is now similar to that of  Lemma~\ref{lm:sommerfeld} 
if we can show that for $|x|>R'$
\begin{equation}\label{eq:omega+muineq}
        \Delta\omegap{\mutfv,A'}(x)
        \leq 2^{7/2}(3\pi)^{-1}\left[\omegap{\mutfv,A'}(x)-\mutfv
        \right]_+^{3/2}
\end{equation}
and
\begin{equation}\label{eq:omega-muineq}
        \Delta\omegam{\mutfv,a'}(x)
        \geq 2^{7/2}(3\pi)^{-1}
        \left[\omegam{\mutfv,a'}(x)-\mutfv\right]_+^{3/2}       
\end{equation}
(in distribution sense). 
The inequality (\ref{eq:omega+muineq}) follows immediately 
from the first inequality in (\ref{eq:omegainequality}).
The inequality (\ref{eq:omega-muineq}) is slightly more complicated.
Note that the definitions of $\omegam{\mutfv,a'}$ and of $\nu'$
imply that $\omegam{\mutfv,a'}(x)=\nu'|x|^{-1}$ if 
$\omegam{\mutfv,a'}(x)<\mutfv$ and 
$\omegam{\mutfv,a'}(x)=\omegam{a'}(x)$ if $\omegam{\mutfv,a'}(x)>\mutfv$.
Thus
if $\omegam{\mutfv,a'}(x)<\mutfv$ we have that $\omegam{\mutfv,a'}$ is
harmonic. Hence (\ref{eq:omega-muineq}) holds in this region. 
If $\omegam{\mutfv,a'}(x)>\mutfv$ then 
$\omegam{\mutfv,a'}(x)=\omegam{a'}(x)$ and (\ref{eq:omega-muineq}) 
follows in this region from the second
inequality in (\ref{eq:omegainequality}).
Finally, since the maximum of two subharmonic functions is also 
subharmonic,
it is clear that the distribution
$\Delta\omegam{\mutfv,a'}$ is a positive measure and
in particular positive on the set (of Lebesgue measure)
zero where $\omegam{\mutfv,a'}(x)=\mutfv$. 
Hence, (\ref{eq:omega-muineq}) holds in distribution sense 
for all $|x|>R'$. 
\hfill\qed\vglue12pt

As an application of the lower bound on $\phitfv$ in 
(\ref{eq:tflowestimate}) we can get an estimate on the
chemical potential $\mutfv$.

\proclaimtitle{Chemical potential estimate}
\proclaim{{C}orollary} \label{cl:muestimate}
With the assumptions and definitions in Theorem~{\rm \ref{thm:sommerfeldmu},} 
in particular{\rm ,} if 
$\mutfv<\liminf_{r\searrow R}\inf_{|x|=r}\phitfv(x)$
we have 
\begin{equation}\label{eq:muestimate}
        (\mutfv)^{3/4}\leq \frac{2^{3/4}}{3\pi^{1/2}}(1+|a(R)|R^{-\zeta})^{1/2}
                \left(\lim_{r\to\infty}
        (4\pi)^{-1}\int_{{\cal S}^2}rV(r\omega)d\omega
        -\int_{\R^3}\rhotfv(y)dy\right).
\end{equation}
\endproclaim 

\demo{Proof}
According to (\ref{eq:tflowestimate}) we have 
$
        \nu(\mutfv)\leq \liminf_{|x|\to\infty}|x|\phitfv(x).
$
Using that $V$ is harmonic and tends to zero 
at infinity we have that for all $r>R$ 
$$
        \liminf_{|x|\to\infty}|x|\phitfv(x)
        \leq(4\pi)^{-1}\int_{{\cal S}^2}rV(r\omega)d\omega
        -\int_{\R^3}\rhotfv(y)dy.
$$
Moreover since, $\mutfv\geq0$ the assumption 
$\mutfv<\liminf_{r\searrow R}\inf_{|x|=r}\phitfv(x)$ implies that 
the spherical average of $V$ is nonnegative.

On the other hand, since 
$\left(1+|a(R)|R^{-\zeta}\right)^{-2}\leq\left(1+a(R)|x|^{-\zeta}\right)^{-2}$
for $|x|\geq R$, we have from (\ref{eq:nudef}) that
$\nu(\mutfv)\geq\nu'$, where
\begin{eqnarray*}
        \nu'&=&\min_{|x|\geq R}
        \max\Bigl\{
        3^42^{-3}\pi^2|x|^{-3}\left(1+|a(R)|R^{-\zeta}\right)^{-2},
        \mutfv|x|\Bigr\}\\&=&
        3\cdot 2^{-3/4}\pi^{1/2}\left(1+|a(R)|R^{-\zeta}\right)^{-1/2}
        (\mutfv)^{3/4}.\\
\noalign{\vskip-36pt}
\end{eqnarray*}
\enddemo
\vglue12pt

This corollary immediately gives a partial converse to 
Theorem~\ref{thm:maxion}.

\proclaimtitle{Upper bound on maximal ionization}
\proclaim{{C}orollary} 
\label{cl:uppermaxion} If $V$ is harmonic and continuous for $|x|>R$ 
and satisfies $V(x)\to0$ as $|x|\to\infty$ 
and if moreover $\mutfv< \liminf_{r\searrow R}\inf_{|x|=r}\phitfv(x)$
then
$$
        \int\rhotfv\leq \lim_{r\to\infty}
        (4\pi)^{-1}\int_{{\cal S}^2} rV(r\omega)d\omega.
$$
In particular{\rm ,}
if $\liminf_{r\searrow R}\inf_{|x|=r}\phitfv(x)>0$
{\rm (}\/which may not necessarily be true\/{\rm )} we have 
$$
    N_c\leq \lim_{r\to\infty}(4\pi)^{-1}\int_{{\cal S}^2} rV(r\omega)d\omega.
$$
\endproclaim 

\numbereddemo{{R}emark} The limit above of course exists since by the harmonicity 
of $V$ and since $V$ tends
to zero at infinity we have that 
$\int_{{\cal S}^2} rV(r\omega)d\omega$ is independent of $r$. 

The difficulty in using Corollaries \ref{cl:muestimate} and 
\ref{cl:uppermaxion} in 
concrete examples lies in establishing the condition
\begin{equation}\label{eq:chemcondition}
        \mutfv<\liminf_{r\searrow R}\inf_{|x|=r}\phitfv(x).
\end{equation}
\enddemo

\section{Estimates on the standard atomic TF theory}
\label{sec:atomictf}

\advance\eqcount by 47

In the usual atomic case
the Coulomb potential $V(x)=Z|x|^{-1}$ is harmonic away from $x=0$
and we can use Corollary~\ref{cl:uppermaxion} for all $R>0$. 
Since $\rhotf*|x|^{-1}$ is a bounded function it 
follows that $\phitf(x)\to\infty$ as $x\to0$. 
The condition (\ref{eq:chemcondition})
is therefore satisfied if $R$ is chosen small enough. 
It therefore follows from Theorem~\ref{thm:maxion}
and Corollary~\ref{cl:uppermaxion} that $N_c=Z$. 
Thus the neutral atom corresponds to $\mutf=0$.

\proclaim{Lemma}\label{lm:ltf3.8} 
Let $\phitf_0$ be the {\rm TF} potential for the neutral atom 
then if $\phitf$ is the potential for a general $\mutf\geq0$ we have
$$
    \phitf_0(x)\leq \phitf(x)\leq\phitf_0(x)+\mutf.
$$
\endproclaim

\vglue-24pt
{\it Proof}.
See Corollary~3.8 (i) and (iii) in \cite{Lieb:tf}.
\hfill\qed\vglue12pt

We now easily get an upper bound agreeing with the
atomic Sommerfeld asymptotics.

\proclaimtitle{Atomic Sommerfeld upper bound}
\proclaim{Theorem} \label{thm:sommerfeldup}
The atomic {\rm TF} potential satisfies the bound
$$
   \phitf(x)\leq \min\{3^42^{-3}\pi^2|x|^{-4}+\mutf,\ Z|x|^{-1}\}.
$$
\endproclaim

{\it Proof}.
This follows immediately from 
and (\ref{eq:tfpotgeneral}) and (\ref{eq:tfupestimate})
together with the fact that
$\rhotf$ is nonnegative.
Simply note that since $\phitf(x)|x|\to Z$ as $x\to0$ 
we have that $A(0,\mutf)=0$ in (\ref{eq:tfupestimate}). 
\hfill\qed

\proclaimtitle{Lower bound on the TF potential}
\proclaim{Lemma} 
\label{lm:firstphitfbound}
In the atomic case we have for all $N>0$ and $Z>0$ 
$$
   \phitf(x)\geq Z|x|^{-1}-\min\left\{N|x|^{-1},\frac{22}{(9\pi)^{2/3}}Z^{4/3}\right\}.
$$
\endproclaim

\demo{Proof}
We have by Newton's theorem
\begin{eqnarray*}
\rhotf*|x|^{-1}&=&
  |x|^{-1}\int_{|y|<|x|}\rhotf(y)dy+\int_{|y|>|x|}\rhotf(y)|y|^{-1}dy\\
  &\leq& \min\left\{N|x|^{-1},\ \int\rhotf(y)|y|^{-1}dy\right\}.
\end{eqnarray*}
{F}rom the Sommerfeld upper bound Theorem~\ref{thm:sommerfeldup} and
the 
TF equation (\ref{eq:tfeqgeneral}) we have 
$$
\rhotf(x)^{2/3}\leq 2(3\pi^2)^{-2/3}\min\{3^42^{-3}\pi^2|x|^{-4},\
Z|x|^{-1}\}.
$$
Hence
$$
\rhotf(x)\leq\min\left\{c_1Z^{3/2}|x|^{-3/2},\,c_2|x|^{-6}\right\},
$$ 
where $c_1:=2^{3/2}(3\pi^2)^{-1}$ and $c_2:=3^52^{-3}\pi$. 
Let $r_0:=(c_2/c_1)^{2/9}Z^{-1/3}$. When $|x|=r_0$ the two functions, in 
the minimum above, are equal.
Thus
\begin{eqnarray*}
  \int\rhotf(y)|y|^{-1}dy&\leq& 4\pi c_1Z^{3/2}\int_0^{r_0}t^{-1/2}dt
  +4\pi c_2\int_{r_0}^\infty t^{-5}dt=\frac{11\pi}{3}c_1^{8/9}c_2^{1/9}
  Z^{4/3}\\
  &=&\frac{22}{(9\pi)^{2/3}}Z^{4/3}.
\end{eqnarray*}
The lemma follows from the definition (\ref{eq:tfpotgeneral}) of
the TF potential.
\enddemo

\proclaimtitle{Atomic Sommerfeld Lower bound}
\proclaim{Theorem} 
\label{thm:sommerfeldlow}
The {\rm TF} potential satisfies
$$
 \phitf(x)\geq\left\{
 \begin{array}{ll}
   Z|x|^{-1}-22(9\pi)^{-2/3}Z^{4/3}, &\mbox{if } |x|\leq \beta_0Z^{-1/3}\\ \\
   \begin{array}{rl}
   \max\Bigl\{&3^42^{-3}\pi^2\left(1+aZ^{-\zeta/3}|x|^{-\zeta}\right)^{-2}|x|^{-4},\\
     &(Z-N)_+|x|^{-1}\Bigr\},
   \end{array}
  &\mbox{if } |x|\geq \beta_0Z^{-1/3}
 \end{array}
\right. ,
$$
where $\beta_0=\frac{(9\pi)^{2/3}}{44}$ and 
$\zeta=(-7+\sqrt{73})/2$ as in Theorem~{\rm \ref{thm:sommerfeldmu}} and
$
  a= 43.7.
$ 
\endproclaim

\demo{Proof} Let
 $R=(9\pi)^{2/3}Z^{-1/3}/44$.
Note that for
$|x|\leq R$ the bound we want to prove is 
identical to the bound in Lemma.~\ref{lm:firstphitfbound}. 

If $N\geq Z$, i.e., $\mutf=0$ the lower bound follows from 
Theorem~\ref{thm:sommerfeldmu} since $a$ is chosen so 
as to make the lower bound continuous at $|x|=R$
and at these points we clearly have $\phitf(x)>0=\mutf$.

For general $N$ the lower bound follows from the case $N=Z$
because of Lemma~\ref{lm:ltf3.8}
and Lemma~\ref{lm:firstphitfbound}.
\enddemo

We end this section by giving a bound on the {\it screened nuclear 
potential} $\Phitf{R}$ at radius $R$ in the atomic TF theory.

 \proclaimtitle{Bound on $\Phitf{R}$}
\proclaim{Lemma}
\label{lm:Phitfrbound}
We have
$$
 \Phitf{|x|}(x)\leq 3^42^{-1}\pi^2|x|^{-4}+\mutf.
$$
\endproclaim

{\it Proof}.
We write $\Phitf{|x|}(x)=\phitf(x)+\int_{|y|>|x|}\rhotf(y)|x-y|^{-1}dy$.
{F}rom Theorem~\ref{thm:sommerfeldup} and the TF equation 
(\ref{eq:tfeqgeneral}) we see that 
$$
   \rhotf(y)= 2^{3/2}(3\pi^2)^{-1}[\phitf(y)-\mutf]_+^{3/2}
    \leq 2^{-3}3^{5}\pi|y|^{-6}
$$
and hence 
\begin{eqnarray*}
  \int_{|y|>|x|}\rhotf(y)|x-y|^{-1}dy&\leq&
  \int_{|y|>|x|}2^{-3}3^{5}\pi|y|^{-6}|x-y|^{-1}dy\\
  &=&\int_{|y|>|x|}2^{-3}3^{5}\pi|y|^{-7}dy=2^{-3}3^{5}\pi^2|x|^{-4}.
\end{eqnarray*}
The lemma follows from Theorem~\ref{thm:sommerfeldup}.
\hfill\qed

\section{Separating the outside from the inside} 
\label{sec:io}
 \advance\eqcount by 47

We shall here control the energy coming from the regions
far from the nucleus. Let $\gammahf$ be an HF minimizer
with $\Tr[\gammahf]=N$. (We are thus assuming that $N$ is such that 
a minimizer exists.)
\advance\theoremcount by 1

\demo{Definition {\rm 6.1. (The localization function)}}  
Fix $0<\lambda<1$
and let\break $G:\R^3\to\R$ be given by
$$
        G(x)=\left\{\begin{array}{ll}
                0&\mbox{if }|x|\leq1\\
                {(\pi/2)(|x|-1)}\left[(1-\lambda)^{-1}-1\right]^{-1}
                &\mbox{if }1\leq|x|\leq(1-\lambda)^{-1}\\
                {\pi}/{2}&\mbox{if }(1-\lambda)^{-1}\leq|x|.
                   \end{array}\right.
$$

We introduce the cut-off radius $r>0$ and define the outside localization 
function $\thetar(x)=\sin G(|x|/r)\index{$\theta_r$}$. Then 
$$
0\leq\thetar(x)
\left\{\begin{array}{ll}=0&\mbox{if $|x|\leq r$}\\
    \leq1&\mbox{if $r\leq|x|\leq(1-\lambda)^{-1}r$}\\
    =1&\mbox{if $(1-\lambda)^{-1}r\leq|x|$}
  \end{array}
\right.
$$
and $|\nabla\thetar(x)|^2+|\nabla (1-\thetar(x)^2)^{1/2}|^2\leq 
({\pi}/{(2\lambda r)})^2$ (since $(1-\lambda)^{-1}-1\geq\lambda$).
\enddemo

We shall consider the HF minimizer restricted to the region $\set{x}{|x|>r}$.
We therefore define the exterior part of the minimizer 
\begin{equation}\label{eq:gammar}
  \gammar=\thetar\gammahf\thetar
\end{equation}
and its density $\rhohfr(x)=\thetar(x)^2\rhohf(x)$.
In order to control $\gammar$ we introduce an auxiliary functional
defined on all density matrices with 
$\Tr\left[-\Delta\gamma\right]<\infty$ (see Remark~\ref{remark:deltagamma}) by 
\begin{equation}\label{eq:eadef}
  \ea(\gamma)=\Tr\left[(-\mfr{1}{2}\Delta-\Phihf{r})
    \gamma\right]+
  \mfr{1}{2}\iint\limits_{|x|\geq r\atop|y|\geq r}
  \uprho_\gamma(x)|x-y|^{-1}\uprho_{\gamma}(y)dx\,dy,\hskip.35in
\end{equation}
where the screened nuclear potential $\Phihf{r}$ 
is defined in (\ref{eq:Phihf}) in Definition~1.1.
Note that the functional $\ea$, in contrast to the HF functional 
$\ehf$ in (\ref{eq:ehf}), does not contain an exchange term.
 
The main result in this section is that $\gammar$ almost
minimizes $\ea$. More precisely, we shall prove the following theorem. 
 
\proclaimtitle{The outside energy}
\proclaim{Theorem} \label{thm:outsideenergy}
For all $0<\lambda<1$ and all $r>0$ we have 
\begin{equation}\label{eq:outsideenergy}
  \ea\left[\gammar\right]\leq  \inf\set{\ea(\upgamma)}{
    \supp\uprho_{\gamma}\subset\set{y}{|y|\geq r},\ 
    \int\uprho_{\gamma}\leq 
    \int\chiplus{r}\rhohf}
  +{\cal R},
\end{equation}
where the error is 
\begin{equation}\label{eq:ioerror}
  {\cal R}=C_\lambda(r)\!\!
  \int\limits_{|x|\geq(1-\lambda)r}\!\!\rhohf(x)dx\nonumber
  +2L_1\hspace{-5ex}\int\limits_{{(1-\lambda)r\leq|x|\leq
      (1-\lambda)^{-1}r}}
  \hspace{-5ex}\left[\Phihf{(1-\lambda)r}(x)\right]_+^{5/2}
  dx+
  \Ex\left(\gammar\right)
  \nonumber \pagebreak
\end{equation}
with 
$$
C_\lambda(r)=\left(\frac{\pi^2}{8\left(\lambda(1-\lambda)r\right)^{2}}
  +\frac{\pi}{r\lambda}\right).
$$
Here $L_1$ is the constant in the Lieb\/{\rm -}\/Thirring inequality 
{\rm (\ref{eq:LT}).}
\endproclaim

\demo{Proof}
Besides $\thetar$ we introduce two other localization functions
$$
        \theta_-=(1-\theta_{r(1-\lambda)}^2)^{1/2}\qquad\hbox{and}\qquad
        \thetam=\left(\theta_{r(1-\lambda)}^2-\thetar^2\right)^{1/2}.
$$
Note that $\theta_-^2+\thetam^2+\thetar^2=1$ and that 
$$(\nabla\theta_-)^2+(\nabla\thetam)^2+(\nabla\thetar)^2\leq 
\left(\pi/(2\lambda(1-\lambda)r)\right)^2.$$
We introduce the inside part of the HF minimizer
\begin{equation}\label{eq:gammai}
        \gammai=\theta_-\gammahf\theta_-.
\end{equation}
We shall prove (\ref{eq:outsideenergy}) by showing that
for all density matrices $\upgamma$ with 
$\supp\uprho_\gamma\subset\set{x}{|x|\geq r}$
and $\int\uprho_{\gamma}\leq\int\chiplus{r}\rhohf$ we have
\begin{equation}\label{eq:outsideenergy1}
        \ea\left[\gammar\right]+\ehf\left[\gammai\right]-{\cal R}
        \leq\ehf\left[\gammahf\right]\leq\ea[\upgamma]+
        \ehf\left[\gammai\right],
\end{equation}
with ${\cal R}$ given by (\ref{eq:ioerror}).
The estimate (\ref{eq:outsideenergy}) 
follows immediately from (\ref{eq:outsideenergy1}).

\demo{{P}roof of the upper bound in {\rm (\ref{eq:outsideenergy1})}} 
Since $\gammahf$ is a minimizer for $\ehf$ under the condition 
$\Tr[\gammahf]\leq N$ (see (\ref{eq:rel}))
we have for any density matrix $\widetilde{\upgamma}$ with 
$\Tr\left[\widetilde{\upgamma}\right]\leq N$
that $\ehf\left(\gammahf\right)\leq\ehf\left(\widetilde{\upgamma}\right)$.
We take
$$
        \widetilde{\upgamma}=\gammai+\upgamma.
$$
Since the support of 
$\uprho_\gamma$ is disjoint from the support of $\theta_-$
we see that $\gammai\upgamma=0$ and hence 
$\widetilde{\upgamma}$ is a density matrix.

Note that 
$$
        \Tr\left[\widetilde{\upgamma}\right]=\Tr\left[\gammai\right]
        +\int\uprho_{\gamma}
        \leq\Tr\left[\gammai\right]+\int\chiplus{r}\rhohf=
        \int(\theta_-^2+\chiplus{r})\rhohf\leq\int\rhohf\leq N.
$$

We shall compute $\ehf\left(\widetilde{\upgamma}\right)$. 
The only terms in $\ehf$ that are not linear in the density matrix 
(and thus do not simply split into a sum of terms  for $\gammai$ and
 $\upgamma$) are the exchange and direct Coulomb energies.
Because of the support properties of $\gammai$ and $\upgamma$ we have that 
$\widetilde{\upgamma}^2=\left(\gammai\right)^2+\upgamma^2$ 
and therefore even the exchange
term satisfies
$$
        \Ex\left(\widetilde{\upgamma}\right)=\Ex\left(\gammai\right)
        +\Ex\left(\upgamma\right)\geq\Ex\left(\gammai\right).
$$

We are thus left with the direct Coulomb energy. For this we 
find that 
$$
        \D\left(\widetilde{\upgamma}\right)=
        \D\left(\gammai\right)+\D\left(\upgamma\right)
        +\int \theta_-^2(y)\rhohf(y)|x-y|^{-1}\uprho_{\gamma}(y)dy.\pagebreak
$$
By the choice of the support of $\theta_-$ we have that 
$$
        \int \theta_-^2(y)\rhohf(y)|x-y|^{-1}\uprho_{\gamma}(x)dx\,dy
        \leq \int\left( \frac{Z}{|x|}
        -\Phihf{r}(x)\right)\uprho_{\gamma}(x)dx.
$$
We have thus proved the upper bound in (\ref{eq:outsideenergy1}).
\enddemo
 
{\it Proof of the lower bound in} (\ref{eq:outsideenergy1}).
Let again the inside part of the HF minimizer be $\gammai$ defined by 
(\ref{eq:gammai}) 
and introduce also the middle part $\gammam=\thetam\gammahf\thetam$. 
Since $\theta_-^2+\thetam^2+\thetar^2=1$  we have from the IMS formula
(\ref{eq:IMS}) that 
\begin{eqnarray}
&&\label{eq:ioklo}\\ 
        \Tr\left[-\mfr{1}{2}\Delta\gammahf\right]&=&\Tr\left[-\mfr{1}{2}\Delta
        \left(\gammai+\gammam
        +\gammar        \right)\right] \nonumber
\\[5pt]
        &&{}-\mfr{1}{2}\Tr\left[\gammahf\left((\nabla\theta_-)^2+(\nabla\thetam)^2
        +(\nabla\thetar)^2\right)
        \right]\nonumber\\[4pt] 
        &\geq&\Tr\left[-\mfr{1}{2}\Delta\left(\gammai+\gammam
        +\gammar        \right)\right]\nonumber\\[4pt]  
        &&{}-({\pi}^2/8)\left(\lambda(1-\lambda)r\right)^{-2}
        \int\limits_{(1-\lambda)r<|x|<r(1-\lambda)^{-1}}\rhohf(x) dx. \nonumber
       \end{eqnarray}

We now come to the lower bounds on the Coulomb terms.
Note that  
\begin{eqnarray*}
        1&=&\left(\theta_-^2(x)+\thetam^2(x)+\thetar^2(x)\right)
        \left(\theta_-^2(y)+\thetam^2(y)+\thetar^2(y)\right)\\[4pt] 
        &\geq&\theta_-^2(x)\theta_-^2(y)+\thetar^2(x)\thetar^2(y)
        +\thetar^2(x)\left(\theta_-^2(y)+\thetam^2(y)\right)\\[4pt] 
        {}&&+\left(\theta_-^2(x)+\thetam^2(x)\right)\thetar^2(y)
        +\thetam^2(x)^2\theta_-^2(y)+\theta_-^2(x)^2\thetam^2(y).
\end{eqnarray*}
Note that $\theta_-^2(x)+\thetam^2(x)\geq \upchi{r}(x)$ and 
$\theta_-^2(x)\geq\upchi{(1-\lambda)r}$
We may therefore estimate the Coulomb kernel from below by
$$
        |x-y|^{-1}\geq \widetilde{V}(x,y),
$$
where
\begin{eqnarray}\label{eq:vtilde}
 \qquad \widetilde{V}(x,y)&:=&\theta_-^2(x)|x-y|^{-1}\theta_-^2(y)+
  \theta^2_r(x)|x-y|^{-1}\theta^2_r(y) \\[4pt]
  &&+\theta_r^2(x)|x-y|^{-1}\upchi{r}(y)+\upchi{r}(x)|x-y|^{-1}\theta_r^2(y) 
  \nonumber\\[4pt]
  && +\thetam^2(x)|x-y|^{-1}\upchi{(1-\lambda)r}(y)+\upchi{(1-\lambda)r}(x)|x-y|^{-1}\thetam^2(y)
  \nonumber.
\end{eqnarray}
The function $\widetilde{V}$ is pointwise positive and symmetric in $x$ and 
$y$.

Recall that $\gammahf$ is a projection onto the subspace spanned by
the orthonormal vectors $u_1,u_2,\ldots,u_N$ \pagebreak and 
\begin{eqnarray}
\noalign{\vskip-4pt}
&& \label{eq:iodelo}\\
\noalign{\vskip-4pt}
        \lefteqn{\D\left(\gammahf\right)-\Ex\left(\gammahf\right)
        =\mfr{1}{4}\sum_{i,j}\iint 
        \frac{\left\|u_i(x)\otimes u_j(y)
        -u_j(x)\otimes u_i(y)\right\|_{\C^2\otimes\C^2}^2}
        {|x-y|}dx\,dy}\hspace{1truecm}
      \nonumber\\
        &\geq& 
        \mfr{1}{4}\sum_{i,j}\iint 
        \left\|u_i(x)\otimes u_j(y)
        -u_j(x)\otimes u_i(y)\right\|_{\C^2\otimes\C^2}^2
        \widetilde{V}(x,y) dx\,dy  \nonumber\\
        &=&\mfr{1}{2}\iint \rhohf(x)\widetilde{V}(x,y)\rhohf(y) dx\,
        dy 
         \nonumber\\&&{}-
        \mfr{1}{2} \iint\Tr_{\C^2}\left|\gammahf(x,y)\right|^2 
        \widetilde{V}(x,y) dx\, dy.\nonumber
\end{eqnarray}
We estimate these two terms independently. We obtain for the
first term in~(\ref{eq:iodelo})
\begin{eqnarray}
      &&  \mfr{1}{2}\iint \rhohf(x)\widetilde{V}(x,y)\rhohf(y) dx\, dy
        =\D\left(\gammai\right)+\D\left(\gammar\right)\label{eq:iodlo}\\
        &&\qquad+\int 
        \left(\frac{Z}{|x|}-\Phihf{r}(x)\right)
        \rhohfr(x)dx
        +\Tr\left[\left(\frac{Z}{|x|}-\Phihf{(1-\lambda)r}(x)\right)
        \gammam\right]. \nonumber
\end{eqnarray}
To estimate the last term in (\ref{eq:iodelo}) we use that 
for all $|x|>r$
$$
        \thetar(x)^2\leq \thetar(x)\leq
        (\pi/2)(|x|/r-1)\left((1-\lambda)^{-1}-1\right)^{-1}.
$$
Thus if  $|y|<r$ we have
$
        |x-y|^{-1}\thetar(x)^2\leq (\pi/2)r^{-1}
        \left((1-\lambda)^{-1}-1\right)^{-1}
$
and hence 
\begin{eqnarray*}
        \lefteqn{\iint\Tr_{\C^2}\left[\left|\gammahf(x,y)\right|^2 \right]
        |x-y|^{-1}\left(\upchi{r}(y)\thetar^2(x)
        +\upchi{(1-\lambda)r}(y)\thetam^2(x)\right) dx\, dy}\\ \\
        &\leq& \frac{\pi}{2r}
        \left((1-\lambda)^{-1}-1\right)^{-1}
        \left(1+(1-\lambda)^{-1}\right)
         \iint\limits_{|y|\leq r
        \atop|x|\geq (1-\lambda)r}
        \Tr_{\C^2}\left[\left|\gammahf(x,y)\right|^2\right] dx\, dy\\ \\
        &\leq& \frac{\pi}{r\lambda}
        \iint\limits_{|y|\leq r
        \atop|x|\geq (1-\lambda)r}
        \Tr_{\C^2}\left[\left|\gammahf(x,y)\right|^2\right] dx\, dy.
\end{eqnarray*}
Moreover, we only increase the last integral if we integrate over 
all $y\in\R^3$. Thus 
\begin{eqnarray*}
        \iint\limits_{|y|\leq r
        \atop|x|\geq (1-\lambda)r}\Tr_{\C^2}\left|\gammahf(x,y)\right|^2 
        dx\, dy
        &\leq& \int\limits_{|x|\geq (1-\lambda)r}\left(\Tr_{\C^2}\int
        \gammahf(x,y)\gammahf(y,x) dy\right)dx\\ \\
        &=& \int\limits_{|x|\geq (1-\lambda)r} 
        \Tr_{\C^2} \left[\left(\gammahf\right)^2(x,x)\right] dx.
\end{eqnarray*}
If we now use that $\left(\gammahf\right)^2=\gammahf$ and that 
$\rhohf(x)=\Tr_{\C^2}\left [\gammahf(x,x)\right] $
we obtain the \pagebreak estimate
\begin{eqnarray}\label{eq:ioelo}
  \mfr{1}{2} \iint\Tr_{\C^2}\left|\gammahf(x,y)\right|^2 
  \widetilde{V}(x,y) dx\, dy
  &\leq&\Ex\left[\gammai\right]+\Ex\left[\gammar\right]\\ 
  \nonumber \\
  &&{}+\frac{\pi}{r\lambda}
  \!\!\int\limits_{|x|\geq (1-\lambda)r}\!\!\!\!\rhohf(x)dx.\nonumber
\end{eqnarray}

If we combine (\ref{eq:ioklo}), (\ref{eq:iodlo}) and 
(\ref{eq:ioelo}) we obtain 
\begin{eqnarray*}
        \ehf\left[\gammahf\right]&\geq& \ehf\left[\gammai\right]
        +\ea\left[\gammar\right]-\Ex\left[\gammar\right]
        +\Tr\left[\left(-\mfr{1}{2}\Delta-\Phihf{(1-\lambda)r}\right)
        \gammam\right] \\ \\
        &&{}-
        \left(\frac{\pi^2}{8\left(\lambda(1-\lambda)r\right)^{2}}
        +\frac{\pi}{r\lambda}\right)
        \int\limits_{|x|\geq(1-\lambda)r}\rhohf(x)dx.
\end{eqnarray*}
Since $0\leq \gammam\leq I$ and the density of $\gammam$
is supported within the set 
$$
        \set{x}{(1-\lambda)r\leq|x|\leq
        (1-\lambda)^{-1}r}
$$ 
we have from the Lieb-Thirring
inequality (\ref{eq:LT}) that 
$$
        \Tr\left[\left(-\mfr{1}{2}\Delta-\Phihf{(1-\lambda)r}\right)
        \gammam\right]\geq -2L_1\int\limits_{(1-\lambda)r\leq|x|\leq
        (1-\lambda)^{-1}r}\left[\Phihf{(1-\lambda)r}(x)\right]_+^{5/2}dx.
$$
The factor of 2 above is due to the spin degrees of freedom.
We have thus proved the lower bound in (\ref{eq:outsideenergy1}).\hfill\qed
\vglue12pt

As a consequence of this theorem and the Lieb-Thirring inequality
(\ref{eq:LTdensity})
we get the following bound.

\proclaimtitle{$L^{5/3}$ bound on $\rhohfr$}
\proclaim{{C}orollary} 
Let $K_1$ denote the constant in the {\rm LT} inequality
{\rm (\ref{eq:LTdensity})} and $e_0${\rm ,} as in {\rm (\ref{eq:ine0}),} 
denote the {\rm TF} energy of a neutral atom with unit nuclear charge and
physical parameter values. Then 
 \begin{equation}\label{eq:l5/3rhohfr}
  \int \rhohfr(y)^{5/3}dy\leq 2 K_1^{-1}{\cal R}+
  \mfr{6}{5}(3\pi^2)^{2/3}K_1^{-2}e_0
  \left[r\sup_{|x|=r}\Phihf{r}(x)\right]_+^{7/3},\hskip.35in
\end{equation}
where $\cal R$ was given in {\rm (\ref{eq:ioerror}).}
\endproclaim 

\demo{Proof}
Since $\Phihf{r}$ is harmonic on the set 
$\{|x|>r\}$ and tends to zero at infinity
we get for all $|y|>r$ that $\Phihf{r}(y)\leq 
|y|^{-1}r\sup_{|x|=r}\Phihf{r}(x)$.
Hence 
\begin{eqnarray*}
 \ea\left[\gammar\right]&\geq&  K_1\int\rhohfr(y)^{5/3}dy
 -\left[r\sup_{|x|=r}\Phihf{r}(x)\right]\int|y|^{-1}\rhohfr(y)dy\\
 &&{}
 +\mfr{1}{2}\iint\rhohfr(y)|y-y'|^{-1}\rhohfr(y')
dy\, dy'.
\end{eqnarray*}
From standard atomic Thomas-Fermi \pagebreak theory it follows  that 
the right-hand side is bounded below by the energy of a neutral  
Thomas-Fermi atom with nuclear charge 
$\left[r\sup_{|x|=r}\Phihf{r}(x)\right]_+$ and with the constant 
$K_1$ in front of the first term. A simple scaling argument
shows that this is 
$$-\frac{3}{10}(3\pi^2)^{2/3}K_1^{-1}
\left[r\sup_{|x|=r}\Phihf{r}(x)\right]_+^{7/3}e_0.
$$

By repeating this argument with only a fraction of
the term $\int (\rhohfr)^{5/3}$ we conclude that 
for all $0<t<1$
$$
  (1-t) K_1\int \rhohfr(y)^{5/3}dy\leq \ea\left[\gammar\right]+
  \mfr{3}{10}(3\pi^2)^{2/3}(tK_1)^{-1}e_0
  \left[r\sup_{|x|=r}\Phihf{r}(x)\right]_+^{7/3}.
$$   
Since $\ea\left[\gammar\right]\leq {\cal R}$ (by choosing the trial
$\gamma=0$ in (\ref{eq:outsideenergy}) ) we get (\ref{eq:l5/3rhohfr}) 
if we choose $t=1/2$.
\enddemo

We still need to show how we can control the exchange term 
$\Ex\left[\gammar\right]$. This is done using a standard 
inequality of Lieb~\cite{Lieb:Coul} (or in an improved version by Lieb 
and Oxford~\cite{LiebOxford}).
They proved the inequality for general wave functions, but we 
need it here only for Hartree-Fock Slater determinants or more
precisely for density matrices.
For completeness we shall give a proof (with a worse constant)
in the simple case we need here.

\proclaimtitle{Exchange inequality}
\proclaim{Theorem} \label{thm:exchineq} 
For any trace class operator $\gamma$ with $0\leq \gamma\leq\I$ we have the estimate
$$
        \Ex\left[\gamma\right]\leq
        1.68 \int\uprho_{\gamma}^{4/3}.
$$
\endproclaim

{\it Proof}.
We shall here present a simple proof that the inequality holds
with 1.68 replaced by 248.3.
To get the much better constant one needs the more
detailed analysis in \cite{LiebOxford}.
We use the representation
$$
        |x|^{-1}=\pi^{-1}\int_0^\infty \upchi{r}*\upchi{r}(x) r^{-5} dr,
$$
where $\upchi{r}$ again denotes the characteristic function of the ball of radius
$r$ centered at the origin.
Thus we may write the exchange energy as
$$
        \Ex\left[\gamma\right]=(2\pi)^{-1}\int_0^\infty\int_{\R^3}
        \Tr[\gamma X_{r,z}\gamma X_{r,z}] r^{-5}dz dr,
$$
where $X_{r,z}$ is the multiplication operator 
$X_{r,z}f(x)=\upchi{r}(x-z)f(x)$.

We now use the two simple estimates
$X_{r,z}\gamma X_{r,z}\leq X_{r,z}^2$ and $X_{r,z}\gamma X_{r,z}
\leq \Tr[\gamma X_{r,z}^2]\I$.
We \pagebreak obtain
$$
        \Tr[\gamma X_{r,z}\gamma X_{r,z}]=
        \Tr[\gamma^{1/2} X_{r,z}(X_{r,z}\gamma X_{r,z})X_{r,z}\gamma^{1/2}]
        \leq \Tr[\gamma X_{r,z}^2]
        = \uprho_{\gamma}*\upchi{r}(z)
$$
and
$$
        \Tr[\gamma X_{r,z}\gamma X_{r,z}=
        \Tr[\gamma^{1/2} X_{r,z}(X_{r,z}\gamma X_{r,z})X_{r,z}\gamma^{1/2}]
        \leq  \Tr[\gamma X_{r,z}^2]^2 = (\uprho_{\gamma}*\upchi{r}(z))^2.
$$

If $\uprho_{\gamma}^*$ denotes the Hardy-Littlewood maximal function we have 
$\uprho_{\gamma}*\upchi{r}(z)\leq (4\pi r^3/3)\uprho_{\gamma}^* (z)$.
Thus with $R(z)=\left((4\pi/3)\uprho_{\gamma}^* (z)\right)^{-1/3}$ 
we can estimate
\begin{eqnarray*}
        \Ex\left[\gamma\right]&\hskip-8pt\leq\hskip-8pt& (2\pi)^{-1}\int_{\R^3} \left(\int_0^{R(z)}
        \left(\frac{4\pi r^3}{3}\uprho_{\gamma}^* (z)\right)^2r^{-5} dr 
        +\int_{R(z)}^\infty
        \frac{4\pi r^3}{3}\uprho_{\gamma}^* (z)r^{-5}dr\right) dz\\
        &\hskip-8pt=\hskip-8pt&(4\pi/3)^{1/3}\int_{\R^3}\left(\uprho_{\gamma}^* (z)\right)^{4/3}dz.
\end{eqnarray*}

If we finally apply the maximal inequality 
$\|f^*\|_p^p\leq (48 p2^p/(\pi(p-1))\|f\|_p^p$, for all $p>1$
(\cite{SteinWeiss}, p.~58) we obtain
\vglue12pt
\hfill ${\displaystyle
        \Ex\left[\gamma\right]\leq
        \frac{384}{\pi}\left(\frac{8\pi}{3}\right)^{1/3}\int
        \uprho_{\gamma}^{4/3}=248.3\int
        \uprho_{\gamma}^{4/3}.
}$
\hfill\qed
\advance\eqcount by 59

\section{Exterior $L^1$-estimate} 
\label{sec:l1}

The aim of this section is to control 
$\int_{|x|>r}\rhohf(x)dx$, for all $r>0$.
As before $\rhohf$ is the density of a Hartree-Fock minimizer
$\gammahf$ with $\Tr[\gammahf]=N$. Thus $\int\rhohf=N$.
 
The difficulty in estimating $\int_{|x|>r}\rhohf(x)dx$ is that  
this quantity cannot be controlled in terms
of the energy $\ehf(\gammahf)$. 
More precisely, $\int_{|x|>r}\uprho_\gamma(x)dx$
can be arbitrarily large even when  $\ehf(\gamma)$
is arbitrarily close to the absolute minimum. 
The simple reason is that ``adding electrons at 
infinity'' will not raise the energy.

Therefore, in order to control $\int_{|x|>r}\rhohf(x)dx$,
we must use the minimizing property of
$\gammahf$.

In contrast, it follows from the Lieb-Thirring inequality
that\break $\int_{|x|>r}\rhohf(x)^{5/3}dx$ can be controlled
in terms of the energy.
By H\"older's inequality it then also follows 
that the integral of $\rhohf$ over any bounded set
can be controlled by the energy. 

The philosophy here will be, to use the minimizing
property of $\gammahf$, to control the integral
of $\rhohf$ over an unbounded set, in terms 
of the integral over a bounded set.

Our main result in this section is stated 
in the next lemma. The proof of the lemma
uses an idea of \pagebreak Lieb~\cite{Lieb:2z+1}.

\proclaimtitle{Exterior $L^1$-estimate}
\proclaim{Lemma} 
\label{lm:extl1}
For all $r>0$ and all $0<\lambda<1$
the density $\rhohf$ of an {\rm HF} minimizer $\gammahf$ 
satisfies the bound 
\begin{eqnarray*}
        \int_{|x|>(1-\lambda)^{-1}r}\rhohf(x)dx&\leq &
        1+2\lambda^{-1}+2\left[\sup_{|x|=(1-\lambda)r}
        |x|\Phihf{(1-\lambda)r}(x)\right]_+\\
        &&{}+\left(K_\lambda r^{-1}\int_{r<|x|<(1-\lambda)^{-1}r}
        \rhohf(x)dx\right)^{1/2},
\end{eqnarray*}
where $K_\lambda:=(\frac{2\lambda}{\pi}+(1-\lambda)^{-1})
\left(\frac{\pi}{2\lambda}\right)^2$.
Here $\Phihf{(1-\lambda)r}$ is the screened nuclear potential 
introduced in 
Definition~{\rm 1.1}.
\endproclaim

{\it Proof}.
Since $\gammahf$ is a minimizer we know that it satisfies the 
Hartree-Fock equations. For example, according to 
Theorem~\ref{thm:HFequations}, $\gammahf$ is a projection 
onto a space spanned by functions $u_1,\ldots,u_N\in L^2(\R^3;\C^2)$ 
satisfying $\HMF u_i =\epsilon_i u_i$, were $\epsilon_i\leq0$.

Let $\Xi\in C^1(\R^3)$ have compact support away from 
$x=0$, be real and satisfy $\Xi(x)^2\leq 1$. Then 
\advance\eqcount by 59
$$
        0\geq\sum_{i=1}^N\epsilon_i\int\left|u_i(x)\right|^2|x|\Xi(x)^2\,dx
        =\sum_{i=1}^N\int u_i(x)^*|x|\Xi(x)^2 \HMF u_i(x) dx.
$$
{F}rom the definition (\ref{eq:HMF}) of the mean field operator
$\HMF$ we obtain 
\begin{eqnarray}
        0&\geq&\sum_{i=1}^N\mfr{1}{2}\int
        \nabla\left(u_i(x)^*|x|\Xi(x)^2\right)
        \cdot\nabla u_i(x)dx
        -Z\int\rhohf\Xi^2\label{eq:|x|HMF}\\
        &&+\iint \left[\rhohf(x)\rhohf(y)-
        \Tr_{\C^2}\left[\left|\gammahf(x,y)\right|^2\right]\right]
        \frac{|y|\Xi(y)^2}{|x-y|}dx\,dy. \nonumber
\end{eqnarray}
We consider separately the different terms above. 
By the IMS formula (\ref{eq:IMSformula})    
\begin{eqnarray}
  \Re\lefteqn{\mfr{1}{2}\int
    \nabla\left(u_i(x)^*|x|\Xi(x)^2\right)
    \cdot\nabla u_i(x)dx}\label{eq:hardy}\\
  &=&\mfr{1}{2}\int \left|\nabla\left(
      \Xi(x)|x|^{1/2}u_i(x)\right)\right|^2
  -\mfr{1}{4}|x|^{-1}\Xi(x)^2
  \left|u_i(x)\right|^2 dx\nonumber\\
  &&-\mfr{1}{2}\int 
  \left(\mfr{1}{2}\nabla\left(\Xi(x)^2\right)
    +|x|\left(\nabla\Xi(x)\right)^2\right)
  \left|u_i(x)\right|^2 dx\nonumber\\
  &\geq&-\mfr{1}{2}\int\left(\mfr{1}{2}\nabla\left(\Xi(x)^2\right)+
    |x|\left(\nabla\Xi(x)\right)^2\right)
  \left|u_i(x)\right|^2 dx,\nonumber
\end{eqnarray}
where we have used Hardy's inequality $\int|\nabla f(x)|^2dx
\geq\mfr{1}{4}\int|x|^{-2}|f(x)|^2dx$.

For the Coulomb terms we estimate it using that
$$
        \rhohf(x)\rhohf(y)-
        \Tr_{\C^2}\left[\left|\gammahf(x,y)\right|^2\right]
        =\mfr{1}{2}\sum_{i,j=1}^N\left\|u_i(x)\otimes u_j(y)
        -u_j(x)\otimes u_i(y)\right\|_{\C^2\otimes\C^2}^2
$$
is nonnegative. Hence
\begin{eqnarray*}
        \lefteqn{\iint \left[\rhohf(x)\rhohf(y)-
        \Tr_{\C^2}\left[\left|\gammahf(x,y)\right|^2\right]\right]
        \frac{|y|\Xi(y)^2}{|x-y|}dx\,dy}\\ \nonumber\\
        &=&\iint \left[\rhohf(x)\rhohf(y)-
        \Tr_{\C^2}\left[\left|\gammahf(x,y)\right|^2\right]\right]
        \frac{|y|\left(1-\Xi(x)^2\right)\Xi(y)^2}{|x-y|}dx\,dy\\ \nonumber\\
        &&{}+\mfr{1}{2}\iint \left[\rhohf(x)\rhohf(y)-
        \Tr_{\C^2}\left[\left|\gammahf(x,y)\right|^2\right]\right]
        \frac{|x|+|y|}{|x-y|}\Xi(x)^2\Xi(y)^2dx\,dy,
\end{eqnarray*}
where we expressed the last term symmetrically in $x$ and $y$.
If we now use the triangle inequality and the fact
$\int\Tr_{\C^2}\left[\left|\gammahf(x,y)\right|^2\right]\Xi(y)^2dy
\leq \rhohf(x)$, which follows from 
$\Xi(x)^2\leq 1$ and $(\gammahf)^2=\gammahf$, we arrive at
\begin{eqnarray}
&&\label{eq:triangle}\\
        \lefteqn{\iint \left[\rhohf(x)\rhohf(y)-
        \Tr_{\C^2}\left[\left|\gammahf(x,y)\right|^2\right]\right]
        \frac{|y|\Xi(y)^2}{|x-y|}dx\,dy}\nonumber\\ 
        &\geq&\iint \left[\rhohf(x)\rhohf(y)-
        \Tr_{\C^2}\left[\left|\gammahf(x,y)\right|^2\right]\right]
        \frac{|y|\left(1-\Xi(x)^2\right)\Xi(y)^2}{|x-y|}dx\,dy 
        \nonumber \\
        &&{}+\mfr{1}{2}\left(\int\rhohf\Xi^2\right)^2-
        \mfr{1}{2}\int\rhohf\Xi^2.\nonumber
\end{eqnarray}

Inserting the inequalities (\ref{eq:hardy}) and
(\ref{eq:triangle}) into (\ref{eq:|x|HMF}) gives
\begin{eqnarray}
&&\label{eq:Xifinal}\\
        0&\geq& -\mfr{1}{2}\int\left(\mfr{1}{2}\nabla\left(\Xi(x)^2\right)+
        |x|\left(\nabla\Xi(x)\right)^2\right)
        \rhohf(x) dx -Z\int\rhohf\Xi^2\nonumber
\\  
 &&{}+\iint \left[\rhohf(x)\rhohf(y)-
        \Tr_{\C^2}\left[\left|\gammahf(x,y)\right|^2\right]\right]
        \frac{|y|\left(1-\Xi(x)^2\right)\Xi(y)^2}{|x-y|}dx\,dy 
        \nonumber\\
 &&{}+\mfr{1}{2}\left(\int\rhohf\Xi^2\right)^2-\mfr{1}{2}\int\rhohf\Xi^2.\nonumber
 \end{eqnarray}

By an approximation argument it is clear that 
we can use (\ref{eq:Xifinal}) for any real function $\Xi$ 
for which $\Xi^2\leq1$ and the function
$\left(\nabla\left(\Xi(x)^2\right)+
        |x|\left(\nabla\Xi(x)\right)^2\right)$ is bounded.
In particular we can choose $\Xi$ identically equal
to $1$, which will recover Lieb's result from \cite{Lieb:2z+1}, 
i.e., $\int\rhohf\leq 2Z+1$.

We shall now choose $\Xi:=\thetar$, where $\thetar$ is the localization 
function given in Definition~6.1.
Then
\begin{equation}
        \mfr{1}{2}\nabla\left(\Xi(x)^2\right)+
        |x|\left(\nabla\Xi(x)\right)^2
        \leq \frac{\pi}{2\lambda r} 
        +(1-\lambda)^{-1}r\frac{\pi^2}{\left(2\lambda r\right)^2}
        =K_\lambda r^{-1}\hskip.4in
        \label{eq:locerror} \pagebreak
\end{equation}
and
\begin{eqnarray*}
        \iint\left[\rhohf(x)\rhohf(y)-
        \Tr_{\C^2}\left[\left|\gammahf(x,y)\right|^2\right]\right]
        \frac{|y|\left(1-\Xi(x)^2\right)\Xi(y)^2}{|x-y|}dx\,dy\\ 
        \geq \iint_{|x|<(1-\lambda)r}
        \left[\rhohf(x)\rhohf(y)-
        \Tr_{\C^2}\left[\left|\gammahf(x,y)\right|^2\right]\right]
        \frac{|y|\thetar(y)^2}{|x-y|}dx\,dy.
\end{eqnarray*}
If we now use that $|x|<(1-\lambda)r$ and $y\in\supp\thetar$ imply
that $|y||x-y|^{-1}\leq \lambda^{-1}$ we obtain 
\begin{eqnarray}\quad\qquad
        \iint\left[\rhohf(x)\rhohf(y)-
        \Tr_{\C^2}\left[\left|\gammahf(x,y)\right|^2\right]\right]
        \frac{|y|\left(1-\Xi(x)^2\right)\Xi(y)^2}{|x-y|}dx\,dy 
        \label{eq:effpotential}\\
        \geq \iint_{|x|<(1-\lambda)r}
        \rhohf(x)\rhohf(y)\frac{|y|\thetar(y)^2}{|x-y|}dx\,dy
        -\lambda^{-1}\int\thetar(y)^2\rhohf(y)\,dy, \nonumber
\end{eqnarray}
where we have also used that 
$\int\Tr_{\C^2}\left[\left|\gammahf(x,y)\right|^2\right]dx
=\rhohf(y)$.
\vglue4pt
If we insert (\ref{eq:locerror}) and (\ref{eq:effpotential})
into (\ref{eq:Xifinal}) we arrive at 
\begin{eqnarray*}
        0&\geq& -\frac{K_\lambda}{2r}\int_{r<|x|<(1-\lambda)^{-1}r}\rhohf(x)dx
        -\int\thetar(y)^2|y|\Phihf{(1-\lambda)r}(y)\rhohf(y)dy\\
        &&{}+\mfr{1}{2}\left(\int\rhohf\thetar^2\right)^2
        -\left(\mfr{1}{2}+\lambda^{-1}\right)\int\rhohf\thetar^2.
\end{eqnarray*}
Using now that $\Phihf{(1-\lambda)r}(y)$ tends to zero at infinity 
and is harmonic 
for $|y|>(1-\lambda)r$, which contains the support of 
$\thetar$, 
we see by a simple comparison argument that
$$
        \thetar(y)^2|y|\Phihf{(1-\lambda)r}(y)
        \leq \thetar(y)^2 \left[\sup_{|x|=(1-\lambda)r}
        |x|\Phihf{(1-\lambda)r}(x)\right]_+.
$$
Thus 
\begin{eqnarray*}
        0&\geq&\left(\int\rhohf\thetar^2\right)^2
        -\left(1+2\lambda^{-1}+2\left[\sup_{|x|=(1-\lambda)r}
        |x|\Phihf{(1-\lambda)r}(x)\right]_+\right)\int\rhohf\thetar^2\\
        &&
        -K_\lambda r^{-1}\int_{r<|x|<(1-\lambda)^{-1}r}\rhohf(x)dx.
\end{eqnarray*}
Finally, in order to arrive at the result of the lemma
we simply use that $0\geq X^2-BX -C$ for $B,C>0$ 
implies $X\leq B+\sqrt{C}$.
\hfill\qed

\vglue-6pt
\section{The semiclassical estimates} 
\label{sec:sc}
\advance\eqcount by 65
\vglue-6pt

In this section we derive the relevant semiclassical estimates.
We do not attempt to give optimal results. We shall be satisfied
with what is needed for the application we have in mind.
In a certain sense it is misleading to refer to the estimates
in this section as {\it semiclassical}. 
Usually, semiclassics refers to the limit
as Planck's constant $\hbar$ tends to zero. One then expands the relevant
physical quantities like energy and density in powers of $\hbar$. 
In our setting there is, however, no semiclassical parameter which could
play the role of  Planck's constant. It is rather that we consider
potentials for which the  semiclassical expressions for the energy and
density are approximately valid. We must then estimate the errors directly
in terms of certain norms of the potential.  
The estimates are semiclassical in the sense that if one introduces
a semiclassical parameter then the errors are of smaller 
order than the leading semiclassical expression.

We are interested in a semiclassical approximation to the sum of the 
negative eigenvalues of a Schr\"odinger operator
$$\index{$h_r$}
        h:=-\mfr{1}{2}\Delta-V,
$$
on $\R^3$.
We shall in this section always assume that the potential 
$V:\R^3\to\R$ is locally in $L^1$ and that its positive part satisfies 
that $[V]_+\in L^{5/2}(\R^3)$. This ensures 
(by the Lieb-Thirring inequality or even by 
Sobolev's inequality) that $h$ is bounded below and can be defined 
as a Friedrichs' extension from the domain $C^\infty_0(\R^3)$.
\footnote{Note that we are not 
including spin. 
The operator $h$ is acting in the space $L^2({\scriptstyle\R}^3)$ 
and {\it not} in the space $L^2({\scriptstyle\R}^3;{\scriptstyle\C}^2)$.}

The semiclassical approximation to the sum of the negative eigenvalues
of $h$ is given by
\begin{equation}\label{eq:scenergy}
        (2\pi)^{-3}\iint_{\mfr{1}{2}p^2-V(x)\leq0}\mfr{1}{2}p^2-V(x)\ dp\, dx
        =-2^{3/2}(15\pi^2)^{-1}\int \left[V(x)\right]_+^{5/2} dx.\quad
\end{equation}
Moreover, the semiclassical approximation to the density, i.e., 
the sum of the absolute square of the eigenfunctions 
corresponding to the negative eigenvalues, is
\begin{equation}\label{eq:scdensity}
        (2\pi)^{-3}\int_{\mfr{1}{2}p^2-V(x)\leq0}1dp
        =2^{3/2}(6\pi^2)^{-1}\left[V(x)\right]_+^{3/2}.
\end{equation}

\phantom{weird}

\numbereddemo{Definition}\label{defi:g} 
For $s>0$, let $g:\R^3\to\R\index{$g$}$ be the ground state of the
Dirichlet-Laplacian for a ball of radius $s$, i.e., the function with
$g(x)=0$ if $|x|>s$ and 
$g(x)=(2\pi s)^{-1/2} |x|^{-1}\sin(\pi|x|/s)$ if $|x|\leq s$. Then
\begin{equation}\label{eq:gproperties}
        0\leq g\leq1,\quad\int g^2=1\quad\int|\nabla g|^2=(\pi/s)^2.
\end{equation}
\enddemo
\pagebreak

\proclaimtitle{Semiclassical approximation}
\proclaim{Lemma} \label{lm:sc}
We assume about the potential that 
$[V]_+,[V-V*g^2]_+\in L^{5/2}(\R^3)$
with $g$ as in Definition~{\rm \ref{defi:g}} above.
Let $e^{(1)}\leq e^{(2)}\leq\ldots<0${\rm ,} denote the negative
eigenvalues of $h=-\mfr{1}{2}\Delta-V$ as an operator
on $L^2(\R^3)$. Then for all $0<\delta<1${\rm ,} all integers $N>0${\rm ,} 
and all $s>0$ we have
\begin{eqnarray}
        \sum_{j=1}^Ne^{(j)}&\geq&
        -2^{3/2}(15\pi^2)^{-1}(1-\delta)^{-3/2}\int \left[V\right]_+^{5/2}
   \label{eq:sclower1}   \\
        &&-\mfr{1}{2}\pi^2s^{-2}N
        -L_1\delta^{-3/2}\left\|\left[V
        -V*g^2\right]_+\right\|_{5/2}^{5/2}
    \nonumber      
\end{eqnarray}
where $L_1$ is the constant in
the  Lieb\/{\rm -}\/Thirring estimate
{\rm (\ref{eq:LT}).} If there are fewer than $N$ negative eigenvalues  
the sum on the left refers simply to the sum over all the negative 
eigenvalues.

On the other hand,{\rm } if also $[V]_+\in L^{3/2}(\R^3)${\rm ,}
we can{\rm ,} for all $s>0${\rm ,} find a density matrix $\gamma$ 
with $\uprho_\gamma(x)=2^{3/2}(6\pi^2)^{-1}\left[V\right]_+^{3/2}*g^2(x)$ 
such that 
\begin{equation}
        \Tr[-\mfr{1}{2}\Delta\gamma]=2^{1/2}(5\pi^2)^{-1}
        \int \left[V\right]_+^{5/2}+\mfr{1}{2}\pi^2s^{-2}\int 2^{3/2}(6\pi^2)^{-1}
        \left[V\right]_+^{3/2}.\qquad 
\label{eq:scupper1}\end{equation}
\endproclaim

\numbereddemo{{R}emark} We are not proving that 
the true density of the projection onto the 
negative eigenvalues of $h$ is approximated 
by the semiclassical expression (\ref{eq:scdensity}). We only claim that
there is a `good' trial density matrix.
In the context where we shall use the semiclassics we shall 
infer the approximation of the true density 
by other means.
\enddemo

{\it Proof  of Lemma~{\rm \ref{lm:sc}}}.
We prove the result using the method of coherent states 
(see Thirring~\cite{Thirring81} and Lieb~\cite{Lieb:tf}).

For $u,p\in\R^3$ let $\Pi_{u,p}\index{\Pi_{u,p}}$ be the one-dimensional 
projection in $L^2(\R^3)$, projecting onto the 
space spanned by the function
$f_{u,p}(x):=\exp(ipx)g(x-u)\index{f_{u,p}}$.
We then have the {\it coherent states identities}
\begin{eqnarray}
        \Tr[\Pi_{u,p}]&=&1,\ \mbox{for all }p,u\label{eq:coherenttr}\\
        (2\pi)^{-3}\iint\Pi_{u,p}dp\,du&=&\I,\ \mbox{on } L^2(\R^3).
        \label{eq:coherentint}
\end{eqnarray}
We also have the identity
\begin{equation}
        \Tr[-\mfr{1}{2}\Delta\Pi_{u,p}] =  \mfr{1}{2}p^2+\mfr{1}{2}\int
        |\nabla g|^2
        =\mfr{1}{2}p^2+\pi^2/(2s^2),\label{eq:PiDelta}
\end{equation}
and for all density matrices $\gamma$
\begin{eqnarray}
        \Tr[-\mfr{1}{2}\Delta\gamma]&=&(2\pi)^{-3}\iint \mfr{1}{2}p^2
        \Tr[\Pi_{u,p}\gamma]dp\, du
        -\pi^2/(2s^2)\Tr[\gamma]\label{eq:gammaDelta}\\
    \quad    \Tr[(V*g^2)\gamma]&=&(2\pi)^{-3}\iint V(u)\Tr[\Pi_{u,p}\gamma]dp\, du\,.
        \label{eq:gammaV}
\end{eqnarray}

{\it Proof of the lower bound} (\ref{eq:sclower1}).
Let $f_1,f_2,\ldots\in L^2(\R^3)$ denote the normalized eigenfunctions
of $h$ corresponding to negative eigenvalues.
It is clear that we may without loss of generality assume that there
are $N$ negative eigenvalues.

We decompose
the operator $h$ as 
$$h=-\mfr{1}{2}(1-\delta)\Delta-V*g^2 
+[-\mfr{1}{2}\delta\Delta
-(V-V*g^2)].$$
We may then write
$\sum_{j=1}^Ne^{(j)}=\sum_{j=1}^N(f_j,hf_j)={\cal A}+{\cal B}$, where
$$
        {\cal A}:=\sum_{j=1}^N(f_j,[-\mfr{1}{2}(1-\delta)\Delta-V*g^2]f_j),
        \quad
        {\cal B}:=\sum_{j=1}^N(f_j,[-\mfr{1}{2}\delta\Delta-(V
        -V*g^2)]f_j).
$$
Hence from (\ref{eq:gammaDelta}) and (\ref{eq:gammaV}) we   obtain
$$
        {\cal A}=(2\pi)^{-3}\iint\left(\mfr{1}{2}(1-\delta)p^2-V(u)\right)
        \sum_{j=1}^N(f_j,\Pi_{u,p}f_j)dp\,du
        -N\pi^2(2s^2)^{-1}.
$$
As a consequence of (\ref{eq:coherenttr}) we have that
$0\leq\sum_{j=1}^N(f_j,\Pi_{u,p}f_j)\leq 1$.

It is therefore clear that
\begin{eqnarray*}
        {\cal A}&\geq& (2\pi)^{-3}\iint_{(1-\delta)\mfr{1}{2}p^2-V(u)\leq0}
        \left((1-\delta)\mfr{1}{2}p^2-V(u)\right)dp\,du-\pi^2(2s^{2})^{-1}N\\
        &=&-2^{3/2}(15\pi^2)^{-1}(1-\delta)^{-3/2}\int \left[V\right]_+^{5/2}
        -\pi^2(2 s^2)^{-1}N.
\end{eqnarray*}

The estimate (\ref{eq:sclower1}) follows by applying the Lieb-Thirring
estimate (\ref{eq:LT}) to conclude that
$
        {\cal B}\geq -L_1\delta^{-3/2}\|\left[V
        -V*g^2\right]_+\|_{5/2}^{5/2}.
$

\demo{{P}roof of the existence of $\gamma$} 
We shall prove that 
$$
        \gamma:= (2\pi)^{-3}\iint_{\mfr{1}{2}p^2-V(u)\leq 0}
        \Pi_{u,p} dp\,du
$$
has the desired properties. {F}rom (\ref{eq:coherentint}) we see that 
$\gamma$ is a density matrix, i.e., $0\leq \gamma\leq \I$.
The density corresponding to $\gamma$ is easily computable
$$
        \uprho_\gamma(x)=\gamma(x,x)=(2\pi)^{-3}
        \iint_{\mfr{1}{2}p^2-V(u)\leq 0}\Pi_{u,p}(x,x) dp\,du
        =2^{3/2}(6\pi^2)^{-1}\left[V\right]_+^{3/2}*g^2(x).
$$
{F}rom (\ref{eq:PiDelta}) we immediately obtain
(\ref{eq:scupper1}).\hfill\qed
\enddemo

Although we shall use the semiclassical approximation in the 
form given in the lemma  we shall for completeness state a 
less technical semiclassical result which follows very easily from
the \pagebreak lemma. 

\proclaimtitle{Semiclassical approximation}
\proclaim{Theorem} \label{thm:sc}
Assume that $0\leq V\in L^{5/2}(\R^3)\cap L^{3/2}(\R^3)$
and $|\nabla V|\in L^{5/2}(R^3)$.
Let $e^{(1)}\leq e^{(2)}\leq\ldots<0${\rm ,} denote the negative
eigenvalues of $h=-\mfr{1}{2}\Delta-V$ as an operator
on $L^2(\R^3)$. Then we have
\begin{eqnarray}\qquad
        \sum_je^{(j)}&\nhs\geq\nhs& -2^{3/2}(15\pi^2)^{-1}\int V(x)^{5/2} dx
        \left\{1+A_{\rm L}\|V\|_{\frac{5}{2}}^{-\frac{3}{2}}
          \|\nabla V\|_{\frac{5}{2}}^{\frac{2}{3}}
        \|V\|_{\frac{3}{2}}^{\frac{1}{2}}\right\}^{\frac{5}{3}}\label{eq:scthmlower1}\\
\noalign{\hbox{and}}
\noalign{\vskip8pt}
\qquad
        \sum_je^{(j)}&\nhs\leq\nhs&
        -2^{3/2}(15\pi^2)^{-1}\int V(x)^{5/2} dx+2^{-1/2}\pi^{-4/3}
        \|V\|_{\frac{5}{2}}\|\nabla V\|_{\frac{5}{2}}^{\frac{2}{3}}
        \|V\|_{\frac{3}{2}}^{\frac{1}{2}},
        \label{eq:scthmupper1}
\end{eqnarray}
where $A_{\rm L}:=\frac{9}{4}2^{-9/10}(15\pi^2)^{3/5}
\left(\frac{2\pi^2}{5}\right)^{1/3}L_0^{1/3}L_1^{4/15}$.
Here $L_0$ and $L_1$ are the constants in
the {\rm CLR} and Lieb\/{\rm -}\/Thirring estimates {\rm (\ref{eq:CLR})} and
{\rm (\ref{eq:LT})} respectively. 
\endproclaim
\demo{Proof}
We may estimate 
\begin{eqnarray*}
        \left|V(u)-V*g^2(u)\right|&\leq&
        \int_{\R^3}\int_0^1|\nabla V(u-ty)||y|g(y)^2dt\,dy\\
        &=&\int_{\R^3}|\nabla V(u-y)||y|\int_0^1t^{-4}g(y/t)^2dtdy\\
        &\leq& (4\pi)^{-1}\int_{|y|\leq s}|\nabla V(u-y)||y|^{-2} dy,
\end{eqnarray*}
where we have used that $\int_0^1t^{-4}g(y/t)^2dt
\leq |y|^{-3}\int_0^\infty t^2 g(t)^2dt =(4\pi)^{-1}|y|^{-3}$ 
(identifying $g$ with a function on $\R_+$).
Hence
\begin{equation}\label{eq:gapproximation}
        \|V-V*g^2\|_{5/2}\leq 
        (4\pi)^{-1}\|\nabla V\|_{5/2}\int_{|y|\leq s}|y|^{-2}dy
        =s\|\nabla V\|_{5/2}.
\end{equation}
For any density matrix $\gamma$,
$\Tr[h\gamma]$ is an upper bound to the sum of the negative eigenvalues of 
$h$. {F}rom (\ref{eq:scupper1}) we find for the density 
matrix constructed in Lemma~\ref{lm:sc} that 
\begin{eqnarray*}
        \Tr[h\gamma]&=&-2^{3/2}(15\pi^2)^{-1}\int \left[V\right]_+^{5/2}\\
        &&{}+2^{3/2}(6\pi^2)^{-1}\int \left[V(u)\right]_+^{3/2}
        \left[V(u)-V*g^2(u)+
        \mfr{1}{2}\pi^2s^{-2}\right]du.
\end{eqnarray*}
The bound (\ref{eq:scthmupper1}) follows from applying 
H\"older's inequality,
(\ref{eq:gapproximation}), and optimizing in $s$.

By the CLR bound (\ref{eq:CLR}) we know that
$h$ has only finitely many negative eigenvalues
and that their number $N$ is bounded by \pagebreak $N\leq L_0\int \left[V\right]_+^{3/2}$.
{F}rom (\ref{eq:sclower1}) and (\ref{eq:gapproximation})
we therefore obtain
\begin{eqnarray*}
        \sum_je^{(j)}
        &\geq&
        -2^{3/2}(15\pi^2)^{-1}(1-\delta)^{-3/2}\int V^{5/2}\\
        &&{}
        -L_0\pi^2(2 s^2)^{-1}\int V^{3/2}
        -L_1\delta^{-3/2}s^{5/2}\int|\nabla V|^{5/2}\\
        &=& -2^{3/2}(15\pi^2)^{-1}(1-\delta)^{-3/2}\int V^{5/2}\\
        &&{}
        -\mfr{9}{4}\left(\mfr{2}{5}\right)^{5/9}
        L_0^{5/9}L_1^{4/9}\pi^{10/9}\delta^{-2/3}
        \left(\int V^{3/2}\right)^{5/9}\left(\int|\nabla V|^{5/2}\right)^{4/9},
\end{eqnarray*}
where we have optimized in the parameter $s$.

We now optimize in the parameter $\delta$. 
Define $\delta'$ by $(1-\delta)^{-3/2}=(1-\delta')^{-2/3}$. (Note that 
$0<\delta<1$ if and only if $0<\delta'<1$.) 
Then $\delta^{-2/3}\leq (4\delta'/9)^{-2/3}$. Thus 
\begin{eqnarray*}
  \sum_je^{(j)}
  &\geq&-2^{3/2}(15\pi^2)^{-1}(1-\delta')^{-2/3}\int V^{5/2}\\&&{}
  -A_1\delta'^{-2/3}\left(\int V^{3/2}\right)^{5/9}
  \left(\int|\nabla V|^{5/2}\right)^{4/9},
\end{eqnarray*}
where 
$
        A_1:=\left(\mfr{9}{4}\right)^{5/3}\left(\mfr{2\pi^2}{5}\right)^{5/9}
        L_0^{5/9}L_1^{4/9}.
$
Using that $$\min_{\delta'}\left[(1-\delta')^{-2/3}a+\delta'^{-2/3}b\right]
= a[1+(b/a)^{3/5}]^{5/3}$$ we arrive at (\ref{eq:scthmlower1}).
\enddemo

We shall need the semiclassical estimates also 
for the operator $h$
restricted to functions on the set $\set{x}{|x|\geq r}$ satisfying 
Dirichlet boundary conditions.

\proclaimtitle{Dirichlet boundary conditions}
\proclaim{Lemma} \label{lm:DBC}
Let the assumptions be as in the beginning of Lemma~{\rm \ref{lm:sc}.}
For $r>0$ let $h_r$ denote the restriction of the 
operator $h=-\mfr{1}{2}\Delta-V$ 
to functions on the set $\set{x}{|x|\geq r}$ satisfying 
Dirichlet boundary conditions.
Denote by $e^{(1)}\leq e^{(2)}\leq \ldots<0$ the negative eigenvalues
of $h$ and by $e^{(1)}_r\leq e^{(2)}_r\leq \ldots<0$
the negative eigenvalues of $h_r$.
Then
$
        \sum_je^{(j)}\leq \sum_je^{(j)}_r.
$
Moreover{\rm ,} if $\gamma$ is a density matrix on $L^2(\R^3)$ 
we may{\rm ,} for all $0<\lambda<1${\rm ,} find a density 
matrix $\widetilde\gamma$ such that $\uprho_{\widetilde\gamma}$
is supported in $\set{x}{|x|\geq r}$ and
$\uprho_{\widetilde\gamma}\leq\uprho_\gamma$ and
$$
\Tr[h_r\widetilde\gamma]\leq \Tr[h\gamma]
+L_1\int_{|x|\leq(1-\lambda)^{-1}r} \left[V\right]_+^{5/2}
+\mfr{1}{2}({\pi}/{(2\lambda r)})^2
\int_{|x|\leq(1-\lambda)^{-1}r}\uprho_\gamma.
$$
\endproclaim

{\it Proof}. That 
the Dirichlet eigenvalues are  upper bounds to the 
eigenvalues on $\R^3$, is a well known simple consequence of
the variational principle.

Let $\thetar$ be the localization function from 
Definition~6.1. We shall choose $\widetilde\gamma
=\thetar\gamma\thetar$.
Then by the IMS formula (\ref{eq:IMS}) 
we have
\begin{eqnarray*}
  \Tr[h\gamma]&=&\Tr[h_r\thetar\gamma\thetar]
  +\Tr[h(1-\thetar^2)^{1/2}\gamma(1-\thetar^2)^{1/2}]\\&&{}
  -\mfr{1}{2}\Tr[\left((\nabla\thetar)^2+
    (\nabla(1-\thetar^2)^{1/2})^2\right)\gamma].
\end{eqnarray*}
By the Lieb-Thirring inequality (\ref{eq:LT}) we have 
$\Tr[h(1-\thetar^2)^{1/2}\gamma(1-\thetar^2)^{1/2}]
        \geq -L_1\int_{|x|\leq(1-\lambda)^{-1}r} \left[V\right]_+^{5/2}$.
Thus the lemma follows from the bound on 
the gradient of $\thetar$ and $(1-\thetar^2)^{1/2}$
given in Definition~6.1.
\hfill\qed
 
\section{The Coulomb norm estimates} 
\label{sec:cn}
\advance\eqcount by 78

In this section we introduce and study the Coulomb norm. 
\numbereddemo{Definition}\label{defi:cn}
For $f,g\in L^{6/5}(\R^3)$ we define the {\it Coulomb inner product}
\begin{equation}\label{eq:Coulombinner}
        D(f,g):=\mfr{1}{2} \iint f(x) |x-y|^{-1} \overline{g(y)}
        dx\,dy
\end{equation}
and the corresponding {\it Coulomb norm}, 
$$
        \|g\|_\rC:=D(g,g)^{1/2}.
$$
\enddemo

By the Hardy-Littlewood-Sobolev estimate
we have 
\begin{equation}\label{eq:HLS}
        \|g\|_\rC\leq \pi^{-1/6}2^{7/6}3^{-1/2}\|g\|_{6/5}.
\end{equation}
In this sharp form the inequality was proved by Lieb~\cite{Lieb:sob}.
Using the Fourier transform we may write
\begin{equation}\label{eq:Dfourier}
        D(f,g)
        =(2\pi)\int \hat{f}(p)\overline{\hat{g}(p)}|p|^{-2} dp,
\end{equation}
from which it follows that the Coulomb norm really is a norm
on $L^{6/5}(\R^3)$.

The following estimate was first used in the context of atomic problems
by Fefferman and Seco~\cite{Fef-Sec}.

 \proclaimtitle{Coulomb norm estimate}
\proclaim{Lemma}\label{lm:cn}
If $f\in L^6(\R^3)$ with $|\nabla f|\in L^2(\R^3)$ and $g\in L^{6/5}(\R^3)$
then 
$$
        \left|\int f\overline g\right|\leq (2\pi)^{-1/2}\|\nabla f\|_2\|g\|_\rC.
$$
\endproclaim

\demo{Proof} Using Plancherel's identity and the representation
(\ref{eq:Dfourier}) we have
\vglue12pt
\hfill ${\displaystyle
        \left|\int f\overline g\right|
        =\left|\int \hat{f}\overline{\hat{g}}\right|
        \leq \||p|\hat{f}(p)\|_2\||p|^{-1}\hat{g}(p)\|_2
        =(2\pi)^{-1/2}\|\nabla f\|_2\|g\|_\rC.
}$\enddemo

We shall next give some simple but very useful consequences
of this estimate. 
\proclaim{{C}orollary} \label{cl:cn} Consider $f\in L^{5/3}(\R^3)\cap L^{6/5}(\R^3)$. 
For all $x\in\R^3$ and $s>0$ we have 
\begin{equation}\label{eq:cn1}
  f*|x|^{-1}
  \leq (25\pi^2 s/16)^{1/5}(\pi s)^{1/5} \|[f]_+\|_{L^{5/3}(\B{x,s})}+
  (2/s)^{1/2}\|f\|_\rC.\qquad
\end{equation} 
For all $x\in\R^3$ and all $\kappa>0$ denote by $A(|x|,\kappa)$ the 
annulus 
$$
 A(|x|,\kappa)= \set{y}{(1-2\kappa)|x|\leq|y|\leq|x|}
$$ 
we then have
\begin{eqnarray}
        \int_{|y|<|x|}f(y)|x-y|^{-1}dy&\leq&
        2^{7/5}\pi^{2/5}(\kappa|x|)^{1/5}
        \left\|[f]_+\right\|_{L^{5/3}\left(A(|x|,\kappa)\right)}
    \label{eq:cn2}
    \\&&{}
        +2^{3/2}\kappa^{-1}|x|^{-1/2}
        \|f\|_\rC.  \nonumber
      \end{eqnarray}
\endproclaim 

\numbereddemo{{R}emark} Note that we do not restrict to $\kappa\leq1/2$. 
We do this to avoid having to check this condition in the 
applications of the corollary. 
\enddemo

{\it Proof of Corollary}~\ref{cl:cn}.
Consider the function $\xi_s:\R^3\to\R$ defined by 
$$
        \xi_s(z):=\left\{\begin{array}{cl}
                        s^{-1},&\mbox{if }|z|\leq s\\
                        |z|^{-1},&\mbox{if }|z|\geq s.\end{array}\right.
$$
It satisfies 
$\|\nabla \xi_s\|_2=(4\pi /s)^{1/2}$. 
Hence from Lemma~\ref{lm:cn} we obtain
\begin{eqnarray*}
  f*|x|^{-1}&\leq& \int_{|y-x|\leq s}[f(y)]_+\left(|x-y|^{-1}
    -s^{-1}\right)dy
  +\int_{\R^3} f(y)\xi_s(x-y)dy\\
  &\leq& (25\pi^4s/16)^{1/5}\|[f]_+\|_{L^{5/3}(\B{x,s})}
  +(2/s)^{1/2}\|f\|_\rC,
\end{eqnarray*}
where we have used that $\int_{|y|<1}(|y|^{-1}-1)^{5/2}dy=\frac{5\pi^2}{4}$.

In order to prove the second half of the corollary we introduce
the function $\Xi_{x,\kappa}:\R^3\to\R^3$ given by 
$$
        \Xi_{x,\kappa}(z):=\left\{\begin{array}{cl}
                        1,&\mbox{if }|z|\leq (1-2\kappa)|x|\\
                        1-(|x|\kappa)^{-1}(|z|-|x|(1-2\kappa)),
                        &\mbox{if }(1-2\kappa)|x|\leq|z|\leq(1-\kappa)|x|\\
                        0,&\mbox{if }(1-\kappa)|x|\leq|z|.\end{array}\right.
$$
Then $|\Xi_{x,\kappa}(z)|\leq1$ and we can estimate
\begin{eqnarray*}
        \int_{|y|<|x|}f(y)|x-y|^{-1}dy&\leq&
        \int_{A(|x|,\kappa)}[f(y)]_+|x-y|^{-1}dy\\
        &&{}+\int_{\R^3}f(y)|x-y|^{-1}\Xi_{x,\kappa}(y)dy\end{eqnarray*}
    \begin{eqnarray*}    &\leq&
        \|[f]_+\|_{L^{5/3}\left(A(|x|,\kappa)\right)}
        \left(\int_{A(|x|,\kappa)}|x-y|^{-5/2}dy\right)^{2/5}\\
        &&{}+(2\pi)^{-1/2}\|f\|_\rC\left(\int\left|\nabla_y
        \left(|x-y|^{-1}\Xi_{x,\kappa}(y)\right)\right|^2dy\right)^{1/2}.
\end{eqnarray*}
It remains to estimate the two integrals above. For the first integral
we find for $\kappa\leq 1/2$
\begin{eqnarray*}
        \int_{A(|x|,\kappa)}|x-y|^{-5/2}dy
        &=&2\pi|x|^{1/2}\int_{r=1-2\kappa}^1\int_{u=-1}^1
        (1-2ru+r^2)^{-5/4}r^2du\,dr\\
        &=&2\pi|x|^{1/2}\alpha_1(\kappa),
\end{eqnarray*}
where
$$
        \alpha_1(\kappa):=4(2\kappa)^{1/2}-(4/3)(2\kappa)^{3/2}
        +(4/3)\sqrt{2}[1-(1-\kappa)^{1/2}(1+2\kappa)]
        \leq 4(2\kappa)^{1/2}.
$$
The last inequality follows from a straightforward careful analysis of
$\alpha_1(\kappa)$.
For the second integral we get 
\begin{eqnarray*}
        \left(\int\left|\nabla_y
        \left(|x-y|^{-1}\Xi_{x,\kappa}(y)\right)\right|^2dy\right)^{1/2}
        &\leq&\left(\int_{|x-y|>\kappa|x|}|x-y|^{-4}dy\right)^{1/2}\\
        &&{}
        +(\kappa|x|)^{-1}\left(\int_{A(|x|,\kappa)}|x-y|^{-2} dy\right)^{1/2}\\
        &=&(4\pi)^{1/2}(\kappa|x|)^{-1/2}(1+\alpha_2(\kappa)^{1/2}),
\end{eqnarray*}
with
$$
        \alpha_2(\kappa):
        =1+(1-\kappa)\ln\left(\frac{1-\kappa}{\kappa}\right)\leq 
        1-\ln\kappa \leq\kappa^{-1},
$$
where we have used that 
$$
        \int_{A(|x|,\kappa)}|x-y|^{-2} dy
        =2\pi|x|\int_{r=1-2\kappa}^1\int_{u=-1}^1
        (1-2ru+r^2)^{-1}r^2 du\,dr=4\pi\kappa |x|\alpha_2(\kappa).
$$
Using
$
        (4\pi)^{1/2}\kappa^{-1/2}(1+\alpha_2(\kappa)^{1/2})
        \leq 4\pi^{1/2}\kappa^{-1}
$
we get  (\ref{eq:cn2}).

The estimate holds also for $\kappa>1/2$ 
since the last term in (\ref{eq:cn2}) can be ignored in this case. 
\hfill\qed

\section{Main estimate} 
\label{sec:me}
\advance\eqcount by 83

We now restrict attention to the case $N\geq Z$.
Throughout the remaining part of this paper
$\rhotf,\phitf,\Phitf{R},\rhohf, \phihf$ 
and $\Phihf{R}$ always refer to the problems with 
particle number $N$. In fact, since $N\geq Z$ the TF functions
correspond to the neutral atom, i.e., $\mutf=0$.
We shall suppress the dependence on $N$ everywhere 
since it is held fixed throughout the discussion. 

From now on we shall no longer explicitly compute the constants
involved in the estimates. We shall use the notation $\const$ to refer
to any universal (in principle explicitly computable) {\it positive}
constant.  Thus
$\const$ does not mean the same constant in all equations or
inequalities.  Even within the same equation we shall use the notation
$\const$ to refer to possibly different universal constants.
Universal constants of particular importance will be given separate
names. We begin by stating the main result of this section.
 
\proclaimtitle{Main estimate}
\proclaim{Theorem} \label{thm:me}
Assume $Z\geq 1$ and $N\geq Z$. There exist universal constants 
$0<\epsilon<4$ and $\CM,\CPhi>0$
such that for all $x\in\R^3$  
we have
\begin{equation}\label{eq:me1}
        \left|\Phihf{|x|}(x)-\Phitf{|x|}(x)\right|
        \leq \CPhi
        |x|^{-4+\varepsilon}+\CM.
\end{equation}
\endproclaim

We shall prove Theorem~\ref{thm:me} by an iterative procedure.
The first step is to control 
``small'' $x$.

 \proclaimtitle{Control of the region close to the nucleus}
\proclaim{Lemma} \label{lm:smallx}
Assume $Z\geq1$ and $N\geq Z$.
For all $\beta>0$ and all
$|x|\leq \beta Z^{-1/3}$ we have
$$
        \left|\Phihf{|x|}(x)-\Phitf{|x|}(x)\right|
        \leq A_\Phi \beta^{49/12-\varepsilon_1}
        |x|^{-4+\varepsilon_1},
$$
where $\varepsilon_1=1/66$ and $A_\Phi>0$ is 
a universal constant.
\endproclaim

 \proclaimtitle{Iterative step}
\proclaim{Lemma}\label{lm:iterate}
Assume $N\geq Z$. For all $\delta,\varepsilon',{\sigma}>0$ with 
$\delta<\delta_0${\rm ,} where $\delta_0$ is some universal constant{\rm ,} there
exists constants
$\epsilon_2,C'_\Phi>0 $ depending only on $\delta$ 
and a constant 
$D=D(\varepsilon',{\sigma})>0$ depending only on 
$\varepsilon',{\sigma}$ with the following property.
For all $R_0<D$ satisfying that
$\beta_0 Z^{-1/3}\leq R_0^{1+\delta}$ 
{\rm (}\/where $\beta_0=\frac{(9\pi)^{2/3}}{44}$ as in 
Theorem~{\rm \ref{thm:sommerfeldlow})} and that  
\begin{equation}\label{eq:phiasp}
        \left|\Phihf{|x|}(x)-\Phitf{|x|}(x)\right|
        \leq {\sigma} |x|^{-4+\varepsilon'}
\end{equation}
holds for all $|x|\leq R_0${\rm ,}
there exists
$R_0'>R_0$ such that 
\begin{equation}\label{eq:itphi}
        \left|\Phihf{|x|}(x)-\Phitf{|x|}(x)\right|
        \leq C'_\Phi |x|^{-4+\varepsilon_2}
\end{equation} 
for all $x$ with $R_0<|x|<R_0'$.
\endproclaim

Lemmas~\ref{lm:smallx} and \ref{lm:iterate} will allow us to 
control small and intermediate $|x|$
to control large $|x|$ we shall need the
following two \pagebreak lemmas. 

\proclaimtitle{Bound on $\int(\rhohf)^{5/3}$}
\proclaim{Lemma} \label{lm:me5/3}
Assume $N\geq Z$. Given $0<\varepsilon',\sigma${\rm ,}
there is a $D>0$ such that if
{\rm (\ref{eq:phiasp})} holds for all $|x|\leq D$ then we have 
\begin{equation}
  \int_{|y|>|x|}\rhohf(y)^{5/3}\,dy\leq \const |x|^{-7},\label{eq:me5/3}
\end{equation}
for all $|x|\leq D$. 
\endproclaim

\proclaimtitle{Bound on $\int\rhohf$}
\proclaim{Lemma} \label{lm:meN}
  Assume that {\rm (\ref{eq:phiasp})}
holds for all $|x|\leq R$ for some $R>0$
  and some $\epsilon',\sigma>0$. 
  Then for $0<r\leq R$ we have  
$$
   \left|\int_{|y|<r}\left(\rhohf(y)-\rhotf(y)\right)dy\right|
   \leq\sigma r^{-3+\epsilon'}
$$
and 
$$
  \int_{|y|>r}\rhohf(y)dy\leq\const(1+\sigma r^{\epsilon'})
  \left(r^{-3}+1\right).
$$
\endproclaim

We shall prove Lemma~\ref{lm:smallx} in Section \ref{sec:smallx}
and Lemmas~\ref{lm:iterate} and \ref{lm:me5/3} in
Section \ref{sec:iteration}. We end this section with the proofs
of Lemma~\ref{lm:meN} and the main estimate Theorem~\ref{thm:me}.
 
\demo{Proof  of Lemma~{\rm \ref{lm:meN}}}
  First note that for $0<r\leq R$ we have
$$
    \int_{|y|<r}\left(\rhotf(y)-\rhohf(y)\right)dy
    =(4\pi)^{-1}r\int\limits_{\omega\in{\cal S}^2}\int\limits_{|y|<r}
    \left(\rhotf(y)-\rhohf(y)\right)|r\omega-y|^{-1}dy\,d\omega.
$$
Thus we have
$$
 \int_{|y|<r}\left(\rhotf(y)-\rhohf(y)\right)dy=
 (4\pi)^{-1}r\,\,\int\limits_{\omega\in{\cal S}^2}
 \Phihf{r}(r\omega)-\Phitf{r}(r\omega)d\omega.
$$
Together with (\ref{eq:phiasp}) this gives the first estimate above.
Moreover, we also have that 
\begin{eqnarray*}
\left|\int_{r/2<|y|<r}\left(\rhotf(y)-\rhohf(y)\right)dy\right|
&\leq& \sup_{|y|=r}\left|\Phihf{r}(y)-\Phitf{r}(y)\right|\\ &
+&\sup_{|y|=r/2}\left|\Phihf{r/2}(y)-\Phitf{r/2}(y)\right| \\&\leq&
\const\sigma r^{-3+\epsilon'}.
\end{eqnarray*}
The TF equation (\ref{eq:tfsystemrho}), and the Sommerfeld estimate in
Theorem~\ref{thm:sommerfeldup} give 
$$\int_{|y|>r/2}\rhotf(y)\,dy\leq\const r^{-3}$$ 
and hence
$$
    \int_{r/2<|y|<r}\rhohf(y)dy\leq \const (1+\sigma r^{\epsilon'})r^{-3}.
$$
{F}rom (\ref{eq:phiasp}), the exterior $L^1$-estimate Lemma~\ref{lm:extl1} (used with 
$\lambda=1/2$ and $r$ replaced by $r/2$), and   
Lemma~\ref{lm:Phitfrbound} (recall that now $\mutf=0$) we immediately 
conclude the estimate on $\int_{|y|>r}\rhohf(y)dy$.
\enddemo

We finally show how to use Lemmas~\ref{lm:smallx}--\ref{lm:meN} 
to prove the main estimate Theorem~\ref{thm:me}.

\demo{Proof of Theorem~{\rm \ref{thm:me}}}
  We first show that we may choose $\delta>0$ small enough such that
  if we choose $\widetilde{R}^{1+\delta}=\beta_0 Z^{-1/3}$ we have for
  all $|x|<\widetilde{R}$ that
  \begin{equation}\label{eq:itphi0}
    \left|\Phihf{|x|}(x)-\Phitf{|x|}(x)\right|
    \leq C''_\Phi |x|^{-4+\frac{\epsilon_1}{2}}
  \end{equation} 
  for a universal constant $C''_\Phi>0$ and with 
  $\epsilon_1$ given in Lemma~\ref{lm:smallx}.
  
  To see this let $\beta>0$ be such that 
  $\left(\beta Z^{-1/3}\right)^{1+\delta}=\beta_0Z^{-1/3}$,
  i.e., $\beta^{1+\delta}=\beta_0Z^{\delta/3}$. We then see from 
  Lemma~\ref{lm:smallx} that for all $|x|<\beta Z^{-1/3}$
  we have
  \begin{eqnarray*}
    \left|\Phihf{|x|}(x)-\Phitf{|x|}(x)\right|
    &\leq& A_\Phi \beta^{\frac{49}{12}-\frac{\epsilon_1}{2}}
    Z^{-\frac{\epsilon_1}{6}}|x|^{-4+\frac{\epsilon_1}{2}}\\
    &=&A_\Phi
 (\beta_0Z^{\delta/3})^{\left(\frac{49}{12(1+\delta)}
     -\frac{\epsilon_1}{2(1+\delta)}\right)}
    Z^{-\frac{\epsilon_1}{6}}|x|^{-4+\frac{\epsilon_1}{2}}.
  \end{eqnarray*}
  Since $\widetilde{R}^{1+\delta}=\beta_0 Z^{-1/3}$, i.e.\ 
  $\widetilde{R}=\beta Z^{-1/3}$, 
  we see that if $\delta$ is small enough and $C''_\Phi$ is chosen
  appropriately then 
  (\ref{eq:itphi0}) holds for all $|x|<\widetilde{R}$.

  We now assume that $\delta$ is also small enough that we may apply  
  Lemma~\ref{lm:iterate}. This gives us constants
  $\epsilon_2,C'_\Phi>0$ and for all $\sigma,\epsilon'>0$ 
  a $D>0$ with the properties stated in
  Lemmas~\ref{lm:iterate} and \ref{lm:me5/3}. 
  We may without loss of generality assume that $D\leq1$.
  Now choose $\sigma=\max\{C'_\Phi,C''_\Phi\}$ and
  $\epsilon'=\min\{\epsilon_1/2,\epsilon_2\}$.
  Note that 
  $\sigma,\epsilon$, and $D$ are now universal constants. 
  We shall  prove that for all $|x|\leq D$ 
  we have 
  \begin{equation}
  \left|\Phihf{|x|}(x)-\Phitf{|x|}(x)\right| 
  \leq \sigma
  |x|^{-4+\varepsilon'}\label{eq:me1-CM}.
\end{equation}
Since we are assuming that $D\leq 1$ it is sufficient to prove 
(\ref{eq:me1-CM}) with $\epsilon'$ replaced by $\epsilon_1/2$ or
$\epsilon_2$.
We have to prove that $D$ belongs to the set
$$
 {\cal M}=\set{0<R\leq1}{\hbox{Inequality }(\ref{eq:me1-CM})\hbox{
     holds for all }|x|\leq R}.
$$ 
If this were not true we would have $D>R_0:=\sup {\cal M}$.
In order to reach a contradiction  we therefore assume this and hence in
particular that $R_0<1$. {F}rom (\ref{eq:itphi0}) it follows that
either $\widetilde{R}>1$ or $\widetilde{R}\in{\cal M}$. 
If $\widetilde{R}>1$ then $R_0=\sup {\cal M}\break =1$ which contradicts our
assumption. On the other hand if $\widetilde{R}\in{\cal M}$ then 
$R_0^{1+\delta}\geq \widetilde{R}^{1+\delta}=\beta_0 Z^{-1/3}$.
It is then an immediate consequence of Lemma~\ref{lm:iterate}
that there exists $R_0'\in{\cal M}$ with $R_0'>R_0$ and this is of
course also a contradiction. This establishes an inequality of the
form (\ref{eq:me1}) for all $|x|\leq D$.

We shall now prove (\ref{eq:me1}) for $|x|>D$. 
We write
$$
\left|\Phihf{|x|}(x)-\Phitf{|x|}(x)\right|\leq
\left|\Phihf{D}(x)-\Phitf{D}(x)\right|+
\left|\int_{D<|y|<|x|}\frac{\left(\rhotf(y)-\rhohf(y)\right)}{|x-y|}\,
  dy\right|.
  $$
  We shall estimate the last term using Lemma~\ref{lm:me5/3} and the
  similar bound 
  $$
  \int_{|y|>|x|}\rhotf(y)^{5/3}\,dy\leq \const |x|^{-7},
  $$ 
  which holds for all $x$ 
  by the Sommerfeld estimate Theorem~\ref{thm:sommerfeldup}
  and the TF equation (\ref{eq:tfsystemrho}). Hence using H\"older's inequality we have
\begin{eqnarray*}
 \left|\,\,\int\limits_{D<|y|<|x|}
   \frac{\left(\rhotf(y)-\rhohf(y)\right)}{|x-y|}dy\right|
 &\leq& \const D^{-21/5}
 \left(\int_{|x-y|<D}\!\!\!\!\!|x-y|^{-5/2}dy\right)^{2/5}\\
 &&+D^{-1}\int_{|y|>D}\left(\rhotf(y)+\rhohf(y)\right)dy.
\end{eqnarray*}
By Lemma~\ref{lm:meN} and the bound 
$\int_{|y|>D}\rhotf(y)\,dy\leq\const D^{-3}$, which is again a
consequence of the Sommerfeld estimate Theorem~\ref{thm:sommerfeldup}
and the TF equation (\ref{eq:tfsystemrho}), we see that this last
expression 
is bounded by a universal constant.

Since $\Phihf{D}(x)-\Phitf{D}(x)$ is harmonic for $|x|>D$
and tends to zero at infinity we have for all $|x|>D$ that 
$$
    \left|\Phihf{D}(x)-\Phitf{D}(x)\right|
    \leq \sup_{|z|=D}\left|\Phihf{D}(z)-\Phitf{D}(z)\right|
    \leq \sigma
  D^{-4+\varepsilon'},
$$
which is also bounded by a universal constant.
Thus (\ref{eq:me1}) holds for all $x$.
\enddemo

\section{Control of the region close to the nucleus: proof of  Lemma~\ref{lm:smallx}} 
\label{sec:smallx}
\advance\eqcount by 89

In order to prove
Lemma~\ref{lm:smallx} we need some basic estimates.

\proclaimtitle{Global $L^{5/3}$ and Coulomb norm estimates}
\proclaim{Lemma} 
\label{lm:5/3cn} 
For all $N$ and $Z$ we have the bound
\begin{equation}
        \int_{\R^3}\rhohf(y)^{5/3}dy\leq\const Z^{7/3}.
        \label{eq:5/3} 
\end{equation}
Moreover{\rm ,} if $Z\geq 1$
\begin{equation}
        \|\rhohf-\rhotf\|_\rC^2\leq \const
        Z^{7(1-\varepsilon_3)/3}\label{eq:||C},
\end{equation}
with $\varepsilon_3:=2/77$. 
\endproclaim

\demo{Proof}
Although we shall only use this result for $N\geq Z$ the proof is almost 
as easy without this restriction, so we treat the more general case
here. 

We first estimate the $L^{5/3}$ norm of $\rhohf$. 
It is easy to see that $\ehf(\gammahf)\leq0$. Thus since
$\D(\gammahf)-\Ex(\gammahf)\geq0$ we have
from the Lieb-Thirring inequality (\ref{eq:LTdensity}) that 
\begin{eqnarray*}
        0\geq\ehf(\gammahf)&\geq &
        \Tr\left[\left(-\mfr{1}{2}\Delta-Z|y|^{-1}\right)\gammahf\right]\\&
        \geq& \int \left(K_1\rhohf(y)^{5/3}-Z|y|^{-1}\rhohf(y)\right)dy.
\end{eqnarray*}
If we use that $\int\rhohf=N$ and the inequality
$ab\leq \mfr{3}{5}a^{5/3}+\mfr{2}{5}b^{5/2}$ we get
for all $\delta>0$ and all $r>0$, 
$$
        0\geq \int (K_1-\mfr{3}{5}\delta^{5/3})\rhohf(y)^{5/3}dy-
        \mfr{2}{5}\int_{|y|<r}(\delta^{-1}Z|y|^{-1})^{5/2}dy 
        -NZr^{-1}.
$$
Choosing $\delta^{5/3}=5K_1/6$ and optimizing in $r$ 
gives  $\int \left(\rhohf\right)^{5/3}\leq \const N^{1/3}Z^2$. 
If we use that, since there exists an HF minimizer with particle number 
$N$, we must have Lieb's bound $N\leq 2Z+1$ (see Theorem~\ref{thm:2z+1}) 
we arrive at (\ref{eq:5/3}).

We turn to the proof of (\ref{eq:||C}).
We rewrite the Hartree-Fock functional (\ref{eq:ehf}) as 
\begin{equation}\label{eq:reehf}
        \ehf(\gamma)=\Tr\left[\left(-\mfr{1}{2}\Delta-\phitf\right)\gamma\right]
        +\|\rhotf-\uprho_\gamma\|_{\rC}^2-D(\rhotf,\rhotf)-\Ex(\gamma), \quad
\end{equation}
where we have used the definition $\phitf(y)=Z|y|^{-1}-\rhotf*|y|^{-1}$ 
and 
$$
        \Tr\left[(\rhotf*|y|^{-1})\gamma\right]
        =2D(\rhotf,\uprho_\gamma)=
        \D(\gamma)+D(\rhotf,\rhotf)-\|\rhotf-\uprho_\gamma\|_{\rC}^2.
$$

{F}rom the semiclassical estimate (\ref{eq:sclower1}) and the fact
that, when $\gamma$ is a density matrix with $\Tr[\gamma]=N$ 
and $h$ is a self-adjoint operator,
then $\Tr[h\gamma]$ is an upper
bound on the sum of the $N$ lowest eigenvalues of $h$, we find 
\begin{eqnarray*}
     &&\hskip-48pt  \Tr\left[\left(-\mfr{1}{2}\Delta-\phitf+\mutf\right)\gammahf\right]\\
        &\geq&-2^{5/2}(15\pi^2)^{-1}(1-\delta)^{-3/2}\int
        \left[\phitf-\mutf\right]_+^{5/2} \\
        &&{}-\mfr{1}{2}\pi^2s^{-2}N
        -2L_1\delta^{-3/2}
        \left\|\left[\phitf-\phitf*g^2\right]_+\right\|_{\frac{5}{2}}^{\frac{5}{2}},
\end{eqnarray*}
for all $0<\delta<1$ and all $s>0$. Recall that the function $g$ was
given in Definition~\ref{defi:g}.  Here we have used the semiclassical 
estimate for the space $L^2(\R^3;\C^2)$. 
The estimate above therefore has an extra factor of $2$ 
in the first and the last term compared to 
(\ref{eq:sclower1}).
Thus 
\begin{eqnarray*}
\lefteqn{\ehf(\gammahf)\geq-2^{5/2}(15\pi^2)^{-1}(1-\delta)^{-3/2}\int
        \left[\phitf-\mutf\right]_+^{5/2}
        -\mutf N -D(\rhotf,\rhotf)}&&\nonumber \\
        &&{}
        +\|\rhotf-\rhohf\|_{\rC}^2-\mfr{1}{2}\pi^2s^{-2}N
        -2L_1\delta^{-3/2}
        \left\|\left[\phitf-\phitf*g^2\right]_+\right\|_{5/2}^{5/2}
        -\Ex(\gammahf).
\end{eqnarray*}

Since $|y|^{-1}-g^2*|y|^{-1}\geq0$ (because the function $|y|^{-1}$ is
superharmonic) we have
$$
        \left\|\left[\phitf-\phitf*g^2\right]_+\right\|_{5/2}^{5/2}\leq
        Z^{5/2}\left\||y|^{-1}-g^2*|y|^{-1}\right\|_{5/2}^{5/2}\leq
        8\pi Z^{5/2} s^{1/2},
$$
where we have used that
$|y|^{-1}-g^2*|y|^{-1}$ is nonnegative, bounded by $|y|^{-1}$, 
and vanishes for $|y|>s$.
If we insert this above and optimize \pagebreak in 
$s$ we obtain
\begin{eqnarray}
&&\\
        \ehf(\gammahf)&\hskip-9pt\geq\hskip-9pt&-2^{5/2}(15\pi^2)^{-1}(1-\delta)^{-3/2}\int
        \left[\phitf-\mutf\right]_+^{5/2}
        \!-\!\mutf N \!-\!D(\rhotf,\rhotf)\nonumber \\
        &\hskip-9pt\hskip-9pt&{}
        +\|\rhotf-\rhohf\|_{\rC}^2
        -\const\delta^{-6/5}N^{1/5}Z^2
        -\Ex(\gammahf).\nonumber
\end{eqnarray}
We choose $\delta:=\mfr{1}{2}Z^{-2/33}$ (this is not optimal). 
Then for $Z\geq1$ we have $\delta\leq 1/2$ and thus 
$(1-\delta)^{-2/3}\leq 1+(2^{5/2}-2)\delta$.
Hence
\begin{eqnarray}\qquad
        \ehf(\gammahf)\geq-2^{5/2}(15\pi^2)^{-1}\int
        \left[\phitf-\mutf\right]_+^{5/2}
        -\mutf N -D(\rhotf,\rhotf)\label{eq:hflower}\\
        +\|\rhotf-\rhohf\|_{\rC}^2- \const Z^{7/3-2/33}
        -\Ex(\gammahf) ,\nonumber
\end{eqnarray}
where we have again used $N\leq 2Z+1$ and the fact that by
(\ref{eq:5/3}) we have  
$$
        2^{5/2}(15\pi^2)^{-1}\int
        \left[\phitf-\mutf\right]_+^{5/2}=\mfr{2}{3}\left[
        (3\pi^2)^{2/3}\mfr{3}{10}\int\left(\rhotf\right)^{5/3}\right]
        \leq\const Z^{7/3}.
$$

On the other hand, since $\gammahf$ minimizes $\ehf$ among all 
density matrices $\gamma$ with $\Tr[\gamma]\leq N$, we can 
find an upper bound to $\ehf(\gammahf)$ by choosing an appropriate 
trial density matrix. 
We choose as trial matrix $\gamma$ an operator
which acts identically on the two spin components. On 
each spin component we choose it to be the density matrix constructed 
in Lemma~\ref{lm:sc} satisfying (\ref{eq:scupper1}) with
$V=\phitf-\mutf$. 
Note that the Thomas-Fermi equation (\ref{eq:tfsystemrho}) and the properties
of $\gamma$ stated in Lemma~\ref{lm:sc} imply that 
$\uprho_\gamma=2^{5/2}(6\pi^2)^{-1}\left[\phitf-\mutf\right]_+^{3/2}*g^2=\rhotf*g^2$
(the extra factor of $2$ compared to Lemma~\ref{lm:sc} is of course 
due to the spin degeneracy). Thus
$\Tr[\gamma]=\int \rhotf\leq N$.

{F}rom (\ref{eq:ehf}) and (\ref{eq:scupper1}) we find, since 
$\Ex(\gamma)\geq0$, that
$$
        \ehf(\gamma)\leq 2^{3/2}(5\pi^2)^{-1}
        \int \left[\phitf-\mutf\right]_+^{5/2}+\mfr{1}{2}\pi^2s^{-2}N
        -\int Z |y|^{-1}\uprho_\gamma(y)dy
        +D(\uprho_\gamma,\uprho_\gamma).
$$
Since $\iint g(x-z)^2|z-w|^{-1}g(y-w)^2dw\,dz\leq |x-y|^{-1}$ 
we see that $D(\uprho_\gamma,\uprho_\gamma)\leq D(\rhotf,\rhotf)$.
Thus from the definition (\ref{eq:phitf}) of $\phitf$ we can write
\begin{eqnarray*}
        \ehf(\gamma)&\leq& 2^{3/2}(5\pi^2)^{-1}
        \int \left[\phitf-\mutf\right]_+^{5/2}
        -\int \left[\phitf(y)-\mutf\right]\rhotf(y)dy- \mutf N\nonumber \\
        &&{} -D(\rhotf,\rhotf)
        +\mfr{1}{2}\pi^2s^{-2}N+\int Z \left(|y|^{-1}-g^2*|y|^{-1}\right)
        \rhotf(y)dy.
\end{eqnarray*}
If we use the TF equation (\ref{eq:tfsystemrho}), the estimate
$\rhotf(y)\leq 2^{3/2}(3\pi^2)^{-1}Z^{3/2}|y|^{-3/2}$ which follows
from the TF equation, and again the facts that 
$|y|^{-1}-g^2*|y|^{-1}$ is nonnegative, 
bounded by $|y|^{-1}$, and vanishes for $|y|>s$,
we obtain after optimizing in $s$
\begin{eqnarray}\qquad
        \ehf(\gamma)&\leq&-2^{5/2}(15\pi^2)^{-1}\int
        \left[\phitf-\mutf\right]_+^{5/2}
        -\mutf N -D(\rhotf,\rhotf)\label{eq:hfupper}\\
        &&{} +\const N^{1/5}Z^2.\nonumber 
\end{eqnarray}

Comparing (\ref{eq:hflower}) and (\ref{eq:hfupper}) and recalling that 
$\ehf(\gamma)\geq\ehf(\gammahf)$ we get that 
$$
        \|\rhotf-\rhohf\|_{\rC}^2\leq \const Z^{7/3-2/15}
        +\const Z^{7/3-2/33}+\Ex(\gammahf).
$$
If we finally use the exchange inequality in Theorem ~\ref{thm:exchineq}
and the estimate (\ref{eq:5/3}) 
we see that 
$$
        \Ex(\gammahf)\leq 1.68\left(\int\left(\rhohf\right)^{5/3}
        \right)^{1/2}
        \left(\int\left(\rhohf\right)\right)^{1/2}
        \leq \const N^{1/2}Z^{7/6}.
$$
Inserting this above and again using $N\leq 2Z+1$ we arrive at 
(\ref{eq:||C}).
\enddemo

\demo{End of proof of Lemma~{\rm \ref{lm:smallx}}}
We write
$$
        \Phihf{|x|}(x)-\Phitf{|x|}(x)
        =\int_{|y|<|x|}\left[\rhotf(y)-\rhohf(y)\right]|x-y|^{-1}dy.
$$
Using the Coulomb norm estimate (\ref{eq:cn2})
we find 
\begin{eqnarray*}
        \left|\Phihf{|x|}(x)-\Phitf{|x|}(x)\right|
        &\leq&
        2^{7/5}\pi^{2/5}(\kappa |x|)^{1/5}
        \max\left\{\|\rhotf\|_{L^{5/3}(\R^3)},
        \,\|\rhohf\|_{L^{5/3}(\R^3)}\right\}
        \\
        &&{}
        +2^{3/2}\kappa^{-1}|x|^{-1/2}\|\rhohf-\rhotf\|_\rC.
\end{eqnarray*}
Thus from Lemma~\ref{lm:5/3cn} and the fact that $\int\left(\rhotf\right)^{5/3}
\leq\const Z^{7/3}$ (which can be seen for instance from the
Sommerfeld estimate Theorem~\ref{thm:sommerfeldup} together with the
TF equation (\ref{eq:tfsystemrho})) we obtain
\begin{eqnarray*}
        \left|\Phihf{|x|}(x)-\Phitf{|x|}(x)\right|
        &\leq& \const(\kappa |x|)^{1/5}Z^{7/5}
          +\const \kappa^{-1}|x|^{-1/2}
        Z^{7(1-\varepsilon_3)/6}\\
        &=&\const |x|^{1/12} Z^{7/6+7(1-\varepsilon_3)/36},
\end{eqnarray*}
where the last equality above follows from choosing the optimal
value for $\kappa$.

Hence if $|x|\leq \beta Z^{-1/3}$ we have
$$
        \left|\Phihf{|x|}(x)-\Phitf{|x|}(x)\right|
        \leq A_{\Phi} \beta^{49/12-7\varepsilon_3/12}
        |x|^{-4+7\varepsilon_3/12}.
$$
The lemma follows since $7\varepsilon_3/12=1/66$.\hfill\qed\enddemo

\section{Proof of  the iterative step Lemma~\ref{lm:iterate} and of 
Lemma~\ref{lm:me5/3}}
\label{sec:iteration}

We begin by fixing some $0<r$
such that (\ref{eq:phiasp}) holds for all $|x|\leq r$.

We shall proceed as for the region close to the nucleus, 
but instead of directly comparing HF and TF.
We shall introduce an intermediate TF theory. 
Namely, the TF theory defined as in  Definition~4.1
from the functional $\etfr:=\etf{\V{r}}$
with the exterior potential $V=\V{r}$ given by 
$$
        \V{r}(y):=\chiplus{r}(y)\Phihf{r}(y)
        =\left\{
        \begin{array}{ll}
        0,&\hbox{if }|y|<r\\
        \Phihf{r}(y),&\hbox{if }|y|\geq r.
        \end{array}\right. 
$$ 
Here again $\chiplus{r}=1-\chiminus{r}$ is the characteristic function 
of the set $\set{x}{|x|\geq r}$.
Note that this potential is harmonic and continuous on $|x|>r$.
Let $\rhotfr$ denote the minimizer for the TF functional,
$\etfr(\rho)$, under the constraint $\int\rho\leq\int\rhohf
\chiplus{r}$.
Denote the corresponding TF potential by 
$$
        \phitfr(y):=\V{r}(y)-\rhotfr*|y|^{-1}
$$ 
and the corresponding chemical potential by $\mutfr$.
We shall prove below (see Lemma~\ref{lm:sommerfeldotf})
that if $r$ is chosen appropriately then $\mutfr=0$.

Note that according to the Thomas-Fermi equation
(\ref{eq:tfeqgeneral}), $\rhotfr$ has support
on the set $\set{y}{|y|\geq r}$.
Since $\V{r}$ on the support of $\rhotfr$ is the potential 
coming from the true HF density for $|y|<r$ we
may interpret $\rhotfr$ as the TF approximation
for only the outside region, i.e., $|y|>r$,  of the atom.
The notation OTF refers to {\it Outside} TF.

 \proclaimtitle{Preliminary bounds on TF and OTF functions}
\proclaim{Lemma}\label{lm:tfotfprelims} 
Assume that $N\geq Z$ then for all $y$
$$
 \phitf(y)\leq 3^42^{-3}\pi^2|y|^{-4}\quad
 \hbox{and}\quad
 \rhotf(y)\leq 3^52^{-3}\pi|y|^{-6}.
$$
For all $|y|\geq \beta_0 Z^{-1/3}$ we have 
$$
\phitf(y)\geq\const |y|^{-4}\quad
 \hbox{and}\quad
 \rhotf(y)\geq \const|y|^{-6}.  
$$
Given $\epsilon',\sigma>0${\rm ,} and $r>0$ such that
{\rm (\ref{eq:phiasp})} holds for all $|x|\leq r$ and $\sigma
r^{\epsilon'}\leq 1$ then for all 
$|y|\geq r$ we have  
$$
\rhotfr(y)\leq \const r^{-6}\quad
\hbox{and}\quad
\phitfr(y)\leq |\V{r}(y)|=|\Phihf{r}(y)|\leq \const r^{-4}.
$$ 
\endproclaim

\demo{Proof}
The upper bounds on the TF functions follow immediately 
from the Atomic Sommerfeld
estimate Theorem~\ref{thm:sommerfeldup})
and the TF equation (\ref{eq:tfsystemrho}) if we recall that
$\mutf=0$.
The lower bounds follow from Theorem~\ref{thm:sommerfeldlow}.

Since $\Phihf{r}$ is harmonic on the set $\{|y|>r\}$ and tends to zero at
infinity it follows from the assumptions on $\epsilon',\sigma,r$ that  
for all $|y|\geq r$ we have
$$
   |\Phihf{r}(y)|\leq \sup_{|z|=r}|\Phihf{r}(z)|
   \leq \const r^{-4},
$$
where in the last inequality 
we have used the iterative assumption (\ref{eq:phiasp})
and Lemma~\ref{lm:Phitfrbound} for the case $\mutf=0$
and the fact that 
$\Phitf{r}\geq\phitf\geq0$ (see e.g.,
Theorem~\ref{thm:sommerfeldlow}).
The inequality  $\phitfr(y)\leq |\V{r}(y)|$ is trivial from the
definition of $\phitfr$.

Finally, from the TF equation (\ref{eq:tfeqgeneral}) we conclude that
for all $|y|\geq r$
\vglue12pt
\hfill ${\displaystyle
\rhotfr(y)\leq \const \V{r}(y)^{3/2}\leq \const r^{-6}.
}$
\enddemo
\advance\eqcount by 95

\proclaimtitle{Preliminary comparison of HF and TF}
\proclaim{Lemma} \label{lmcnrtftorp}\hskip-8pt
Assume that $N\geq Z$. Given $\epsilon',\sigma>0${\rm ,} and $r>0$ such that
{\rm (\ref{eq:phiasp})} holds for all $|x|\leq r$ then
\begin{equation}
   \int\chiplus{r}\left(\rhotf
          -\rhohf\right) \leq {\sigma} 
    r^{-3+\varepsilon'}.
\end{equation}
\endproclaim

\demo{Proof}
We have 
$$
        \int_{|y|<r}\left(\rhotf(y)-\rhohf(y)\right)\, dy
        =(4\pi)^{-1}r\int_{{\cal S}^2}
        \left(\Phihf{r}(r\omega)-\Phitf{r}(r\omega)\right)\, d\omega
$$
where $d\omega$ denotes the surface measure of the unit sphere
${\cal S}^2$. Thus according to (\ref{eq:phiasp}) we have 
$$
        \left|\int_{|y|<r}\left(\rhotf(y)-\rhohf(y)\right)\, dy\right|\leq
       {\sigma} r^{-3+\varepsilon'}.
$$
Since
$\int\rhotf\leq N=\int\rhohf$ we have
\vglue12pt
\hfill ${\displaystyle
  \int\left(\chiplus{r}\rhotf-\chiplus{r}\rhohf\right)
  \leq\int_{|y|<r}\left(\rhohf(y)-\rhotf(y)\right)\, dy\leq
    \sigma r^{-3+\varepsilon'}.
}$ 
\enddemo

For $|x|>r$ we may write
\begin{equation}\label{eq:AAA}
 \Phihf{|x|}(x)-\Phitf{|x|}(x)
 ={\cal A}_1(r,x)
 +{\cal A}_2(r,x)+{\cal A}_3(r,x),
\end{equation}
where 
\begin{eqnarray}
 {\cal A}_1(r,x)&=&\phitfr(x)-\phitf(x),\label{eq:A1defi}\\
 {\cal A}_2(r,x) &=&\int_{|y|>|x|}
 \left[\rhotfr(y)-\rhotf(y)\right]|x-y|^{-1}dy,\label{eq:A2defi}\\
 {\cal A}_3(r,x)&=&
 \int_{r<|y|<|x|}\left[\rhotfr(y)-\rhohf(y)\right]|x-y|^{-1}dy.
 \label{eq:A3defi}
\end{eqnarray}
      
We turn first to estimating ${\cal A}_1$ and ${\cal A}_2$.
Thus we need to control the difference between the full TF approximation
and the TF approximation for the outside region. 
Our strategy is to first prove that $\phitfr(x)$ and $\phitf(x)$ are close 
on the set $\{|x|=r\}$. An application of the Sommerfeld estimates
in Theorem~\ref{thm:sommerfeldmu} will then give excellent control on 
the difference $\phitfr(x)-\phitf(x)$ for all $|x|>r$. 
Controlling the behavior on the set $\{|x|=r\}$ is difficult and 
we begin with a weak estimate on the difference between 
$\rhotfr$ and $\rhotf$. In fact, we first estimate the 
difference in Coulomb norm. 

 \proclaimtitle{Coulomb norm comparison of TF and OTF}
\proclaim{Lemma}\label{lm:TFOTF} \hskip-8pt
Assume $N\!\geq\! Z$. Given constants $\varepsilon',{\sigma}>0$ 
there exists  a constant  $D>0$ depending only on 
$\varepsilon',{\sigma}$
such that for all $r$ with $\beta_0Z^{-1/3}\leq r\leq D$
for which  
{\rm (\ref{eq:phiasp})} holds for all $|x|\leq r$ we have 
\begin{equation}\label{eq:cnorm}
        \|\rhotfr-\chiplus{r}\rhotf\|_\rC^2\leq \const{\sigma} 
                r^{-7+\varepsilon'}.
\end{equation}
Here again
$\chiplus{r}$ denotes the characteristic function of the set 
$\set{y}{|y|\geq r}$. Moreover{\rm ,}
\begin{equation}\label{eq:cmu}
   \mutfr\leq \const {\sigma}^{1/2}r^{-4+\frac{\varepsilon'}{2}}.
\end{equation}
\endproclaim

\vglue-8pt
{\it Proof}.
To prove this we make a perturbation analysis of the TF functional.
We introduce the perturbation  potential
\begin{equation}\label{eq:perturbationpotential}
        W(x)=\Phihf{r}(x)-\Phitf{r}(x).
\end{equation}
Then for all $|x|>r$
$$
        \Phitf{r}(x)=\V{r}(x)-W(x).
$$
Note that $W$ is harmonic for $|x|>r$ and tends to zero at infinity.
Hence,  since we assume that the iterative assumption 
(\ref{eq:phiasp})
holds for $|x|=r$, we have
\begin{equation}\label{eq:iterativeassumption}
        \sup_{|x|\geq r}|W(x)|=
        \sup_{|x|=r}|W(x)|\leq {\sigma} r^{-4+\epsilon'}.
\end{equation}
We claim that there exist two functions $W_1$, $W_2$ 
with $\supp W_1\subset \set{x}{|x|<3r}$
and $\supp W_2\subset\set{x}{|x|>2r}$ such that $W(x)=W_1(x)+W_2(x)$
and 
\begin{eqnarray}
  \sup_{|x|\geq r}|W_1(x)|&\leq& \sup_{|x|=r}|W(x)|\leq {\sigma}
  r^{-4+\epsilon'}\label{eq:W1}\\
  \int|\nabla W_2|^2&\leq&4\pi r \sup_{|x|=r}|W(x)|^2
  \leq 4\pi
  {\sigma}^2 r^{-7+2\epsilon'}\label{eq:W2}.
\end{eqnarray}
In order to prove this we let
$$
        F(x)=\left\{\begin{array}{ll}
                0&\mbox{if }|x|< 2r\\
                (|x|-2r)r^{-1}, 
                &\mbox{if }2r \leq|x|\leq 3r\\
                1&\mbox{if }|x|>3r .
                   \end{array}\right.
$$
Set $W_1(x)=(1-F(x))W(x)$ and $W_2(x)=F(x)W(x)$.
The first estimate (\ref{eq:W1}) follows immediately from
(\ref{eq:iterativeassumption}). 
By a simple integration by parts, similar to the one used to prove the
IMS formula (\ref{eq:IMSformula}), we obtain (note
that $W(x)$ behaves like $c|x|^{-1}$ and $|\nabla W(x)|$
behaves like $c|x|^{-2}$ at infinity so there are no 
contributions from infinity
to the integration by parts)
\begin{eqnarray*}
    \int|\nabla W_2|^2&=&\int|\nabla F|^2 |W|^2-\int|F|^2W\Delta W
    =\int|\nabla F|^2 |W|^2\\
    &\leq& \sup_{|x|\geq r}|W(x)|^2\int_{2r\leq |x|\leq 3r}r^{-2}dx=
    4\pi r \sup_{|x|\geq r}|W(x)|^2,
\end{eqnarray*}
where the second equality follows since $W$ is harmonic on the support 
of $F$. The estimate (\ref{eq:W1}) now also follows from 
(\ref{eq:iterativeassumption}). 

We are now ready to estimate the TF densities.
We shall use $\chiplus{r}\chiminus{R}\rhotf$ for some $R\geq r$
(possibly $R$ is infinity) 
as a trial density in $\etfr$.

Since $\rhotfr$ minimizes $\etfr\left(\uprho\right)+\mutfr\int\uprho$
and $\mutfr=0$ unless $\int\rhotfr=\int\chiplus{r}\rhohf$
we have
\begin{equation}\label{eq:etfrtrial}
  \mutfr\left(\int\chiplus{r}\rhohf-\int\chiplus{r}\chiminus{R}\rhotf\right)
  \leq \etfr\left(\chiplus{r}\chiminus{R}\rhotf\right)- 
  \etfr\left(\rhotfr\right).
\quad
\end{equation}
We write the right side as
\begin{eqnarray}
&&\label{eq:etfrsplit}\\
        \etfr\left(\chiplus{r}\chiminus{R}\rhotf\right)- 
        \etfr\left(\rhotfr\right)
        &=&\etfr\left(\chiplus{r}\chiminus{R}\rhotf\right)-
                \etfr\left(\chiplus{r}\rhotf\right) \nonumber\\ &&{}+
                \etfr\left(\chiplus{r}\rhotf\right)
                -\etfr\left(\rhotfr\right).\nonumber
\end{eqnarray}
We have for the first two terms  
\begin{eqnarray}
      && \etfr\left(\chiplus{r}\chiminus{R}\rhotf\right)-
                \etfr\left(\chiplus{r}\rhotf\right) \label{eq:etfrfirst}\\ &&\qquad\quad=
        \int\phitf\rhotf\chiplus{R}+ \|\chiplus{R}\rhotf\|_\rC^2\nonumber\\
    &&\qquad \qquad   +\ \int\left(\Phihf{r}-\Phitf{r}\right)\chiplus{R}\rhotf
        -\mfr{3}{10}(3\pi^2)^{2/3}\int \chiplus{R}(\rhotf)^{5/3}   \nonumber\\ 
        &&\qquad\quad\leq\ \int\phitf\rhotf\chiplus{R}+ \|\chiplus{R}\rhotf\|_\rC^2
        +{\sigma}r^{-4+\varepsilon'}
        \int\chiplus{R}\rhotf,\nonumber
\end{eqnarray}
where we have used (\ref{eq:iterativeassumption}). For the last two
terms in (\ref{eq:etfrsplit}) we find
 \begin{eqnarray}
&& \label{eq:etfrsecond}\\
 &&\hskip-12pt\etfr(\chiplus{r}\rhotf)-\etfr(\rhotfr)
        =\int W(\rhotfr-\chiplus{r}\rhotf)-
        \|\rhotfr-\chiplus{r}\rhotf\|_\rC^2\nonumber\\
        &&\mbox{}+\int_{|y|\geq r}\biggl(
        \Bigl[\mfr{3}{10}(3\pi^2)^{2/3}\rhotf(y)^{5/3}
        -\phitf(y)\rhotf(y)\Bigr]\nonumber\\ 
        &&\phantom{\mbox{}+\int_{|y|\geq r}}-
        \Bigl[\mfr{3}{10}(3\pi^2)^{2/3}\rhotfr(y)^{5/3}-
        \phitf(y)\rhotfr(y)\Bigr]\biggr)dy.\nonumber 
\end{eqnarray}
Using that $\rhotf$ satisfies the TF equation (\ref{eq:tfsystemrho})
and that $\mutf=0$
we see that for fixed $y$ the expression
$$
        \mfr{3}{10}(3\pi^2)^{2/3}t^{5/3}-\phitf(y)t,
        \qquad t\geq0
$$
takes its minimal value for $t=\rhotf(y)$. 
Hence we conclude that
the last integral above is negative. Thus combining
(\ref{eq:etfrtrial})--(\ref{eq:etfrsecond})
we have
\begin{eqnarray*}
 &&\mutfr\left(\int\chiplus{r}\rhohf
      -\int\chiplus{r}\chiminus{R}\rhotf\right)
    \leq\int W(\rhotfr-\chiplus{r}\rhotf)-
    \|\rhotfr-\chiplus{r}\rhotf\|_\rC^2  \\ 
  &&\qquad
  +\int\phitf\rhotf\chiplus{R}
  + \|\chiplus{R}\rhotf\|_\rC^2
  +{\sigma}r^{-4+\varepsilon'}
  \int\chiplus{R}\rhotf.
\end{eqnarray*}
Lemma~\ref{lm:tfotfprelims} implies that $\int\phitf\rhotf\chiplus{R}$, 
$\|\chiplus{R}\rhotf\|_\rC^2\leq\const R^{-4}\int\chiplus{R}\rhotf$.
We thus  \pagebreak arrive at
\begin{eqnarray}&& \label{eq:cmain} 
\\
   \mutfr\left(\int\chiplus{r}\rhohf
       -\int\chiplus{r}\chiminus{R}\rhotf\right)
     &\nhs\leq\nhs&\int W(\rhotfr-\chiplus{r}\rhotf)-
        \|\rhotfr-\chiplus{r}\rhotf\|_\rC^2
\nonumber  \\ 
      &\nhs\nhs&{}
        \!\!\!\!+\left(\const R^{-4}
          +{\sigma}r^{-4+\varepsilon'}\right)
        \int\rhotf\chiplus{R}. \nonumber
\end{eqnarray}
This is the main estimate from which we shall derive the estimates of the 
lemma. We shall do this by choosing different values for $R$.
One for the estimate (\ref{eq:cnorm}) another for 
(\ref{eq:cmu}).
We shall choose $R$  such that 
\begin{equation}\label{eq:Rchoice}
   \int\chiplus{r}\chiminus{R}\rhotf\leq\int\chiplus{r}\rhohf.
\end{equation}
 Let $R_{\max}$ denote the largest possible $R$ for which this holds. 
Then 
\begin{equation}\label{eq:Rmax}
\int\chiplus{R_{\max}}\rhotf=
\left(\int\chiplus{r}\rhotf-\int\chiplus{r}\rhohf\right)_+
\leq {\sigma}r^{-3+\varepsilon'},
\end{equation}
where the last inequality follows from
Lemma~\ref{lmcnrtftorp}. 
By Lemma~\ref{lm:tfotfprelims} we have for all $R\geq
r\geq\beta_0Z^{-1/3}$ that
\begin{equation}\label{eq:Rbound}
 \int\chiplus{R}\rhotf \geq \const R^{-3}.
\end{equation}
We shall now make the assumption that 
$D$ is chosen so small that if  $r\leq D$ then
${\sigma}r^{\epsilon'}\leq1$.
Thus from (\ref{eq:Rmax}) and (\ref{eq:Rbound}) we conclude that
\begin{equation}\label{eq:Rmaxbound}
        R_{\max}^{-4}\leq \const {\sigma}^{4/3} r^{-4
            +\frac{4}{3}\varepsilon'}\leq \const r^{-4}.
\end{equation}
{F}rom (\ref{eq:Rmax}) and (\ref{eq:cmain}) with $R=R_{\max}$ 
we get, again using the above assumption on $D$, that
\begin{equation}\label{eq:coulombw}
   \|\rhotfr-\chiplus{r}\rhotf\|_\rC^2\leq\int W(\rhotfr-\chiplus{r}\rhotf)
        +\const {\sigma}r^{-7+\varepsilon'}.\hskip.4in
\end{equation}
We estimate the 
integral on the right by dividing it in two parts
\begin{eqnarray*}
&& \int W(\rhotfr-\chiplus{r}\rhotf)
    \leq \int|W_1|(\rhotfr+\chiplus{r}\rhotf)
    +\int W_2(\rhotfr-\chiplus{r}\rhotf) \\
  &&\qquad\leq\  {\sigma}r^{-4+\epsilon'}\int_{r<|x|<3r}\rhotfr(x)
  +\chiplus{r}(x)\rhotf(x)dx+\int W_2(\rhotfr-\chiplus{r}\rhotf),
\end{eqnarray*}
where we have also used (\ref{eq:W1}).
{F}rom Lemma~\ref{lm:tfotfprelims} we arrive at 
\begin{equation}\label{eq:intW} 
  \int W(\rhotfr-\chiplus{r}\rhotf)
  \leq \const {\sigma}r^{-7+\epsilon'}
  +\int W_2(\rhotfr-\chiplus{r}\rhotf).\hskip.25in
\end{equation}
The last term in this estimate we now control using the 
Coulomb norm estimate Lemma~\ref{lm:cn}. Note that $\rhotf$ and $\rhotfr$ 
both belong to $L^{6/5}$ since they are in $L^{5/3}\cap L^{1}$.
We find from (\ref{eq:W2}) that 
\begin{equation}\label{eq:W22}
   \int W_2(\rhotfr-\chiplus{r}\rhotf)\leq 
  \const{\sigma}r^{-\frac{7}{2}+\epsilon'}     
   \|\rhotfr-\chiplus{r}\rhotf\|_\rC.\hskip.5in
\end{equation}
Inserting the last two estimates into (\ref{eq:coulombw})
gives (recall that ${\sigma}r^{\epsilon'}\leq1$)
$$
\|\rhotfr-\chiplus{r}\rhotf\|_\rC^2
\leq \const {\sigma}r^{-7+\epsilon'}
+\const \left({\sigma}r^{-7+\epsilon'}\right)^{1/2}
\|\rhotfr-\chiplus{r}\rhotf\|_\rC
$$
 and (\ref{eq:cnorm}) follows immediately from this.

We now return to (\ref{eq:cmain}) and make a new choice for $R$.
It follows from the first inequality in (\ref{eq:Rmaxbound}) that that we can
find $R$ with $R\leq R_{\max}$ 
satisfying 
$$
R^{-4}=\const{\sigma}^{1/2}r^{-4+\epsilon'/2}.
$$
Note that, since 
$$
R^{-4}\leq \const{\sigma}^{1/2}D^{\epsilon'/2}r^{-4},
$$
we can 
choose $D$ small enough to ensure that $r\leq R$. 
{F}rom (\ref{eq:Rbound}) we have
\begin{equation}\label{eq:intchiplusR}
\int\chiplus{R}\rhotf\geq \const
  {\sigma}^{3/8}r^{-3+\frac{3}{8}\epsilon'}.
\end{equation}
It follows from (\ref{eq:Rmax}) and (\ref{eq:Rbound}) that we may
assume that the constant in the definition of $R$ is chosen such as
to ensure that
$
\int\chiplus{R}\rhotf\geq2\int\chiplus{R_{\max}}\rhotf.
$
Thus since (see (\ref{eq:Rmax}))
$$
        \int\chiplus{r}\rhohf
       -\int\chiplus{r}\chiminus{R}\rhotf\geq
        -\int\chiplus{R_{\max}}\rhotf +\int\chiplus{R}\rhotf
$$
we have
$$
\mfr{1}{2}\int\chiplus{R}\rhotf \leq\int\chiplus{r}\rhohf
-\int\chiplus{r}\chiminus{R}\rhotf.
$$
Thus from (\ref{eq:intchiplusR}), (\ref{eq:intW}), (\ref{eq:W22}), (\ref{eq:cnorm}),
and (\ref{eq:cmain}) we conclude (\ref{eq:cmu}).
\hfill\qed\vglue12pt
 
Using these fairly weak estimates we shall now show that 
the outside TF potential and density satisfy Sommerfeld
type estimates.

\proclaimtitle{Sommerfeld estimates for OTF}
\proclaim{Lemma} \label{lm:sommerfeldotf}
Assume $N\geq Z$. Given constants $\varepsilon',{\sigma}>0$ 
there exists a constant  $D>0$ depending only on 
$\varepsilon',{\sigma}$
such that for all $r$ with $\beta_0Z^{-1/3}\leq r\leq D$
for which 
{\rm (\ref{eq:phiasp})} holds for all $|x|\leq r$ then $\mutfr=0$ and for
all $|x|\geq r$ we have 
\begin{eqnarray}
        \phitfr(x)&\leq& 3^42^{-3}\pi^2|x|^{-4}
        \left(1+Ar^\zeta|x|^{-\zeta}\right)\label{eq:phitfrup}\\
\noalign{\hbox{and}}
        \phitfr(x)&\geq&
        3^42^{-3}\pi^2|x|^{-4}\left(1+ar^\zeta|x|^{-\zeta}\right)^{-2},
        \label{eq:phitfrlow}
\end{eqnarray}
where $a$ and $A$ are universal constants {\rm (}\/but not necessarily
positive\/{\rm )} and $a>-1$.
Here as before $\zeta=(-7+\sqrt{73})/2\approx0.77$.
\endproclaim

\demo{Proof}
Note first that the potential $\V{r}$ satisfies the assumptions 
in Theorem~\ref{thm:sommerfeldmu} with $R=r$. Hence if we can show 
that $\mutfr< \inf_{|x|=r} \phitfr(x)$
the potential $\phitfr$ will satisfy the Sommerfeld estimates
described in the theorem. In order to control $\phitfr(x)$ for $|x|=r$
we note that for all $|x|\geq r$ we have 
\begin{equation}\label{eq:phitfrbound}
  \phitfr(x)=\phitf(x)+\left(\chiplus{r}\rhotf-\rhotfr\right)*|x|^{-1}
  +W(x),
\end{equation}
where as in (\ref{eq:perturbationpotential}),
$W=\Phihf{r}-\Phitf{r}$. According to the Coulomb norm 
Corollary~\ref{cl:cn} we get for all $s>0$ that 
\begin{eqnarray*}
    \lefteqn{\left|\left(\chiplus{r}\rhotf-\rhotfr\right)*|x|^{-1}\right|}&&
    \\ 
    &\leq&\const s^{1/5}\max\{\|\chiplus{r}\rhotf\|_{L^{5/3}(\B{x,s})},
    \|\rhotfr\|_{L^{5/3}(\B{x,s})}\}  \\ &&{}
    +\const s^{-1/2}\|\chiplus{r}\rhotf-\rhotfr\|_\rC.
\end{eqnarray*}
{F}rom Lemma~\ref{lm:tfotfprelims}
we see that 
$$    
    \max\{\|\chiplus{r}\rhotf\|_{L^{5/3}(\B{x,s})},
    \|\rhotfr\|_{L^{5/3}(\B{x,s})}\}\leq
    \const
    r^{-6} s^{9/5},
$$
where we have assumed that $D$ is such that ${\sigma}r^{\epsilon'}\leq1$.
Inserting this and the estimate (\ref{eq:cnorm}) 
from Lemma~\ref{lm:TFOTF}
above we obtain that for all $|x|\geq r$
\begin{eqnarray}
&& \label{eq:(rhotf-rhotfr)bound}\\
  \left|\left(\chiplus{r}\rhotf-\rhotfr\right)*|x|^{-1}\right|
  &\leq& \const r^{-6}s^2
  +\const s^{-1/2}\left(
    {\sigma}r^{-7+\epsilon'}\right)^{1/2} \nonumber
\\
  &=&\const \sigma^{2/5} r^{-4+2\epsilon'/5}\nonumber
\end{eqnarray}
where we have optimized in $s$
in order to get the last expression.
Thus from (\ref{eq:phitfrbound}), the iterative assumption (\ref{eq:phiasp}), 
and Lemma~\ref{lm:tfotfprelims} we obtain
\begin{eqnarray*}
  \inf_{|x|=r}\phitfr(x)&\geq&\inf_{|x|=r}\phitf(x)
  -\const \sigma^{2/5} r^{-4+2\epsilon'/5}\\
  &\geq&\const r^{-4}-\const\sigma^{2/5} r^{-4+2\epsilon'/5}.
\end{eqnarray*}
We have used that since $\sigma r^{\epsilon'}\leq 1$ 
the error from (\ref{eq:(rhotf-rhotfr)bound}) is worse than the error
from (\ref{eq:phiasp}). Note that the constant in front of $r^{-4}$ 
above is positive.

{F}rom 
Lemma~\ref{lm:TFOTF}  we know that 
$
  \mutfr\leq \const {\sigma}^{1/2}r^{-4+{\epsilon'}/{2}}
$
Hence we may choose $D$ such that 
$\mutfr<\inf_{|x|=r}\phitfr(x)$.

As above we of course also have
$$
  \sup_{|x|=r}\phitfr(x)\leq \const r^{-4}+\const\sigma^{2/5}
  r^{-4+2\epsilon'/5}.
$$
Thus $\const r^{-4}\leq \inf_{|x|=r}\phitfr(x)
\leq\sup_{|x|=r}\phitfr(x)\leq\const r^{-4}$.

That $\mutfr=0$ 
follows from Corollary~\ref{cl:muestimate} as follows.
By harmonicity of $\V{r}$ we have
$$
   \lim_{r'\to\infty}(4\pi)^{-1}\int_{{\cal S}^2}r'\V{r}(r'\omega)d\omega
   =(4\pi)^{-1}\int_{{\cal S}^2}r\Phihf{r}(r\omega)d\omega
   =Z-\int\chiminus{r}\rhohf.
$$
Thus  from Corollary~\ref{cl:muestimate} we have if
$\mutfr\ne0$, i.e., if $\int\rhotfr=\int\chiplus{r}\rhohf$ that
$$
   0<(\mutfr)^{3/4}\leq \const\left( Z-\int\chiminus{r}\rhohf-\int\rhotfr\right)
   =\const (Z-\int\rhohf),
$$
which is a contradiction since $\int\rhohf=N\geq Z$.

The estimates (\ref{eq:phitfrup}) and (\ref{eq:phitfrlow})
now follow from Theorem~\ref{thm:sommerfeldmu}. 
\enddemo

We are now ready to give the bounds on ${\cal A}_1$ 
and ${\cal A}_2$ defined in (\ref{eq:A1defi}) and~(\ref{eq:A2defi}).

\proclaimtitle{Control of ${\cal A}_1$ and ${\cal A}_2$}
\proclaim{Lemma} \label{lm:A1A2}
Assume $N\geq Z$. Given constants $\varepsilon',{\sigma}>0$  
there exists a constant  $D>0$ depending only on 
$\varepsilon',{\sigma}$
such that for all $r$ with $\beta_0Z^{-1/3}\leq r\leq D$
for which 
{\rm (\ref{eq:phiasp})} holds for all $|x|\leq r$
we have  for all $|x|\geq r$ that
\begin{equation}\label{eq:A1}
         \left|{\cal A}_1(r,x)\right|
        \leq \const r^{\zeta}|x|^{-4-\zeta}
\end{equation}
and
\begin{equation}\label{eq:A2}
  \left|{\cal A}_2(r,x)\right|
  \leq \const r^{\zeta}|x|^{-4-\zeta},
\end{equation}
with $\zeta=(-7+\sqrt{73})/2\approx0.77$.
\endproclaim 

\demo{Proof}
Combining  
Theorems~\ref{thm:sommerfeldup}, \ref{thm:sommerfeldlow},
with Lemma~\ref{lm:sommerfeldotf} and recalling that $\mutf=0$
immediately gives the bound on ${\cal A}_1$. 

If we use the TF equation (\ref{eq:tfeqgeneral})
we obtain from Lemma~\ref{lm:sommerfeldotf}
that for all $|y|\geq r$
$$
\left|\rhotfr(y)-3^52^{-3}\pi |y|^{-6}\right|\leq \const
r^{\zeta}|y|^{-6-\zeta}.
$$
Of course we similarly have from 
Theorems~\ref{thm:sommerfeldup}, \ref{thm:sommerfeldlow},
and the TF equation (\ref{eq:tfsystemrho}) that
\begin{eqnarray*}
  \rhotf(y)&\leq& 3^52^{-3}\pi |y|^{-6}
  \\
\noalign{\hbox{and}}
       \rhotf(y)&\geq& 3^52^{-3}\pi |y|^{-6}-\const
       Z^{-\zeta/3}|y|^{-6-\zeta}.
\end{eqnarray*}
Since $r\geq \beta_0 Z^{-1/3}$
we conclude that for all $|y|>r$
\begin{equation}\label{eq:|rhorhotf|1}
   \left|\rhotf(y)-\rhotfr(y)\right|
   \leq \const r^{\zeta}|y|^{-6-\zeta}.
\end{equation}
Thus
$$
|{\cal A}_2|\leq\const  \int_{|y|>|x|}
r^{\zeta}|y|^{-7-\zeta}\,dy
$$
which gives the bound in (\ref{eq:A2}).
\enddemo

We turn now to estimating ${\cal A}_3$.
This requires estimating the difference between
$\rhotfr$ and $\rhohf$. 
We again begin by estimating this difference in the Coulomb
norm. More precisely, we estimate in Coulomb norm the difference
between the ``outside'' TF density $\rhotfr$ 
and the ``outside'' HF density $\chiplus{r}\rhohfr$. 
This is done through a semiclassical analysis of the 
exterior region $\{|x|>r\}$.

 \proclaimtitle{Coulomb norm comparison of HF and OTF}
\proclaim{Lemma}\label{lm:cnhfotf} \hskip-8pt
Assume $N\!\geq\! Z$. Given constants $\varepsilon',{\sigma}>0$  
there exists a constant  $D>0$ depending only on 
$\varepsilon',{\sigma}$
such that for all $r$ with $\beta_0Z^{-1/3}\leq r\leq D$
for which 
{\rm (\ref{eq:phiasp})} holds for all $|x|\leq r$
we have
\begin{equation}\label{eq:cnhfotf}
   \|\rhotfr-\chiplus{r}\rhohf\|_{\rC}\leq \const
   r^{-\frac{7}{2}+\frac{1}{6}}
\end{equation}
and
\begin{equation}\label{eq:me5/3hf}
 \int\left(\chiplus{r}\rhohf\right)^{5/3}\leq \const r^{-7}.
\end{equation}
\endproclaim

{\it Proof}.
 Let $\gamma$ be the density matrix on 
$L^2(\R^3;\C^2)$, but diagonal in spin, constructed in 
the semiclassical approximation
Lemma~\ref{lm:sc} for the potential $V=\phitfr$.
Note that from Lemma~\ref{lm:sommerfeldotf} we have $\phitfr(y)\geq0$ for
$|y|\geq r$ and from its definition $\phitfr(y)\leq 0$ for $|y|<r$. 
{F}rom Lemma~\ref{lm:sc} we have that
\begin{equation}\label{eq:rhogammaotf}
  \uprho_\gamma=2^{3/2}(3\pi^2)^{-1}\left[\phitfr\right]_+^{3/2}*g^2=\rhotfr*g^2,
\end{equation}
where we have used
the TF equation (\ref{eq:tfeqgeneral}) and the fact $\mutfr=0$ proved
in Lemma~\ref{lm:sommerfeldotf}. Here  
$g$ was given in Definition~\ref{defi:g}. {F}rom
Lemma~\ref{lm:sc} we also have
\begin{eqnarray}
 \Tr\left[-\mfr{1}{2}\Delta\gamma\right]&=& \frac{2^{3/2}}{5\pi^2}
 \int\left[\phitfr\right]_+^{5/2}
 +\frac{2^{1/2}}{3}s^{-2}\int 
        \left[\phitfr\right]_+^{3/2} \label{eq:deltagammaotf}
\\
 &\leq&\mfr{3}{10}(3\pi^2)^{2/3}\int(\rhotfr)^{5/3}+\const
 s^{-2}r^{-3}.\nonumber
\end{eqnarray}
(Note the factor of 2 in the formulas above 
compared to Lemma~\ref{lm:sc}. This is due to the fact that there was
no spin in Lemma~\ref{lm:sc}.) The last inequality above follows from 
the Sommerfeld estimate for OTF given in Lemma~\ref{lm:sommerfeldotf}.

According to Lemma~\ref{lm:DBC} we may, for all
$0<\lambda'<1$,  choose a density matrix 
$\widetilde\gamma$ such that its density
$\uprho_{\widetilde\gamma}$ has support in 
$\{|x|\geq r\}$ and such that 
$\uprho_{\widetilde\gamma}\leq\uprho_\gamma$ and
\begin{eqnarray*}
  \Tr\left[\left(-\mfr{1}{2}\Delta-\V{r}\right)\widetilde\gamma\right]
  &\leq&\Tr\left[\left(-\mfr{1}{2}\Delta-\V{r}\right)\gamma\right]
  +\const\int_{|x|\leq(1-\lambda')^{-1}r} \left[\V{r}\right]_+^{5/2}\\
  &&{}
  +\const(\lambda' r)^{-2}\int_{|x|\leq(1-\lambda')^{-1}r}\uprho_\gamma.
\end{eqnarray*}
If we make the assumption that $D$ is chosen to ensure that 
$\sigma r^{\epsilon'}\leq1$ we may use the estimate on $\V{r}$ 
from Lemma~\ref{lm:tfotfprelims} and 
use the same lemma to conclude that $\rho_\gamma(y)\leq \const r^{-6}$ for all $y$.
If we recall that $\V{r}$ has support for $|x|\geq r$ we get that
\begin{eqnarray*}
\Tr\left[\left(-\mfr{1}{2}\Delta-\V{r}\right)\widetilde\gamma\right]
  &\leq&\Tr\left[\left(-\mfr{1}{2}\Delta-\V{r}\right)\gamma\right]
  +\const\left(\lambda' r^{-7}+\lambda'^{-2}r^{-5}\right)\\
  &=&\Tr\left[\left(-\mfr{1}{2}\Delta-\V{r}\right)\gamma\right]
  +\const r^{-7+2/3},
\end{eqnarray*}
where we have made the choice $\lambda'=r^{2/3}$ and assumed 
that $D$ is such that $\lambda'<1/2$.

Since $\int\uprho_{\widetilde\gamma}\leq \int\uprho_\gamma=
\int\rhotfr\leq\int\chiplus{r}\rhohf$ we see from 
Theorem~\ref{thm:outsideenergy} that in terms of the 
auxiliary functional $\ea{}$ defined in (\ref{eq:eadef})
\begin{eqnarray*}
   \ea\left[{\gammar}\right]\leq\ea\left[\widetilde{\gamma}\right]+
        {\cal R}
        &\leq&\Tr\left[\left(-\mfr{1}{2}\Delta-\V{r}\right)\gamma\right]
  +\const r^{-7+2/3}\\&&{}+
  \mfr{1}{2}\iint\uprho_\gamma(x)|x-y|^{-1}\uprho_{\gamma}(y)dx\,dy+
        {\cal R},
\end{eqnarray*}
where $\gammar$ and ${\cal R}$ were defined in (\ref{eq:gammar}) and
(\ref{eq:ioerror}) respectively in terms
of a parameter $0<\lambda<1$ (different from the $\lambda'$ used
above).
We have here used that 
$\Phihf{r}\uprho_{\widetilde{\gamma}}=\V{r}\uprho_{\widetilde{\gamma}}$,
since $\uprho_{\widetilde{\gamma}}$ has support in 
$\{|x|\geq r\}$, and that 
$$
  \mfr{1}{2}\iint\uprho_{\widetilde{\gamma}}(x)|x-y|^{-1}
  \uprho_{\widetilde{\gamma}}(y)dx\,dy
  \leq \mfr{1}{2}\iint\uprho_{{\gamma}}(x)|x-y|^{-1}
  \uprho_{{\gamma}}(y)dx\,dy
$$
since 
$
\uprho_{\widetilde{\gamma}}\leq \uprho_{{\gamma}}
$.
     
Since $|x|^{-1}$ is superharmonic we have
$$
   \iint g(x-z)^2|z-w|^{-1}g(y-w)^2\,dz\,dw\leq |x-y|^{-1}
$$ 
and we conclude that 
\begin{eqnarray*}
   \ea\left[{\gammar}\right]
        &\leq&\Tr\left[\left(-\mfr{1}{2}\Delta-\V{r}\right)\gamma\right]
  +\const r^{-7+2/3}\\&&{}+
  \mfr{1}{2}\iint\rhotfr(x)|x-y|^{-1}\rhotfr(y)dx\,dy+
        {\cal R}.
\end{eqnarray*}

{F}rom (\ref{eq:rhogammaotf}) and (\ref{eq:deltagammaotf}) we find
that
\begin{eqnarray}\quad
   \ea\left[{\gammar}\right]
   &\nhs\leq\nhs&\etfr(\rhotfr)+ \int \V{r}\left(\rhotfr-\rhotfr*g^2\right)
   +\const s^{-2}r^{-3} \label{eq:calrs}\\
        &\nhs\nhs&{}+\const r^{-7+2/3}+
        {\cal R}.\nonumber
\end{eqnarray}
We have  
$$
   \int \V{r}\left(\rhotfr-\rhotfr*g^2\right)=
   \int \left(\V{r}- \V{r}*g^2\right)\rhotfr.
$$
Now since $\V{r}(y)$ is harmonic for $|y|>r$ 
we conclude that $\V{r}*g^2(y)=\V{r}(y)$ for $|y|>r+s$.
Hence we get from 
Lemma~\ref{lm:tfotfprelims} that 
\begin{equation}\label{eq:vrs}
  \int \V{r}\left(\rhotfr-\rhotfr*g^2\right)\leq \const r^{-4}\int_{|y|<r+s}
  \rhotfr\leq \const r^{-8}s.\quad
\end{equation}
We insert this into 
(\ref{eq:calrs}) and arrive at 
\begin{eqnarray}
&&\label{eq:meeaupper1}\\
   \ea\left[{\gammar}\right]&\leq&
   \etfr(\rhotfr)
   +\const( s^{-2}r^{-3}+ r^{-8}s)
        +\const r^{-7+2/3}+{\cal R}\nonumber\\
        &=&\etfr(\rhotfr)+\const r^{-7+2/3}+{\cal R},\nonumber
\end{eqnarray}
with the choice $s=r^{5/3}$. 

We shall now estimate  $\cal R$. 
We shall choose the $\lambda$ used to define $\gammar$ and 
$\cal R$  in such a way that $\lambda\leq 1/2$.
We then see that 
the constant $C_{\lambda}(r)$ in Theorem~\ref{thm:outsideenergy}
satisfies $C_{\lambda}(r)\leq \const (\lambda r)^{-2}$,
since $\lambda\leq 1/2$ and $r\leq 1$. 
We see from Lemma~\ref{lm:meN} that
\begin{equation}\label{eq:l1rhohfr}
  \int_{|y|>(1-\lambda)r}\rhohf\leq \const
  r^{-3},
\end{equation}
where we have used that $r, {\sigma}r^{\epsilon'}\leq1$.

Moreover, from Lemma~\ref{lm:tfotfprelims} with $r$ replaced by 
$(1-\lambda)r$ we have 
$$
   \int\limits_{(1-\lambda)r<|y|<(1-\lambda)^{-1}r}
   \left[\Phihf{(1-\lambda)r}(y)\right]_+^{5/2}dy\leq \const r^{-7}\lambda.
$$
Hence we have 
$$
   {\cal R}\leq\const \lambda^{-2} r^{-5} +
   \const r^{-7}\lambda+\Ex\left[\gammar\right].
$$
If we now use the exchange 
inequality in Theorem~\ref{thm:exchineq} and (\ref{eq:l1rhohfr}) we get
(recall that $\rhohfr=\thetar^2\rhohf$ is the density 
corresponding to $\gammar$)
\begin{eqnarray*}
    \Ex\left[\gammar\right] &\leq&
    \const\int\left(\rhohfr\right)^{4/3}\leq 
    \const\left(\int\rhohfr\right)^{1/2}
    \left(\int\left(\rhohfr\right)^{5/3}\right)^{1/2}\\
    &\leq& \const r^{-3/2}
    \left({\cal R}+r^{-7} \right)^{1/2},
\end{eqnarray*}
where we have also used that according to 
(\ref{eq:l5/3rhohfr}) and Lemma~\ref{lm:tfotfprelims} we have
\begin{equation}\label{eq:me5/3rhohfr}
  \int\left(\rhohfr\right)^{5/3}\leq \const {\cal R}+\const r^{-7}. 
\end{equation} 
We may therefore conclude that
\begin{equation}\label{eq:Restimate}
    {\cal R}\leq \const r^{-7}(r^2\lambda^{-2}+\lambda)+\const r^{-5}.
\end{equation}
We may use (\ref{eq:me5/3rhohfr}) and (\ref{eq:Restimate}) to prove 
(\ref{eq:me5/3hf}). Recall that $\rhohfr(y)=\rhohf(y)$ if 
$|y|>(1-\lambda)^{-1}r$. Now (\ref{eq:me5/3hf}) follows if we simply observe that 
(\ref{eq:me5/3rhohfr}) and (\ref{eq:Restimate})
hold  with $r$ replaced by $r/2$ and $\lambda=1/2$. We shall 
make a possible different choice of $\lambda$  \pagebreak below.

We shall now prove a lower bound on $\ea\left[{\gammar}\right]$.
We write 
\begin{eqnarray*}
  \ea\left[\gammar\right]&=&
  \Tr\left[\left(-\mfr{1}{2}\Delta-\Phihf{r}\right)\gammar\right]
  +\mfr{1}{2}\iint\rhohfr(x)|x-y|^{-1}\rhohfr(y)dx\,dy\\
   &=&\Tr\left[\left(-\mfr{1}{2}\Delta-\Phihf{r}
       +\rhotfr*|x|^{-1}\right)\gammar\right]
   +\|\rhohfr-\rhotfr\|_{\rC}^2\\{}&&
   -\mfr{1}{2}\iint\rhotfr(x)|x-y|^{-1}\rhotfr(y)dx\,dy.
\end{eqnarray*}
 If we use that on the support of $\rhohfr$ we have $\Phihf{r}=\V{r}$ 
we may write this as 
\begin{eqnarray}
  \ea\left[\gammar\right]&=&\Tr\left[\left(-\mfr{1}{2}\Delta-\phitfr\right)
    \gammar\right]
  +\|\rhohfr-\rhotfr\|_{\rC}^2\label{eq:meealower}
\\{}&&
  -\mfr{1}{2}\iint\rhotfr(x)|x-y|^{-1}\rhotfr(y)\,dx\,dy.\nonumber
\end{eqnarray}
The trace may be bounded below by the sum of the first 
$N'$ negative eigenvalues
of the operator $-\mfr{1}{2}\Delta-\phitfr $, 
where $N'$ is the smallest integer larger than $\Tr[\gammar]=\int\rhohfr$.
{F}rom Lemma~\ref{lm:sc} (again with an extra factor of $2$ 
due to spin)
we therefore have that for all $s>0$ and all $0<\delta<1$ 
\begin{eqnarray}\qquad
    \Tr\left[\left(-\mfr{1}{2}\Delta-\phitfr
       \right)\gammar\right]
  &\geq& -\frac{2^{5/2}}{(15\pi^2)}(1-\delta)^{-3/2}
  \int \left(\phitfr\right)_+^{5/2}\label{eq:hfotfsc}\\
        &&{}-\pi^2s^{-2}\left(\int\rhohfr+1\right)\nonumber\\ &&{}
        -2L_1\delta^{-3/2}\left\|\left[\phitfr
        -\phitfr*g^2\right]_+\right\|_{5/2}^{5/2}.\nonumber
\end{eqnarray}
We first estimate the last term. Since $\rhotfr*|x|^{-1}$ is
superharmonic we have by the mean value property
that $\rhotfr*|x|^{-1}\geq \rhotfr*|x|^{-1}*g^2$. Thus we have 
$$
  \phitfr-\phitfr*g^2 =\V{r}-\V{r}*g^2+\rhotfr*|x|^{-1}*g^2-
  \rhotfr*|x|^{-1}\leq\V{r}-\V{r}*g^2.
$$
The same argument which led to (\ref{eq:vrs})  gives
that $\V{r}(y)-\V{r}*g^2(y)=0$ unless
$r-s\leq |y|\leq r+s$. Since by Lemma~\ref{lm:tfotfprelims} we have 
$|\V{r}(y)|\leq \const r^{-4}$ (recall that $\V{r}$ is supported on
$\{|y|\geq r\}$) we obtain
$$
   \left\|\left[\phitfr
        -\phitfr*g^2\right]_+\right\|_{5/2}^{5/2}
    \leq \const r^{-8}s,
$$
if we assume that $s\leq r$. 
(Note that $s$ here does not have to be chosen as in the upper bound).

{F}rom (\ref{eq:l1rhohfr}) we also get that $\int\rhohfr
\leq\int\rhohf\chiplus{r}\leq \const r^{-3}$.

Finally from the Sommerfeld estimate (\ref{eq:phitfrup}) and
the fact that $\phitfr(x)$ is positive only if
$\V{r}(x)>0$, i.e., only if $|x|\geq r$, we find
that 
$$
   \int \left(\phitfr\right)_+^{5/2}\leq \const r^{-7}.
$$
We therefore see from (\ref{eq:hfotfsc}) 
and the TF equation (\ref{eq:tfeqgeneral}) (recall that $\mutfr=0$, 
by Lemma~\ref{lm:sommerfeldotf}) that 
if $0<\delta<1/2$ then 
\begin{eqnarray*}
  &&\hskip-.75in   \Tr\left[\left(-\mfr{1}{2}\Delta-\phitfr
       \right)\gammar\right] \\
  &\geq&\mfr{3}{10}(3\pi^2)^{2/3}\int\left(\rhotfr\right)^{5/3}
    -\int\phitfr\rhotfr\\{}&&
    -\const\left(\delta r^{-7}+\delta^{-3/2}r^{-8}s\right)
    -\const s^{-2}r^{-3}\\
    &=&\mfr{3}{10}(3\pi^2)^{2/3}\int\left(\rhotfr\right)^{5/3}
    -\int\phitfr\rhotfr -\const 
    r^{-7+\frac{1}{3}},
\end{eqnarray*}
where we have chosen $\delta=r^{-2/5}s^{2/5}$ and $s=r^{11/6}$, 
which agrees with $s\leq r$.

If we insert this last estimate into (\ref{eq:meealower})
we obtain
$$
  \ea\left[\gammar\right]\geq \etfr\left(\rhotfr\right)
  +\|\rhohfr-\rhotfr\|_{\rC}^2-\const r^{-7+\frac{1}{3}}.
$$
If we compare this with (\ref{eq:meeaupper1}) we see that 
$$
   \|\rhohfr-\rhotfr\|_{\rC}^2\leq\const r^{-7+\frac{1}{3}}
  +{\cal R}.
$$
Finally, we use the Hardy-Littlewood-Sobolev
inequality (\ref{eq:HLS}) and (\ref{eq:me5/3hf}) to conclude that 
\begin{eqnarray*}
    \lefteqn{\|\chiplus{r}\rhohf-\rhohfr\|_{\rC}}&&\\ &\leq& 
    \const\|\chiplus{r}\rhohf-\rhohfr\|_{6/5}
    \leq \const\left(\int_{r<|y|<(1-\lambda)^{-1}r} 
      \rhohf(y)^{6/5}dy\right)^{5/6}\\
    &\leq& \const\left(\int_{r<|y|} 
      \rhohf(y)^{5/3}dy\right)^{3/5}
    \left(\int_{r<|y|<(1-\lambda)^{-1}r} 1dy\right)^{7/30}\\
    &\leq&\const \lambda^{7/30} r^{-7/2}.
\end{eqnarray*}
We thus get from (\ref{eq:Restimate}) that
\begin{eqnarray*}
   \lefteqn{\|\chiplus{r}\rhohf-\rhotfr\|_{\rC}\leq
   \|\chiplus{r}\rhohf-\rhohfr\|_{\rC}+\|\rhohfr-\rhotfr\|_{\rC}}&&
   \\ &\leq&
   \const r^{-\frac{7}{2}+\frac{1}{6}}+
   \const r^{-7/2}(r\lambda^{-1}+\lambda^{1/2}+\lambda^{7/30})
\end{eqnarray*}
which gives (\ref{eq:cnhfotf}) if we choose $\lambda=\min\{1/2, r^{5/7}\}$.
\hfill\qed\vglue12pt

We may now estimate ${\cal A}_3$ defined in (\ref{eq:A3defi}).

 \proclaimtitle{Controlling ${\cal A}_3$}
\proclaim{Lemma}\label{lm:A3}\hskip-8pt
Assume $N\geq Z$. Given constants $\varepsilon',{\sigma}\!>\!0$ 
there exists a constant  $D>0$ depending only on 
$\varepsilon',{\sigma}$
such that for all $r$ with $\beta_0Z^{-1/3}\leq r\leq D$
for which 
{\rm (\ref{eq:phiasp})} holds for all $|x|\leq r$
we have  for all $|x|\geq r$ that
\begin{equation}\label{eq:A3}
         \left|{\cal A}_3(r,x)\right|
        \leq \ \const(|x|/r)^{1/12}r^{-4+\frac{1}{36}}.
\end{equation}
\endproclaim 

\demo{Proof}
We shall use the Coulomb norm estimate (\ref{eq:cn2})
with $f=\rhotfr-\chiplus{r}\rhohf$.
We then immediately see from  
Lemma~\ref{lm:cnhfotf}, and the fact, 
which follows from Lemma~\ref{lm:sommerfeldotf} and the 
TF equation (\ref{eq:tfeqgeneral}),  that 
$\rhotfr(y)\leq \const|y|^{-6}$ , that 
$$
    \left|{\cal A}_3(r,x)\right|\leq \const
    (\kappa |x|)^{1/5}r^{-21/5}+
    \const\kappa^{-1}|x|^{-1/2}r^{-\frac{7}{2}+\frac{1}{6}}.
$$
This gives (\ref{eq:A3}) if we choose 
$
   \kappa= \const
   (r/|x|)^{7/12}r^{\frac{5}{36}}.
$
\enddemo

\demo{End of proof of Lemma~{\rm \ref{lm:me5/3}}}
For $|x|\geq \beta_0 Z^{-1/3}$ the estimate in 
(\ref{eq:me5/3}) follows from  (\ref{eq:me5/3hf}). 
For $|x|\leq \beta_0 Z^{-1/3}$ we get from (\ref{eq:5/3}) that
\vglue12pt
\hfill ${\displaystyle
\int_{|y|>|x|}\rhohf(y)^{5/3}\,dy\leq \int_{\R^3}\rhohf(y)^{5/3}\,dy
\leq\const Z^{7/3}\leq \const |x|^{-7}.
}$\hfill\qed
\enddemo
\vglue12pt
{\it End of proof of the iterative Lemma~{\rm \ref{lm:iterate}}}.
Let $D>0$ depending on ${\sigma},\epsilon'$ be the 
smaller of the values $D$ occurring in Lemmas~\ref{lm:A1A2} and
\ref{lm:A3}. 
We may without loss of generality assume that $D\leq1$. 

Given $\delta>0$.
We consider $R_0<D$ satisfying $\beta_0Z^{-1/3}\leq R_0^{1+\delta}$
and we assume that (\ref{eq:phiasp}) holds for all $|x|\leq R_0$.

Set $R_0'=R_0^{1-\delta}$ and $r=R_0^{1+\delta}$.
Then we have $\beta_0Z^{-1/3}\leq r\leq R_0<D$ and we can therefore apply
Lemmas~\ref{lm:A1A2} and \ref{lm:A3}. Moreover $R'_0>R_0$.
In order to prove (\ref{eq:itphi}) for $R_0<|x|<R'_0$
we use (\ref{eq:AAA}) and Lemmas~\ref{lm:A1A2} and \ref{lm:A3}.
We obtain that for all $|x|\geq r$ 
$$
 \left|\Phihf{|x|}(x)-\Phitf{|x|}(x)\right|
 \leq\const 
 \left(r/|x|\right)^\zeta|x|^{-4}+\const \left(|x|/r\right)^{1/12}r^{-4+\frac{1}{36}}.
$$
Moreover, for all $R_0<|x|<R_0'$ we have 
$$
|x|^{\frac{2\delta}{1-\delta}}\leq
\frac{r}{|x|}\leq |x|^\delta
$$
\vglue-9pt
\noindent and thus
$$
\left|\Phihf{|x|}(x)-\Phitf{|x|}(x)\right|
\leq\const \left( |x|^{-4+\zeta\delta}
 +|x|^{-4+\frac{1}{36}-
     \frac{73\delta}{9(1-\delta)}}\right).
$$
It follows that if $\delta$ is small enough there exist 
$\epsilon,C'_\Phi>0$ such that (\ref{eq:itphi}) is satisfied. \hfill\qed

\section{Proving the main results Theorems~\ref{thm:potentialestimate}, 
\ref{thm:radius}, \ref{thm:maxN}, and \ref{thm:ie}}
\label{sec:provingmain}
\advance\eqcount by 139

The main result Theorem ~\ref{thm:maxN} on the maximal number 
of electrons $N$ is a simple consequence of Lemma~\ref{lm:meN} and 
Theorem~\ref{thm:me}.  

\demo{Proof of Theorem~{\rm \ref{thm:maxN}}}
We may of course assume that $N\geq Z$ and that $Z\geq 1$
(otherwise the result follows from Lieb's bound
Theorem~\ref{thm:2z+1}).  Then $\int\rhotf=Z$. 
We can then use Lemma~\ref{lm:meN} with
$R$ chosen so small that $\CM\leq \CPhi R^{-4+\epsilon}$,
because then (\ref{eq:phiasp}) holds with $\sigma=2\CPhi$ and
$\epsilon'=\epsilon$.
We conclude from Lemma~\ref{lm:meN} that 
$$
\int\rhohf\leq \int_{|x|<R}\rhotf(x)\,dx+
\sigma R^{-3+\epsilon'}+\const(1+\sigma R^{\epsilon'})(1+R^{-3})
\leq Z+\const,
$$
since now $R,\sigma$, and $\epsilon'$ are universal constants. 
We have thus  
concluded the result of Theorem~\ref{thm:maxN}.
\enddemo

The asymptotics of the radius of an infinite atom 
given in Theorem~\ref{thm:radius} is a
simple consequence of the main estimate Theorem~\ref{thm:me}
and the Sommerfeld asymptotics.

\demo{Proof  of Theorem~{\rm \ref{thm:radius}}}
\label{proof:radius}
Note that in the neutral case $N=Z$ we have from the main 
estimate Theorem~\ref{thm:me} that  
\begin{eqnarray*}
 \left|\int_{|x|>R}\rhotf(x)-\rhohf(x)\,dx\right|&=&
 \left|\int_{|x|<R}\rhohf(x)-\rhotf(x)\,dx\right|\\ &=&
 \left|(4\pi)^{-1}R\int_{{\cal S}^2} \Phitf{R}(R\omega)
 -\Phihf{R}(R\omega)\,d\omega\right|\\
 &\leq& \CPhi R^{-3+\varepsilon}+\CM R.
\end{eqnarray*}
Theorem~\ref{thm:radius} now easily follows from TF equation
(\ref{eq:tfsystemrho}) and the Sommerfeld laws
Theorem~\ref{thm:sommerfeldup} and \ref{thm:sommerfeldlow} for the
case $N=Z$, i.e., $\mutf=0$.
\enddemo

The potential estimate in Theorem~\ref{thm:potentialestimate}
is somewhat more difficult to prove.

\demo{ Proof of Theorem~{\rm \ref{thm:potentialestimate}}}
As in the proof of the main estimate Theorem~\ref{thm:me}
we separately treat small $|x|$, intermediate $|x|$, 
and large $|x|$.

We first consider small $|x|$. Note that 
$$
 \phihf(x)-\phitf(x)
 =\int\left(\rhotf(y)-\rhohf(y)\right)|x-y|^{-1} dy.
$$
Thus using the Coulomb norm estimate (\ref{eq:cn1}) we obtain
\begin{eqnarray*}
  \lefteqn{\left|\int\left(\rhotf(y)-\rhohf(y)\right)
      |x-y|^{-1} dy\right|}&&\\
  &\leq& \const s^{1/5}
  \max\left\{\|\rhotf\|_{L^{5/3}(\B{x,s})},
  \|\rhohf\|_{L^{5/3}(\B{x,s})}\right\}
  \\{}&&+
  \const s^{-1/2}\|\left(\rhotf-\rhohf\right)\|_\rC.
\end{eqnarray*}
If we use Lemma~\ref{lm:5/3cn} and optimize in $s$ we arrive at 
\begin{equation}\label{eq:smphihfphitf}
 \left|\phihf(x)-\phitf(x)\right|\leq
 \const Z^{4/3-\epsilon_3}\leq 
 \const |x|^{-4+3\epsilon_3-4\delta+3\delta\epsilon_3}\qquad
\end{equation}
for some universal $\epsilon_3$, if $|x|^{1+\delta}<\beta_0 Z^{-1/3}$. 

We now turn to intermediate $|x|$.
We shall choose a $D>0$ such that, with the 
notation of Theorem~\ref{thm:me}, we have 
$\CM\leq \CPhi D^{-4+\epsilon}$.
Then by Theorem~\ref{thm:me} we have that (\ref{eq:phiasp}) holds 
for all $|x|\leq D$ with $\sigma=2\CPhi$ and $\epsilon'=\epsilon$. 
We may now assume that $D\leq 1$ and that $D$ is smaller than the  values for $D$
in Lemmas~\ref{lm:A1A2} and \ref{lm:cnhfotf} corresponding 
to the above choices of $\sigma$ and $\epsilon'$. 

Consider 
$$(\beta_0 Z^{-1/3})^{\frac{1}{1+\delta}}\leq
|x|<D^{\frac{1}{1+\delta}}.
$$
Set
$r=|x|^{1+\delta}$. Then $|x|\geq r$ and $\beta_0 Z^{-1/3}\leq r\leq D$. 
We shall use the notation from Section~\ref{sec:iteration}.
We write 
$$
 \phihf(x)-\phitf(x)=\phihf(x)-\phitfr(x)+\phitfr(x)-\phitf(x).
$$
The difference between the last two terms was defined in 
(\ref{eq:A1defi}) to be ${\cal A}_1(r,x)$ and this was estimated in
Lemma~\ref{lm:A1A2}.
 
We have
$$
 \phihf(x)-\phitfr(x)=\int\left(\rhotfr(y)-\chiplus{r}(y)\rhohf(y)\right)
 |x-y|^{-1}dy.
$$ 
Exactly as above, for small $|x|$, we now use 
Theorem~\ref{thm:me}, Lemma~\ref{lm:cnhfotf} and the Coulomb norm
estimate (\ref{eq:cn1}) to conclude that 
$$
 \left|\phihf(x)-\phitfr(x)\right|\leq 
 \const r^{-4+\frac{1}{21}}\leq \const |x|^{\left(-4+\frac{1}{21}\right)(1+\delta)},
$$ 
If we combine this with (\ref{eq:A1}) from
Lemma~\ref{lm:A1A2} we obtain
\begin{equation}\label{eq:intphihfphitf}
  \left|\phihf(x)-\phitf(x)\right|
  \leq\const\left(|x|^{-4+\zeta\delta}
    +|x|^{\left(-4+\frac{1}{21}\right)(1+\delta)}\right).\hskip.5in
\end{equation}
Combining (\ref{eq:smphihfphitf})
and (\ref{eq:intphihfphitf}) we see that by choosing $\delta$ small
enough we have proved (\ref{eq:mainresult})
for all $|x|\leq D^{\frac{1}{1+\delta}}$.

We turn to $|x|\geq D^{\frac{1}{1+\delta}}$, i.e., $|x|$ greater than
some universal constant. 
Here we may write 
$$
 \left|\phihf(x)-\phitf(x)\right|
 \leq \left|\Phihf{|x|}(x)-\Phitf{|x|}(x)\right|
 +\left|\int_{|y|>|x|}\left(\rhotf(y)-\rhohf(y)\right)|x-y|^{-1} dy\right|.
$$
The first term is controlled by the main estimate
Theorem~\ref{thm:me}. 
If we use that according to Lemma~\ref{lm:me5/3} we have 
$\int_{|y|>|x|}\rhohf(y)^{5/3}dy\leq \const$ and that the 
same estimate holds for the TF density (see
Lemma~\ref{lm:tfotfprelims}) we may
estimate the second term above as follows. Using 
H\"older's inequality we have 
\begin{eqnarray*}
 \left|\,\,\int\limits_{|x|<|y|}\!\!\!\!\!
   \left(\rhotf(y)-\rhohf(y)\right)|x-y|^{-1}dy\right|
 &\nhs\leq\nhs& \const 
 \left(\int_{|x-y|<1}\!\!\!\!\!|x-y|^{-5/2}dy\right)^{2/5}\\
 &\nhs\nhs&{}+\int_{|x|<|y|}\left(\rhotf(y)+\rhohf(y)\right)dy
 \leq \const,
\end{eqnarray*}
where the last estimate follows from 
Lemma~\ref{lm:meN}.
\enddemo

We  end the paper by giving the proof of the 
bound on the ionization energy in Theorem~\ref{thm:ie}.

\demo{Proof of Theorem~{\rm \ref{thm:ie}}}
\label{proof:ie}
Since the HF energy is a nonincreasing function of $N$ we have
that $0\leq E^\HF(Z-1,Z)-E^{\HF}(Z,Z)$.
In order to prove an upper bound we shall construct a trial density
matrix $\gamma$ for $\ehf$ with $\Tr[\gamma]\leq Z-1$.
We then clearly have that $\ehf\left[\gamma\right]\geq E^\HF(Z-1,Z)$. 
Let $\theta_-$ be given in terms of appropriate $r,\lambda>0$  as 
in the beginning of the proof of Theorem~\ref{thm:outsideenergy}.
We then choose as our trial matrix
$$
        \gammai=\theta_-\gammahf\theta_-,
$$
where $\gammahf$ is the HF
minimizer with $\Tr\left[\gammahf\right]=Z$. 
According to the definition of $\theta_-$ we have
$$
 \Tr\left[\gammai\right]\leq Z-\int_{|y|>r}\rhohf(y)dy.
$$
We choose  $\lambda=1/2$.
Let $R>0$ be such that $\CM=\CPhi R^{-4+\varepsilon}$. 
We shall now choose $r$ satisfying $r\leq R$. 
Then according to Theorem~\ref{thm:me} we have that 
(\ref{eq:phiasp}) holds for $|x|\leq r$ with $\sigma=2\CPhi$ and
$\epsilon'=\epsilon$. {F}rom Lemma~\ref{lm:meN} we therefore conclude
that 
\begin{eqnarray*}
 \int_{|y|>r}\rhohf(y)dy&=&\int\rhohf-\int_{|y|<r}\rhohf(y)dy\\
 &=&\int_{|y|<r}\rhotf(y)-\rhohf(y)\, dy+\int_{|y|>r}\rhotf(y)\,dy\\
 &\geq&\int_{|y|>r}\rhotf(y)\,dy-\const r^{-3+\epsilon}
\end{eqnarray*}
where we have used that $\int\rhohf=\int\rhotf=Z$.

We may of course assume that $Z$ is larger than some fixed universal 
constant. For $Z$ less than a universal constant, 
the total energy $E^{\HF}(Z,Z)$ and hence the ionization energy 
$E^{\HF}(Z-1,Z)-E^{\HF}(Z,Z)$ are bounded by universal constants
(see Theorem~\ref{thm:HFenergy}). We can therefore assume that 
$\beta_0Z^{-1/3}< R$ and we shall choose $\beta_0Z^{-1/3}<r$. 
It then follows from
Theorem~\ref{thm:sommerfeldlow}, the TF equation
(\ref{eq:tfsystemrho}) (recall that we consider the case $\mutf=0$ and
$N=Z$)
that $\rhotf(y)\geq\const |y|^{-6}$ for all
$|y|\geq r$. Hence

\begin{eqnarray*}
 \int_{|y|>r}\rhohf(y)dy&\geq& \const r^{-3} -\const
 r^{-3+\epsilon}.
\end{eqnarray*}
Thus we may choose $r$ to be a small enough universal number    
(assuming that $Z$ is large enough to allow $\beta_0Z^{-1/3}<r$)
to ensure that
$\int_{|y|>r}\rhohf(y)dy\geq1$ and hence 
that $ \Tr\left[\gammai\right]\leq Z-1$.

{F}rom the estimate (\ref{eq:outsideenergy1}) in the proof
of Theorem~\ref{thm:outsideenergy} we have
$$
 \ehf\left[\gammai\right]\leq\ehf\left[\gammahf\right]-
 \ea\left[\gammar\right]+{\cal R}=E^{\HF}(Z,Z)-
 \ea\left[\gammar\right]+{\cal R}
$$
where as before $\gammar=\thetar\gammahf\thetar$, 
$\cal R$ is given in (\ref{eq:ioerror}), and 
the functional $\ea$ was defined in (\ref{eq:eadef}).

It remains to prove that 
\begin{equation}\label{eq:gammarR}
 -\ea\left[\gammar\right]+{\cal R}\leq
 \const.
\end{equation}

As in (\ref{eq:Restimate}) we conclude that 
$
 {\cal R}\leq \const r^{-7}\leq\const
$, where we have used that $r$ is a universal constant.

In order to estimate $\ea\left[\gammar\right]$ 
we note that, since $\Phihf{r}(y)$  is harmonic for $|y|>r$ 
and tends to 0 at infinity, we have that for all $|y|\geq r$ 
\begin{eqnarray*}
 \Phihf{r}(y)\leq |y|^{-1}r\sup_{|x|=r}\Phihf{r}(x)
 &\nhs\leq\nhs& |y|^{-1}r\sup_{|x|=r}|\Phitf{r}(x)|
  + |y|^{-1}\left(\CPhi r^{-3+\varepsilon}+r\CM\right) \\
 &\nhs\leq\nhs& \const r^{-3} |y|^{-1},
\end{eqnarray*}
where we have used the main estimate Theorem~\ref{thm:me}
and the bound  on $\Phitf{r}$ in
Lemma~\ref{lm:Phitfrbound} with $\mutf=0$. 
If we use that $r$ is some universal constant we get 
$$
 \Phihf{r}(y)\leq \const |y|^{-1}.
$$
Using the Lieb-Thirring inequality (\ref{eq:LTdensity})
we see from the definition (\ref{eq:eadef}) of the auxiliary 
functional $\ea$ that  
\begin{eqnarray*}
  \ea\left[\gammar\right]&\geq&
  K_1\int\rhohfr(y)^{5/3}dy
  -\const \int\frac{\rhohfr(y)}{|y|}dy\\&&{}
  +\mfr{1}{2}\iint
  \rhohfr(x)|x-y|^{-1}\rhohfr(y)dx\,dy.
\end{eqnarray*}
Here again $\rhohfr=\theta_r^2\rhohf$ is the density corresponding to 
$\gammar$.
It follows from standard atomic TF theory 
that 
$$
  \inf_{\rho\geq 0}\left\{ K_1\int\rho(y)^{5/3}dy
  -\const \int\frac{\rho(y)}{|y|}dy
  +\mfr{1}{2}\iint
  \rho(x)|x-y|^{-1}\rho(y)dx\,dy\right\}
$$
is some universal constant. Hence 
$
 \ea\left[\gammar\right]\geq -\const
$
and we have proved (\ref{eq:gammarR}).
\enddemo

\end{document}